\documentclass[12pt]{article}

\pdfoutput=1 

\usepackage[T1]{fontenc} 

\usepackage{CJK}
\usepackage{mathtools}
\usepackage{tikz}
\usepackage{amssymb}
\usepackage{epsfig}
\usepackage{bbm}
\usepackage{bbold}
\usepackage{tabularx,booktabs}
\usepackage{graphicx}
\usepackage{hyperref}
\usepackage{romannum}
\usepackage{subfigure}
\usepackage{graphics,graphpap}
\usepackage{graphicx}
\usepackage{amsmath,exscale,amsthm,bm}
\usepackage{graphicx}
\usepackage{yfonts}
\usepackage{mathrsfs}
\usepackage[toc]{appendix}
\usetikzlibrary{trees}
\usetikzlibrary{positioning,arrows}
\usetikzlibrary{decorations.pathmorphing}
\usetikzlibrary{decorations.markings}
\usetikzlibrary{decorations.text}
\usetikzlibrary{fit,chains,calc,shapes.geometric,intersections}
\usepackage{natbib}
\setcitestyle{square,comma,numbers,sort&compress}
\usepackage{circledsteps}

\begin{document}

\def\mc#1{\mathcal#1}
\def\a{\alpha}
\def\b{\beta}
\def\c{\chi}
\def\d{\delta}
\def\e{\epsilon}
\def\f{\phi}
\def\g{\gamma}
\def\h{\eta}
\def\i{\iota}
\def\j{\psi}
\def\k{\kappa}
\def\la{\lambda}
\def\m{\mu}
\def\n{\nu}
\def\o{\omega}
\def\p{\pi}
\def\q{\theta}
\def\r{\rho}
\def\s{\sigma}
\def\t{\tau}
\def\u{\upsilon}
\def\x{\xi}
\def\z{\zeta}
\def\D{\Delta}
\def\F{\Phi}
\def\G{\Gamma}
\def\J{\Psi}
\def\L{\Lambda}
\def\O{\Omega}
\def\P{\Pi}
\def\Q{\Theta}
\def\S{\Sigma}
\def\U{\Upsilon}
\def\X{\Xi}

\def\ve{\varepsilon}
\def\vf{\varphi}
\def\vr{\varrho}
\def\vs{\varsigma}
\def\vq{\vartheta}

\newcommand{\vev}[1]{\langle #1 \rangle}
\def\dg{\dagger}                                     
\def\ddg{\ddagger}                                   
\def\wt#1{\widetilde{#1}}                    
\def\mt{\widetilde{m}_1}
\def\mti{\widetilde{m}_i}
\def\mtj{\widetilde{m}_j}
\def\rt{\widetilde{r}_1}
\def\mtt{\widetilde{m}_2}
\def\mttt{\widetilde{m}_3}
\def\rtt{\widetilde{r}_2}
\def\mb{\overline{m}}
\def\VEV#1{\left\langle #1\right\rangle}        
\def\be{\begin{equation}}
\def\ee{\end{equation}}
\def\ds{\displaystyle}
\def\ra{\rightarrow}

\def\bea{\begin{eqnarray}}
\def\eea{\end{eqnarray}}
\def\NO{\nonumber}
\def\Bar#1{\overline{#1}}
\def\ylz{\textcolor{red}}

\title{
\vspace*{1mm}
{\bf Impact of flavour coupling \\ on $SO(10)$-inspired leptogenesis}

\author{
{\Large Pasquale Di Bari and  Xubin Hu}
\\
{\it Physics and Astronomy}, 
{\it University of Southampton,} \\
{\it  Southampton, SO17 1BJ, U.K.} 
\\
}}
\maketitle \thispagestyle{empty}
\pagenumbering{arabic}

\begin{abstract}
We discuss the impact of flavour coupling on the predictions of low energy neutrino parameters
from $SO(10)$-inspired leptogenesis  (SO10INLEP). The  right-handed (RH) neutrino mass spectrum is strongly hierarchical so that
successful leptogenesis relies  on  generating the asymmetry from
next-to-lightest RH neutrino decays ($N_2$-leptogenesis) and  
 circumventing the lightest RH neutrino washout. These two conditions yield
distinctive predictions such as a lower bound on the lightest neutrino mass $m_1 \gtrsim 1\,{\rm meV}$. We first review 
the status of SO10INLEP, noticing how cosmological observations are now testing a 
particular neutrino mass window, $m_1 \simeq (10$--$30)\,{\rm meV}$, where only the first octant is allowed 
and a large range of values for the Dirac phase is excluded.  
Including flavour coupling, we find that the lower bound relaxes to $m_1 \gtrsim 0.65\,{\rm meV}$. Moreover,
new muon-dominated solutions appear slightly relaxing the upper bound on the atmospheric mixing angle. 
We also study the impact on strong thermal SO10INLEP (ST-SO10INLEP) scenario where, 
in addition to successful leptogenesis,  one can washout a large pre-existing asymmetry.
Contrarily to naive expectations, for which flavour coupling could jeopardise the scenario
allowing a large pre-existing asymmetry to survive unconditionally, we show, and explain analytically, 
that ST-SO10INLEP is still viable within almost the same allowed region of parameters. There is even a slight relaxation of the 
$m_1$ viable window from (9--30)meV to (4--40)meV for a $10^{-3}$ pre-existing asymmetry. 
The new results from atmospheric neutrinos, mildly favouring normal ordering and first octant, 
are now in nice agreement with the  predictions of ST-SO10INLEP. 
Intriguingly, the predicted $0\nu\beta\beta$ signal is starting to be within the reach of KamLAND-Zen. 

\end{abstract}

\newpage

\section{Introduction}

We are entering an important stage for the prospect of testing high scale thermal leptogenesis \cite{fy} 
within minimal type I seesaw models \cite{seesaw}, the minimal scenario of leptogenesis \cite{Blanchet:2012bk}.  
Colliders have found no evidence of new physics at the TeV scale and below so far,  
placing strong constraints (and doubts, certainly from us) on low scale leptogenesis scenarios. 
At the same time, low energy neutrino experiments continue to progress steadily and in the next years 
they will be able to provide information on the neutrino mixing unknowns and to test the absolute neutrino mass scale 
in regions of the parameter space that are quite important for high scale leptogenesis scenarios. 
Moreover,  the discovery of gravitational waves \cite{LIGOScientific:2016aoc}
is stimulating many new ideas on how to test  high scale leptogenesis \cite{Dror:2019syi,DiBari:2020bvn,DiBari:2021dri,Fu:2022lrn}.
For these reasons, the often proclaimed statement for which  high scale leptogenesis scenarios 
are untestable is today outdated.

An example of well motivated, testable, high scale leptogenesis scenario is  SO10INLEP. 
This relies on quite minimal assumptions on the Dirac neutrino mass matrix, 
so-called $SO(10)$-inspired conditions \cite{SO10inspired}. They are 
typically realised within various grandunified models, in particular within $SO(10)$ models. 
With these assumptions, and barring the highly fine-tuned compact spectrum case \cite{Akhmedov:2003dg,compact, decrypting}, 
the emerging spectrum of RH neutrinos is strongly hierarchical. 
In particular the lightest RH neutrino, $N_1$, has, typically,
a mass $M_1 \sim (10^5$--$10^6)\,{\rm GeV}$, well below the lower bound for successful leptogenesis \cite{di,cmb}.  
On the other hand, the next-to-lightest RH neutrino, $N_2$, turns out to have a mass with just the correct order-of-magnitude,
$M_2 \sim 10^{11}\,{\rm GeV}$, to produce an asymmetry able to explain the observed value, 
realising successful $N_2$-leptogenesis \cite{geometry}. 
In order for the  $C\!P$ asymmetries of $N_2$ to be sufficiently large, 
the existence of a third heavier RH neutrino species, $N_3$, is necessary to produce 
the needed interference in $N_2$ decays. This nicely fits within the properties of
$SO(10)$-models that indeed, notoriously, predict the existence of three RH neutrino species, 
one for each of the three family $SO(10)$ fermionic sixteenth-plet representations. 
When $SO(10)$-inspired conditions are imposed, typically one obtains $M_3 \sim (10^{14}$--$10^{16}$)\,{\rm GeV}. 
However, for successful thermal leptogenesis it is sufficient that the reheat 
temperature $T_{\rm RH} \gtrsim M_2 \sim 10^{11}\,{\rm GeV}$, since only the $N_2$'s need 
to thermalise, while $N_3$ plays a role at the virtual level only, in the interfering loop diagrams for $N_2$-decays.
Still this is a crucial role, considering that this interference is the source of the $N_2$-$C\!P$ asymmetries and, ultimately,
of the observed baryon asymmetry. For this reason,  when $N_3$ decouples, in the limit $M_3 \gg 10^{16}\,{\rm GeV}$, 
all flavoured $C\!P$ asymmetries of $N_2$ tend to vanish. This yields an upper bound on $M_3$
and, consequently, from the seesaw formula, a lower bound on the lightest neutrino mass \cite{riotto1}. 
Although one can have a sizeable asymmetry generated from $N_2$-decays at very high temperatures, $T \sim M_2$,
for long time such a strongly hierarchical RH neutrino spectrum has been regarded as an obstacle
to reproduce the correct observed baryon asymmetry within SO10INLEP.
The reason is that, within an unflavoured description of SO10INLEP, 
the washout from $N_1$ at $T \sim M_1$ is necessarily strong  and
a successful $N_2$-leptogenesis scenario  cannot be realised for such a strongly hierarchical spectrum. 
However, when flavour effects are included \cite{flavoureffects,flavourlep},  
the calculation of the $B-L$ asymmetry  splits into three different contributions, 
one for each charged lepton flavour \cite{vives}. The strong washout from $N_1$ also splits into three weaker 
components. In this way an allowed region in the space of low energy neutrino parameters opens up 
and successful SO10INLEP can be attained \cite{riotto1,riotto2}.  
The asymmetry is dominantly produced in the tauon flavour. Subdominant muon-dominated flavour solutions also exist, 
while electron-dominated solutions just fall short of reproducing the 
observed asymmetry.\footnote{They appear marginally in a supersymmetric version \cite{susy}. We should also mention that
there is a second way to circumvent the $N_1$ washout, even when this is strong in all three flavours \cite{susy}. If one the
ligthest Dirac neutrino mass is at least two orders of magnitude lower than the up quark mass, then $M_1$ gets lower than the 
sphaleron freeze-out temperature $T_{\rm sph}^{\rm off} \simeq 132\,{\rm GeV}$ \cite{DOnofrio:2014rug}. In this case
the washout acts only on the lepton asymmetry but not on the observed baryon asymmetry.} 

It is interesting that successful SO10INLEP is realised only within a region of the entire low energy neutrino parameter space. 
This yields constraints, and predictions, to be confronted with the experimental results \cite{riotto2}.
A first important feature of SO10INLEP predictions is that inverted ordered
neutrino masses (IO) are very marginally allowed \cite{riotto2,decrypting}, since it necessarily requires the atmospheric neutrino mixing angle  
to be in the second octant and a lower bound on the lightest neutrino mass. For $\theta_{23} \lesssim 50^{\circ}$, 
as suggested by latest neutrino oscillation data global analyses at $3\sigma$ \cite{nufit24},
 one has a lower bound on the lightest neutrino mass $m_1 \gtrsim 20\,{\rm meV}$, 
 corresponding to $\sum_i m_i \gtrsim 130\,{\rm meV}$. This is in tension with 
the cosmological upper bound on the sum of neutrino masses obtained
combining CMB anisotropy, baryon acoustic oscillation observations and supernovae data
\cite{Planck:2018vyg,Allali:2024aiv,Naredo-Tuero:2024sgf,DESI:2025ejh}
\be\label{upperbm}
\sum_i m_i < 120\,{\rm meV} \;\; (95\% \, {\rm C.L.}) \,  .
\ee
For this reason,  at $3\s$, SO10INLEP is viable only for normal ordering (NO) and 
this is in nice  agreement with the latest results from atmospheric neutrino oscillation experiments, that tend to favour NO 
at $92.3\%$ C.L. \cite{Super-Kamiokande:2023ahc}. This is strengthened by the most recent global analyses 
disfavouring IO at $\sim 2.5 \sigma$ \cite{nufit24}.
Therefore, in the following, we will focus exclusively on NO and we will always refer to the 
cosmological upper bound on the lightest neutrino mass  
\be\label{upperbm1}
m_1 < 30\,{\rm meV} \;\; (95\% \, {\rm C.L.}) \,  ,
\ee
derived from Eq.~(\ref{upperbm}) in the case of NO.\footnote{In the IO case the
upper bound Eq.~(\ref{upperbm}) would imply an even more stringent upper bound on the lightest neutrino mass,
$m_3 < 16\,{\rm meV}$ (95\% ${\rm C.L.}$). The significant difference between the upper bound on the lightest neutrino 
mass in the NO and IO cases, shows that the upper bound 
Eq.~(\ref{upperbm}) is essentially ruling out quasi-degenerate neutrinos at 95$\%$ C.L. .}

As mentioned, the most interesting constraint from successful SO10INLEP is the existence of the lower bound on the lightest neutrino mass \cite{riotto1,riotto2}.
This depends in a non-trivial way on the $C\!P$-violating Dirac phase $\delta$ and the atmospheric mixing angle $\theta_{23}$ \cite{DiBari:2020plh}. 
There is a large region of the plane $\delta$ versus $\theta_{23}$
that is already incompatible with SO10INLEP at 95\% C.L.
since the cosmological upper bound on neutrino masses Eq.~(\ref{upperbm}) is violated.
In the semi-hierarchical regime, for  $10\,{\rm meV} \lesssim m_1 \lesssim 30\,{\rm meV}$, 
most stringent constraints hold within SO10INLEP. 
In particular, SO10INLEP is incompatible with the atmospheric mixing angle in the second octant.  
On the other hand, this is also a very  interesting neutrino mass regime,  since only for $m_1$ in the  (10--30)\,meV range 
ST-SO10INLEP can be realised \cite{ST}. This is a special scenario where, in addition to the condition of
successful leptogenesis, one can also wash out a pre-existing asymmetry as large as $10^{-1}$ \cite{problem}. 
For a hierarchical RH neutrino spectrum, without even imposing
$SO(10)$-inspired conditions, successful strong thermal leptogenesis can only be realised for
quite special conditions: a tauon-dominated $N_2$-leptogenesis scenario \cite{problem}
and for $m_1 \gtrsim 10\,{\rm meV}$ \cite{sophie}.
It is then quite non trivial that within successful SO10INLEP,  one can also satisfy
the conditions for strong thermal leptogenesis for a subset of the solutions. In particular,
they can only be satisfied for $m_1 \lesssim 30\,{\rm meV}$, just coinciding with the
current cosmological upper bound Eq.~(\ref{upperbm1}). 
This is another important point showing how we are now just entering a 
crucial stage in testing SO10INLEP and its strong thermal case ST-SO10INLEP.

For this reason, it is of great importance to understand how solid and accurate these theoretical predictions are. 
The constraints on the low energy neutrino parameters have been derived in various papers and with independent codes.
They have also been derived analytically, expressing  the RH neutrino mass spectrum and mixing matrix 
in terms of the low energy neutrino parameters \cite{decrypting,full}. In this way one obtains  
an analytical expression of the final baryon asymmetry in terms of the low energy neutrino parameters. However, the calculation of the asymmetry has been done within a  set of approximations that neglects 
different effects. Three of them are at the level of field theory description:
\begin{itemize}  
\item Flavour coupling effects from spectator processes \cite{Barbieri:1999ma,Buchmuller:2001sr,fuller}.
\item Running of parameters from radiative corrections.
\item Partial equilibration of spectator processes \cite{Garbrecht:2014kda,Garbrecht:2019zaa}.
\end{itemize}
Two additional ones concern the kinetic theory description. 
The asymmetry has been so far calculated solving simple momentum integrated 
rate equations in the fully flavoured regime, neglecting: 
\begin{itemize}
\item Momentum dependence (requiring solution of the full Boltzmann equations) \cite{Garayoa:2009my}.
\item Decoherence  (requiring solution of density matrix equation) \cite{Barbieri:1999ma,flavoureffects}.
\end{itemize}
In this paper, encouraged by the current agreement of (ST-)SO10INLEP predictions with neutrino oscillation experiment results and by the
fact that absolute neutrino mass scale experiments are starting to test a crucial regime, as discussed above, 
we start  implementing these effects, studying the impact of flavour coupling effects on (ST-)SO10INLEP.  
This is certainly 
the most urgent extension of the calculation of the asymmetry within (ST-)SO10INLEP to be considered. 
In particular, in the case of ST-SO10INLEP one can even legitimately suspect that flavour coupling could jeopardise the whole scenario. This is because
a contribution from a large pre-existing asymmetry might survive until the present, 
leaking from one flavour to another, thus escaping both $N_2$ and $N_1$ washout.  
As we will discuss in detail, and anticipated in the abstract, this
does not happen and, ultimately, even in the ST-SO10INLEP case, the account of flavour coupling  yields just some slight corrections that, however,
might prove to be important when atmospheric neutrino mixing angle and Dirac phase will be measured with errors $\sim 1^\circ$
and $\sim 10^\circ$, respectively, by DUNE+T2HK \cite{Ballett:2016daj}. 
At the same time we will highlight some new important aspects of (ST-)SO10INLEP predictions.
 When these are confronted  with the latest results from global analyses on neutrino mixing parameters,
we show how the current best fit for NO and first octant, nicely agrees with (ST-)SO10INLEP predictions,
removing the tension that was existing with previous results \cite{chianese} favouring second octant for $\theta_{23}$.  Moreover, as we mentioned, 
even without imposing strong thermal leptogenesis, for $m_1 \sim (10$--$30)\,{\rm meV}$ the second octant 
is not compatible with SO10INLEP constraints. 
In addition, there are large ranges of $\delta$ that are excluded even for $m_1 \lesssim 10\,{\rm meV}$. 
Therefore, during next years, neutrino oscillation experiments 
will either be able to increase the statistical significance of the agreement or 
of course, will be able to rule out SO10INLEP. It is then important, not to draw incorrect conclusions, to start
a systematic study of theoretical uncertainties to reduce them at the level of the expected experimental errors.  

This is the main aim of the paper, that is structured in the following way. 
In Section 2 we review SO10INLEP, discussing the assumptions and the equations to calculate the asymmetry, 
describing the results that we obtained ignoring flavour coupling (we have re-derived then once more confirming previous analyses and updating the experimental constraints in the plots) and comparing the  predictions  to the latest experimental constraints. 
In Section 3 we discuss flavour coupling and how this modifies the calculation of the final asymmetry. We show the results for some benchmark points  
and compare them to the case when flavour coupling is ignored. We also point out how the calculation of the asymmetry becomes much more CPU time consuming. For that reason, the derivation of the hyper-surface in the space of parameters, giving the allowed region, becomes much more challenging to derive. This is indeed so convoluted  to require at least one million of points for a precise determination. The use of the analytical expressions for the 
RH neutrino masses and mixing matrix \cite{decrypting,full} greatly help in this respect. 
We show the results of the scatter plots giving both the three-dimensional projection of the hypersurface in the space  
($\delta,\theta_{23},m_1$), as previously done in \cite{DiBari:2020plh} neglecting flavour coupling, and different two-dimensional projections. 
We also highlight the different flavour dominance for each point, confirming that, even including flavour coupling effects, 
the bulk of the solutions, approximately $90\%$,  are tauonic solutions, while muonic solutions represent a subdominant $10\%$ contribution. 
However, when flavour coupling is included, we show that a very small $0.1\%$ contribution of electronic solutions appear
and also new muonic solutions.  These new  muonic solutions fall in a region of the parameter space that would otherwise be excluded.\footnote{A special subset of the muonic solutions that we obtain  were also found within a specific Pati-Salam grandunified model with 
discrete flavour symmetries \cite{DiBari:2015oca}. However, this special subset is now excluded by 
the latest neutrino oscillation experimental results since they require $\theta_{23} \simeq 54^\circ$.}
In Section 4 we discuss ST-SO10INLEP, showing how the calculation of the evolution of a large pre-existing asymmetry,  and its relic value, 
is modified by the account of flavour coupling. Here, we show how the allowed region not only survives but even, for
some values of $\delta$, allows slightly larger values of $\theta_{23}$. Moerover, the allowed range of $m_1$ values also slightly englarges.
Finally, in Section 5, we draw some final remarks on the 
importance of (ST-)SO10INLEP and of our results within the current quest for new physics. 

\section{$SO(10)$-inspired leptogenesis}

In this section we review general features of SO10INLEP, neglecting flavour coupling. We briefly discuss the set of assumptions 
that define SO10INLEP and then  we show  how the requirement of successful SO10INLEP produces constraints on the low energy neutrino parameters
that we compare with the latest neutrino oscillation experimental results.
In this way we can discuss the status of SO10INLEP showing that there is currently a non-trivial 
compatibility between SO10INLEP predictions and low energy neutrino experimental data.
We also show how with the new cosmological upper bound on neutrino masses, we are starting to test
a very interesting neutrino mass range for SO10INLEP. 

\subsection{Seesaw mechanism and low energy neutrino data} 

Inspired by $SO(10)$ models, we extend the SM introducing three RH neutrinos $N_{R 1}, N_{R 2}$ and $N_{R 3}$
with Yukawa couplings $h$ and a Majorana mass term $M$. In the flavour basis, where both
charged lepton mass matrices $m_{\ell}$ and $M$ are diagonal, one can write   
the leptonic mass terms  generated after spontaneous symmetry breaking 
by the Higgs expectation value $v =174\,{\rm GeV}$ as ($\a=e,\m,\t$ and $I=1,2,3$)
\be
- {\cal L}^{m}_{{\ell}+\nu} = \,  \overline{\a_L} \, D_{m_{\ell}}\,\a_R + 
                              \overline{\nu_{\a L}}\,m_{D\a I} \, N_{R I} +
                               {1\over 2} \, \overline{N^{c}_{R I}} \, D_{M} \, N_{R I}  + \mbox{\rm h.c.}\, ,
\ee
where $D_{m_{\ell}} \equiv {\rm diag}(m_e,m_{\m},m_{\t})$ is the diagonal charged lepton mass matrix, 
$D_{M}\equiv {\rm diag}(M_1,M_2,M_3)$ is the diagonal Majorana mass matrix
and $m_{D}=h v$ is the neutrino Dirac mass matrix.  
In the seesaw limit, for $M \gg m_D$, the mass spectrum splits into two sets of Majorana neutrino eigenstates,
a light set, $\nu_1$, $\nu_2$ and $\nu_3$, with masses $m_{1} \leq m_{2} \leq m_3$, given by the seesaw formula \cite{seesaw}
\be\label{seesaw}
D_m =  U_\nu^{\dagger} \, m_D \, {1\over D_M} \, m_D^T  \, U_\nu^{\star}  \,   ,
\ee
with $D_m = {\rm diag}(m_1,m_2,m_3)$, and a heavy set, $N_1$, $N_2$ and $N_3$, with masses almost coinciding 
with the three $M_I$ in $D_M$. The matrix $U_\nu$, diagonalising the light neutrino mass
matrix $m_{\nu} = -m_D\,M^{-1}\,m_D^T$ in the weak basis, can be identified
with the PMNS lepton mixing matrix. 
This can be parameterised in terms of the usual three mixing angles $\theta_{ij}$, the Dirac phase $\d$
and the Majorana phases $\rho$ and $\s$, as
\be
U_\nu=  \left( \begin{array}{ccc}
c_{12}\,c_{13} & s_{12}\,c_{13} & s_{13}\,e^{-{\rm i}\,\d} \\
-s_{12}\,c_{23}-c_{12}\,s_{23}\,s_{13}\,e^{{\rm i}\,\d} &
c_{12}\,c_{23}-s_{12}\,s_{23}\,s_{13}\,e^{{\rm i}\,\d} & s_{23}\,c_{13} \\
s_{12}\,s_{23}-c_{12}\,c_{23}\,s_{13}\,e^{{\rm i}\,\d}
& -c_{12}\,s_{23}-s_{12}\,c_{23}\,s_{13}\,e^{{\rm i}\,\d}  &
c_{23}\,c_{13}
\end{array}\right)
\, {\rm diag}\left(e^{i\,\rho}, 1, e^{i\,\sigma}
\right)\,   ,
\ee
where $s_{ij}\equiv \sin\theta_{ij}$ and $c_{ij}\equiv \cos\theta_{ij}$.
As we said, successful SO10INLEP  cannot be realised within IO, compatibly with the latest upper bounds on neutrino masses and 
atmospheric neutrino mixing angle. For this reason we only consider NO. 
In this case latest global analyses of neutrino oscillation experiment results, including atmospheric neutrino
data from Super-Kamiokande and IceCube collaborations,  find for the mixing angles and 
leptonic Dirac phase $\d$ the following best fit values, $1\s$ errors  and $3\s$ intervals \cite{nufit24}: 
\bea\label{expranges}
\theta_{13} & = &  8.56^{\circ}\pm 0.11^{\circ} \in [8.19^{\circ}, 8.89^{\circ}] \,  , \\ \nonumber
\theta_{12} & = &  {33.68^{\circ}}^{+0.73^\circ}_{-0.70^\circ} \in [31.63^{\circ}, 35.95^{\circ}]  \,  , \\ \nonumber
\theta_{23} & = &  {43.3^{\circ}}^{+1.0^{\circ}}_{-0.8^\circ} \in [41.3^{\circ}, 49.9^{\circ}]  \,  ,  \\ \nonumber
\d & = &  {-148^{\circ}} ^{+26^{\circ}}_{-41^{\circ}} \in  [-236^{\circ}, 4^{\circ}]  \, .
\eea 
As one can notice, there is a $3\s$ exclusion interval , $\d\ni [4^\circ, 134^{\circ}]$, 
for the Dirac phase that disfavours $\sin\d > 0$. 
Neutrino oscillation experiments are also sensitive to squared neutrino mass differences, 
finding  for the solar neutrino mass scale 
\be
m_{\rm sol}\equiv \sqrt{m^{\, 2}_2 - m_1^{\, 2}} = (8.65\pm 0.11)\,{\rm meV} \,  ,
\ee
and for the atmospheric neutrino mass scale 
\be
m_{\rm atm}\equiv \sqrt{m^{\, 2}_3 - m_1^{\, 2}} = (50.1\pm 0.2)\,{\rm meV}  \,   .
\ee
The sum of the two scales yields a lower bound on the sum of the neutrino masses:
\be
\sum_i m_i \geq m_{\rm sol} + m_{\rm atm} = (58.75 \pm 0.25)\, {\rm meV} \geq 58.25\,{\rm meV}  \;\; (95\% {\rm C.L.}) \,  .
\ee
No neutrinoless double beta ($0\nu\b\b$) decay signal has been detected so far.  This implies that
there are no experimental constraints on the Majorana phases and that experiments can only 
place an upper bound on the effective  $0\nu\b\b$ neutrino mass $m_{ee} \equiv |m_{\nu ee}|$. 
The most stringent one has been set by the KamLAND-Zen collaboration, that found \cite{KamLAND-Zen:2024eml}
\be\label{upperbmee}
m_{ee} \leq (28 \mbox{--} 122)\,{\rm meV}  \;\;   (90\%\, {\rm C.L.}) \,  ,
\ee
where the range accounts for nuclear matrix element uncertainties.  This translates into an upper bound 
on the lightest neutrino mass $m_1 \leq (84 \mbox{--} 353) \, {\rm meV}$ (90\%\, {\rm C.L.}). When this is compared with the upper bound Eq.~(\ref{upperbm1}) from cosmological observations (assuming $\Lambda$CDM), one can clearly see how the latter is much more stringent.\footnote{It is even more stringent
than it looks like, considering that the cosmological upper bound is at $95\%$ C.L. and the
upper bound from $0\nu\b\b$  at $90\%$ C.L. .}
However, a  future $0\nu\b\b$ positive signal with a measurement of  $m_{ee}$ would not just provide information 
on the absolute neutrino mass scale but also some partial information on the Majorana phases. 

Finally, the KATRIN experiment has recently placed the upper bound 
\be\label{upperbmnue}
m_{\nu_e} \lesssim 0.45\,{\rm eV} \;\;\;  (90\% \,{\rm C.L.})
\ee 
on the effective electron neutrino mass \cite{Katrin:2024tvg}.
Since this falls in the quasi-degenerate limit,  it translates into an equal upper bound on $m_1$.

\subsection{Combining seesaw and $SO(10)$-inspired conditions}

Let us now briefly review seesaw models when $SO(10)$-inspired conditions \cite{SO10inspired} are imposed on the neutrino Dirac mass matrix. 
This can be diagonalised with the so-called singular value decomposition (sometimes also referred to as bi-unitary parameterisation) as
\be\label{svd}
m_D = V^{\dagger}_L \, D_{m_D} \, U_R \,  ,
\ee
where $D_{m_D} \equiv {\rm diag}(m_{D1},m_{D2},m_{D3})$ and 
$V_L$ and $U_R$ are two unitary matrices acting respectively on the
left-handed (LH) and RH neutrino fields and operating the transformation from the 
weak basis (where $m_{\ell}$ is diagonal) to the Yukawa basis (where $m_D$ is diagonal).  

Parameterising the neutrino Dirac masses $m_{Di}$ in terms of the up quark masses,\footnote{For the values of the 
up-quark masses at the scale of leptogenesis ($\sim 10^{11}\,{\rm GeV}$), we adopt  
$(m_{\rm up},m_{\rm charm}, m_{\rm top})=(1\,{\rm MeV}, 400\,{\rm MeV}, 100\,{\rm GeV})$ \cite{fusaokakoide}.}
\be
m_{D 1} = \a_1 \, m_{\rm up} \, ,  \;  
m_{D 2} = \a_2\, m_{\rm charm} \, ,  \; 
m_{D 3} =\a_3 \, m_{\rm top} \,  ,
\ee
we impose {\em $SO(10)$-inspired conditions} \cite{SO10inspired,Akhmedov:2003dg,riotto1} defined as
\begin{equation}
\mbox{\rm (i)} \;  \a_i = {\cal O}(1) \, ; \;\;\;\;  \;\;\;\;\;\;\;\;\;  \mbox{\rm (ii)} \;\;  I\leq V_L \lesssim V_{CKM} \,   .
\end{equation}
With the latter we imply that  using for the matrix $V_L$ the same parameterisation 
as for the leptonic mixing matrix $U_\nu$, the three
mixing angles $\theta_{12}^L$, $\theta_{23}^L$ and $\theta_{13}^L$
do not have values  much larger than the three mixing angles
in the CKM matrix and in particular 
$\theta_{12}^L \lesssim \theta_{c} \simeq 13^{\circ}$,
where $\theta_c$ is the Cabibbo angle.

Let us now define $M \equiv U^{\star}_R\,D_M\,U^{\dagger}_R$ and 
$\widetilde{m}_{\nu} \equiv V_L\,m_{\nu}\,V_L^T$. These are, respectively,  the Majorana mass matrix and  
the light neutrino mass matrix  in the Yukawa basis.  In this way
from the seesaw formula Eq.~(\ref{seesaw}),  using the singular value decomposed 
form Eq.~(\ref{svd}) for $m_D$, we obtain an expression for the inverse Majorana mass matrix
in terms of low energy neutrino parameters and theory parameters constrained by $SO(10)$-inspired conditions:
\be\label{invM}
M^{-1} \equiv  U_R \, D_M \, U_R^T = - D_{m_D}^{-1} \, \widetilde{m}_{\nu} \, D_{m_D}^{-1} \,  .
\ee
From this, diagonalising the matrix on the RH side of Eq.~(\ref{invM}),
one can derive expressions for the three RH neutrino masses  and the RH neutrino mixing matrix $U_R$
in terms of $m_{\nu}$, $V_L$ and the three $\a_i$'s. 

From the analytical procedure discussed in \cite{Akhmedov:2003dg,decrypting,full},
one finds simple expressions for the three RH neutrino masses, 
\be\label{RHspectrum}
M_1    \simeq    {\a_1^2 \,m^2_{\rm up} \over |(\widetilde{m}_\nu)_{11}|} \, , \;\;
M_2  \simeq     {\a_2^2 \, m^2_{\rm charm} \over m_1 \, m_2 \, m_3 } \, {|(\widetilde{m}_{\nu})_{11}| \over |(\widetilde{m}_{\nu}^{-1})_{33}|  } \,  ,  \;\;
M_3  \simeq   \a_3^2\, {m^2_{\rm top}}\,|(\widetilde{m}_{\nu}^{-1})_{33}| \, ,
\ee
and for the RH neutrino mixing matrix
\be\label{RHnumix}
U_R \simeq  
\left( \begin{array}{ccc}
1 & -{m_{D1}\over m_{D2}} \,  {\widetilde{m}^\star_{\nu 1 2 }\over \widetilde{m}^\star_{\nu 11}}  & 
{m_{D1}\over m_{D3}}\,
{ (\widetilde{m}_{\n}^{-1})^{\star}_{13}\over (\widetilde{m}_{\n}^{-1})^{\star}_{33} }   \\
{m_{D1}\over m_{D2}} \,  {\widetilde{m}_{\nu 12}\over \widetilde{m}_{\nu 11}} & 1 & 
{m_{D2}\over m_{D3}}\, 
{(\widetilde{m}_{\n}^{-1})_{23}^{\star} \over (\widetilde{m}_{\n}^{-1})_{33}^{\star}}  \\
 {m_{D1}\over m_{D3}}\,{\widetilde{m}_{\nu 13}\over \widetilde{m}_{\nu 11}}  & 
- {m_{D2}\over m_{D3}}\, 
 {(\widetilde{m}_\nu^{-1})_{23}\over (\widetilde{m}_\nu^{-1})_{33}} 
  & 1 
\end{array}\right) 
\,  D_{\Phi} \,  ,
\ee
with the three phases in 
$D_{\phi} \equiv {\rm diag}(e^{-i \, {\Phi_1 \over 2}}, e^{-i{\Phi_2 \over 2}}, e^{-i{\Phi_3 \over 2}})$ 
given by \cite{full}
\bea
\Phi_1 & =  & {\rm Arg}[-\widetilde{m}_{\nu 11}^{\star}] \,  , \\
\Phi_2 & = & {\rm Arg}\left[{\widetilde{m}_{\nu 11}\over (\widetilde{m}_{\nu}^{-1})_{33}}\right] -2\,(\rho+\s) - 2\,(\rho_L + \s_L) \, , \\
\Phi_3 & =  & {\rm Arg}[-(\widetilde{m}_{\nu}^{-1})_{33}] \,  .
\eea
One can also derive an  expression for the 
orthogonal matrix. Starting from its definition
$\O = D_m^{-{1\over 2}}\, U_\nu^{\dagger} \, m_D \,  D_M^{-{1\over 2}} $ \cite{Casas:2001sr}
that, using Eq.~(\ref{svd}), becomes \cite{riotto1}
\be
\O= D_m^{-{1\over 2}}\, U_\nu^{\dagger} \, V_L^{\dagger} \, D_{m_D} \, U_R \, D_M^{-{1\over 2}} \,  
\ee
or, in terms of its matrix elements ($\a=e,\m,\t$; $i=1,2,3$; $I=1,2,3$),\footnote{The different labels denote the three different
sets of lepton flavours. While the charged lepton neutrino flavours and the neutrino mass eigenstates form two orthonormal bases,
the heavy neutrino lepton flavours cannot in order for the $C\!P$ asymmetries not to all vanish and have successful leptogenesis.}
\be\label{Oij}
\O_{iJ} \simeq {1\over \sqrt{m_i \, M_J}} \, 
\sum_{l=1}^{3} \, m_{D l} \, U^{\star}_{\nu\, \a i}\,V^{\star}_{L\,l \a}\,U_{R \, l J} \,  ,
\ee
one finds \cite{full}
\be\label{Omegaapp}
\hspace{-9mm}\O \simeq 
\left( \begin{array}{ccc}
i\, {(\widetilde{m}_{\nu}\,W^{\star})_{11} \over \sqrt{m_1 \, \widetilde{m}_{\nu 11}}} & 
\sqrt{m_2\,m_3\,(\widetilde{m}_{\nu}^{-1})_{33} \over \widetilde{m}_{\nu 11}}\,
\left(W^{\star}_{2 1} - W^{\star}_{31}\,{{(\widetilde{m}_{\nu}^{-1})_{23}}\over (\widetilde{m}_{\nu}^{-1})_{33}}\right) & 
{W^{\star}_{31}\over \sqrt{m_1\,(\widetilde{m}_{\nu}^{-1})_{33}}} \\
i\, {(\widetilde{m}_{\nu}\,W^{\star})_{12} \over \sqrt{m_2 \, \widetilde{m}_{\nu 11}}} & 
\sqrt{m_1\,m_3\,(\widetilde{m}_{\nu}^{-1})_{33} \over \widetilde{m}_{\nu 11}}\,
\left(W^{\star}_{22} - W^{\star}_{32}\,{{(\widetilde{m}_{\nu}^{-1})_{23}}\over (\widetilde{m}_{\nu}^{-1})_{33}}\right)  
& {W^{\star}_{32}\over \sqrt{m_2\,(\widetilde{m}_{\nu}^{-1})_{33}}}  \\
i\, {(\widetilde{m}_{\nu}\,W^{\star})_{13} \over \sqrt{m_3 \, \widetilde{m}_{\nu 11}}}  & 
\sqrt{m_1\,m_2\,(\widetilde{m}_{\nu}^{-1})_{33} \over \widetilde{m}_{\nu 11}}\,
\left(W^{\star}_{2 3} - W^{\star}_{3 3}\,{{(\widetilde{m}_{\nu}^{-1})_{23}}\over (\widetilde{m}_{\nu}^{-1})_{33}}\right)  
& {W^{\star}_{33}\over \sqrt{m_3\,(\widetilde{m}_{\nu}^{-1})_{33}}}  
\end{array}\right)  \,   ,
\ee
where we defined $W \equiv V_L \, U_\nu$. The validity of these expressions clearly breaks down in the close vicinity
of the two level crossings where either  $(\widetilde{m}_{\nu}^{-1})_{33}$, or $\widetilde{m}_{\nu 11}$,
or  both, vanish. However, as discussed in \cite{decrypting,full}, the first two cases cannot yield successful leptogenesis.
The compact spectrum scenario discussed in the introduction correspond to a situation close to the third case.  
In all three cases the orthogonal matrix entries diverge, a clear indication of the fine tuning that is needed in the the seesaw formula to 
satisfy neutrino oscillation experimental data. 

\subsection{$SO(10)$-inspired leptogenesis}

Let us now discuss how the matter-antimatter asymmetry of the universe can be calculated within $SO(10)$-inspired leptogenesis.
This can be expressed in terms of the baryon-to-photon number ratio, whose measured value 
from {\em Planck} data (including lensing) combined with external data sets \cite{Planck:2018vyg}, is found
\be\label{etaBexp}
\eta_{B}^{\rm exp} = (6.13 \pm 0.04)\, \times 10^{-10}  \,   . 
\ee
In general, the  final asymmetry is given by the sum of two terms,\footnote{Strictly speaking one should also include a third term, the asymmetry generated after
all RH neutrino have decayed, for example by electroweak baryogenesis. However, such a contribution would be controlled by completely independent new physics so one can only assume that a post-leptogenesis production is absent or in any case negligible. From this point of view, in line with what we we wrote in the introduction, the lack of new physics at colliders supports a solution of the matter-antimatter asymmetry puzzle at high scale.
Of course, one could consider a situation where two baryogenesis mechanisms both contribute to the final asymmetry but such a situation would be quite fine-tuned and not particularly attractive.}
\be\label{2terms}
N_{B-L}^{\rm f} = N_{B-L}^{\rm p,f} + N_{\rm B-L}^{\rm lep,f}  \,  .
\ee
The first term is the relic value of a possible pre-existing asymmetry
and the second is the asymmetry generated from the decays of the seesaw neutrinos.
The baryon-to-photon number ratio is then also, in general,  
the sum of two contributions, $\eta_B^{\rm p}$ and $\eta_B^{\rm lep}$,
respectively. Typically, one assumes  that the initial pre-existing asymmetry, 
generated after or during inflation and prior to leptogenesis, is negligible. 
In this way, one finds solutions respecting just successful leptogenesis, i.e.,  
for which $N_{\rm B-L}^{\rm lep,f}$  reproduces the observed asymmetry.

However, some external mechanism might have generated
a large value of the initial pre-existing asymmetry, $N_{B-L}^{\rm p,i}$,
between the end of inflation and the onset of leptogenesis.
In the absence of any washout, this would translate into 
a sizeable value of $\eta_B^{\rm p}$ comparable or greater than $\eta_B^{\rm exp}$.
In this case, then one should also add a strong thermal leptogenesis condition, requiring that the initial pre-existing
asymmetry is efficiently washed out by seesaw neutrino inverse processes. 
In Section 4 we discuss how a subset of the solutions satisfying successful SO10INLEP
can also simultaneously satisfy this strong thermal leptogenesis condition realising ST-SO10INLEP \cite{ST}.
We also show how this subset is not disrupted by the inclusion of flavour coupling but just slightly modified.
For the time being, we will simply assume that the initial pre-existing asymmetry contribution is absent or in any case negligible.

The  contribution to $\eta_B$ from leptogenesis, can be calculated as \cite{Buchmuller:2004nz}
\be\label{etaBlep}
\eta_B^{\rm lep} =a_{\rm sph}\,{N_{B-L}^{\rm lep,f}\over N_{\g}^{\rm rec}} \simeq 
0.96\times 10^{-2}\,N_{B-L}^{\rm lep,f} \,  .
\ee
This expression accounts for sphaleron conversion \cite{manton,kuzmin,klebnikov,Harvey:1990qw} and photon dilution.
The second numerical expression holds when the abundances $N_X$ are normalised in a portion of comoving volume such that  the ultra-relativistic 
thermal equilibrium abundance of a RH neutrino $N_{N_i}^{\rm eq}(T \gg M_i) =1$. 
Successful leptogenesis requires $\eta_B^{\rm lep} = \eta_B^{\rm exp}$,
where $\eta_B^{\rm exp}$ is the measured value given in Eq.~(\ref{etaBexp}).


The final $B-L$ asymmetry from leptogenesis can be calculated, neglecting flavour coupling, 
as the sum of the three contributions, one from each (charged lepton) flavour, explicitly 
\cite{vives,bounds,fuller,density}:
\bea\label{twofl} \nonumber
N_{B-L}^{\rm lep, f} & \simeq &
\left[{K_{2e}\over K_{2\tau_2^{\bot}}}\,\ve_{2 \tau_2^{\bot}}\kappa(K_{2 \tau_2^{\bot}}) 
+ \left(\ve_{2e} - {K_{2e}\over K_{2\tau_2^{\bot}}}\, \ve_{2 \tau_2^{\bot}} \right)\,\kappa(K_{2 \tau_2^{\bot}}/2)\right]\,
\, e^{-{3\pi\over 8}\,K_{1 e}}+ \\ \nonumber
& + &\left[{K_{2\mu}\over K_{2 \tau_2^{\bot}}}\,
\ve_{2 \tau_2^{\bot}}\,\kappa(K_{2 \tau_2^{\bot}}) +
\left(\ve_{2\mu} - {K_{2\mu}\over K_{2\tau_2^{\bot}}}\, \ve_{2 \tau_2^{\bot}} \right)\,
\kappa(K_{2 \tau_2^{\bot}}/2) \right]
\, e^{-{3\pi\over 8}\,K_{1 \mu}}+ \\
& + &\ve_{2 \tau}\,\kappa(K_{2 \tau})\,e^{-{3\pi\over 8}\,K_{1 \tau}} \,  .
\eea
In this expression the (six) $K_{I\a}$ are the {\em flavoured decay parameters}, defined as
\be
K_{I\a} \equiv {\G_{I\a}+\overline{\G}_{I\a}\over H(T=M_I)}= 
{|m_{D\a I}|^2 \over M_I \, m_{\star}} \,  ,
\ee
where  $\Gamma_{I\a}=\Gamma (N_I \ra \phi^\dagger \, l_\alpha)$ 
and $\bar{\Gamma}_{I \a}=\Gamma (N_I \ra \phi \, \bar{l}_\alpha)$ are the
zero temperature limit of the flavoured decay rates into $\a$ leptons
and anti-leptons in the three-flavoured regime, $m_{\star}\equiv 16\,\pi^{5/2}\, \sqrt{g_{\star}}/(3 \sqrt{5})\,(v^2 / M_{\rm Pl}) \simeq 1.08 \,{\rm meV}$
is the equilibrium neutrino mass, $H(T)=\sqrt{g^{SM}_{\star}\,8\,\pi^3/90}\,T^2/M_{\rm P}$ is the expansion rate
and $g_{\star}^{SM}=106.75$ is the number of ultra-relativistic degrees of freedom in the standard model.
The  quantities 
\be
\ve_{2\a} \equiv -(\G_{2\a}-\overline{\G}_{2\a})/(\G_2 + \overline{\G}_2)
\ee
are the $N_2$-flavoured $C\!P$ asymmetries, with $\G_2 \equiv \sum_{\a} \G_{2\a}$ and $\overline{\G}_{2}\equiv \sum_{\a} \, \overline{\G}_{2\a}$.
Finally,  we defined $K_{2\tau_2^{\bot}}\equiv K_{2e}+K_{2\mu}$ and 
$\ve_{2 \tau_2^{\bot}} \equiv \ve_{2e} + \ve_{2\mu}$.

Using the bi-unitary parameterisation Eq.~(\ref{svd}) for $m_D$,  
the flavoured decay parameters can be expressed as\footnote{One can also express 
the decay parameters in terms of the orthogonal matrix:
\be
K_{I\a} = \left|\sum_j\,\sqrt{m_j\over m_{\star}}\,U_{\nu\, \a j}\,\O_{j I}\right|^2 \, ,
\hspace{5mm}\mbox{\rm and}\hspace{5mm}
K_I = \sum_I \, {m_j \over m_{\star}} \, |\O_{jI}|^2 \, .
\ee
However, in SO10INLEP, the bi-unitary parameterisation clearly provides the primary way to calculate
all leptogenesis parameters, the $K_{I\a}$'s and the $\ve_{2\a}$'s, since, as one can see from Eqs.~(\ref{RHspectrum})
and (\ref{RHnumix}), one gets directly RH neutrino masses $M_I$ and RH neutrino mixing matrix $U_R$ in terms 
of low energy neutrino parameters in $m_{\nu}$, neutrino Dirac masses $m_{Di}$ and parameters in $V_L$. 
}
\be\label{KialVL}
K_{I\a} = {\sum_{k,l} \, 
m_{Dk}\, m_{Dl} \,V_{L k\a} \, V_{L l \a}^{\star} \, U^{\star}_{R kI} \, U_{R l I} 
\over M_I \, m_{\star}}\,  .
\ee

The flavoured $C\!P$ asymmetries can be calculated using \cite{crv}
\be\label{eps2a}
\ve_{2\a} \simeq
\overline{\ve}(M_2) \, \left\{ {\cal I}_{23}^{\a}\,\x(M^2_3/M^2_2)+
\,{\cal J}_{23}^{\a} \, \frac{2}{3(1-M^2_2/M^2_3)}\right\}\, ,
\ee
where
\be
\overline{\ve}(M_2) \equiv {3\over 16\,\pi}\,{M_2\,m_{\rm atm} \over v^2} \, , 
\hspace{3mm} \xi(x)=\frac{2}{3}x\left[(1+x)\ln\left(\frac{1+x}{x}\right)-\frac{2-x}{1-x}\right] \,  ,
\ee
\be
{\cal I}_{23}^{\a} \equiv   {{\rm Im}\left[m_{D\a 2}^{\star}
m_{D\a 3}(m_D^{\dag}\, m_D)_{2 3}\right]\over M_2\,M_3\,\mtt\,m_{\rm atm} }\,   
\hspace{5mm}
\mbox{\rm and}
\hspace{5mm}
{\cal J}_{23}^{\a} \equiv  
{{\rm Im}\left[m_{D\a 2}^{\star}\, m_{D\a 3}(m_D^{\dag}\, m_D)_{3 2}\right] 
\over M_2\,M_3\,\mtt\,m_{\rm atm} } \,{M_2\over M_3}   ,
\ee
with $\mtt \equiv (m_D^{\dag}\, m_D)_{2 2}/M_2$.  However, the first term is important only when RH neutrino masses are quasi-degenerate,
close to crossing level solutions.  In our case such solutions are falling in region of parameter spaces that are now excluded by experiments\footnote{For example, a crossing level solution where $\widetilde{m_\nu}^{-1}_{33}=0$ appears at $m_1 \simeq 0.1 \,{\rm eV}$ for the 
range of values of $\theta_{23}$ we are considering. In any case, in the code we use the full expression for $\ve_{2\alpha}$, also including the 
contribution from the self-energy diagram to the $C\!P$ asymmetries.} 
and one can reliably use the approximate expression \cite{full}
\be\label{ve2a}
\ve_{2\a} \simeq {3 \over 16\, \pi}\, {M_2 \over M_3}   {{\rm Im}\left[m_{D\a 2}^{\star}
m_{D\a 3}(m_D^{\dag}\, m_D)_{2 3}\right]\over v^2 \, (m^\dagger_D\,m_D)_{22} }\,   .
\ee
Again, using the bi-unitary parameterisation, one obtains:
\be\label{ve2al}
\ve_{2\a} \simeq {3 \over 16\, \pi \, v^2}\,
{|(\widetilde{m}_{\nu})_{11}| \over m_1 \, m_2 \, m_3}\,
{\sum_{k,l} \,m_{D k} \, m_{Dl}  \,  {\rm Im}[V_{L k \a }  \,  V^{\star}_{L  l \a } \, 
U^{\star}_{R k 2}\, U_{R l 3} \,U^{\star}_{R 3 2}\,U_{R 3 3}] 
\over |(\widetilde{m}_{\nu}^{-1})_{33}|^{2} + |(\widetilde{m}_{\nu}^{-1})_{23}|^{2}}   \,  .
\ee
Finally, for the efficiency factors at the production $\k(K_{2\a})$ we can use a 
simple analytic expression valid for initial thermal abundance \cite{flavourlep}
\be\label{kappa}
\k(K_{2\a}) = 
{2\over z_B(K_{2\a})\,K_{2\a}}
\left(1-e^{-{K_{2\a}\,z_B(K_{2\a})\over 2}}\right) \,  , \;\; z_B(K_{2\a}) \simeq 
2+4\,K_{2\a}^{0.13}\,e^{-{2.5\over K_{2\a}}} \,   .
\ee
Since  all solutions are characterised by strong wash-out at the production
(either $K_{2\t} \gg 1$ or $K_{2 \tau_2^{\bot}} \gg 1$ respectively for tauon and muon-dominated solutions),
the final asymmetry does not depend on the initial $N_2$ abundance anyway.\footnote{For $\tau_B$ solutions,
that we introduce in the next section, one can find values of $K_{2\tau}$ as low as $K_{2\tau}\simeq 1$. 
However, solutions with $K_{2\tau} \lesssim 3$, where there is some
dependence on the initial conditions and larger theoretical uncertainties, 
are realised only for values $m_1 \gtrsim 30\,{\rm meV}$, now disfavoured by the cosmological 
upper bound Eq.~(\ref{upperbm1}).} 

We have now all the analytical expressions to calculate the final asymmetry in a fast and accurate way.
Notice that in general the asymmetry within minimal leptogenesis with three RH neutrinos would depend on 18
quantities: $\ve_{I\alpha}, K_{I\alpha}$ ($I=1,2,3$ and $\a = e, \mu, \tau$). However, from
Eq.~(\ref{twofl}) one can see that in SO10INLEP, since it realises a $N_2$-leptogenesis scenario, 
the asymmetry depends only on nine quantities, $\ve_{2\a}\, K_{2\a} , K_{1\a} \, (\a=e,\mu,\tau)$, while there is no dependence
on  $\ve_{1\a},\ve_{3\a},K_{3\a} (\a=e,\mu,\tau)$.  These nine quantities can be expressed in terms of the seesaw parameters.
As we have seen, using the bi-unitary parameterisation and the seesaw formula, one obtains Eqs.~(\ref{RHspectrum})
and (\ref{RHnumix}) for the three Majorana masses and the RH neutrino mixing matrix $U_R$
in terms of the six parameters in $V_L$, the nine parameters in the  low energy neutrino matrix $m_\nu$
and the three  Dirac neutrino masses $m_{Di}$.  Therefore, in this way, the three RH neutrino masses and the six parameters
in $U_R$ are traded off with the 9 low energy neutrino parameters. Moreover, the $\ve_{2\a}$'s depend, approximately, only
on $m_{D2}$ and the flavoured decay parameters do not depend on the Dirac neutrino masses $m_{D 1}, m_{D2}$ and $m_{D3}$. 
In this way, ultimately, one obtains for the final baryon-to-photon number ratio an expression of the form
\be
\eta_{B}^{\rm lep} \simeq \eta_{B}^{\rm lep}(m_{\rm sol},m_{\rm atm},\theta_{12},\theta_{13};
\theta_{23},\delta,m_1;\rho,\sigma; V_L, \alpha_2) \,  .
\ee
Notice that we have grouped the parameters in the following way:
\begin{itemize}
\item The first four low energy neutrino parameters are those measured 
accurately and precisely enough that one has can assume small Gaussian errors.
In the scatter plots, these parameters have been Gaussianly randomly generated.
However the errors are so small that just fixing them at the best fit values would 
practically produce the same results;
\item We have then the three neutrino unknowns $\theta_{23}$, $\delta$ and $m_1$. Here we have been very conservative
not to contaminate the predictions from SO10INLEP with the partial experimental information we have. 
We have uniformly randomly generated the atmospheric neutrino mixing angle in the range $\theta_{23} =[38^\circ, 52^\circ]$
and the neutrino oscillation $C\!P$ violating phase in the full range $\delta= [-\pi,\pi]$.  Finally we have 
uniformly logarithmically generated the lightest neutrinos mass in the range $\log(m_1/{\rm eV}) = [-4,0]$.
\item The Majorana phases have been simply randomly uniformly generated in the full range $[0,2\pi]$.
\item The 6 parameters in $V_L$ have been uniformly randomly generated. The three mixing angles 
have been capped to the CKM matrix values, following $SO(10)$-inspired conditions, explicitly we
adopted: $\theta_{12}^L \leq  13^{\circ} \simeq \theta_{12}^{CKM} \equiv \theta_c$, 
$\theta_{23}^L \leq  2.4^{\circ}\simeq \theta_{23}^{CKM}$, $\theta_{13}^L \leq  0.2^{\circ}\simeq \theta_{13}^{CKM}$.
In any case the results do not depend on a precise choices of these upper bounds \cite{ST}. Finally, 
for the parameter $\a_2$, we have used throughout the paper a kind of maximum standard value, $\a_2 \leq 5$.  
\end{itemize} 
   
\subsection{Successful leptogenesis condition and solutions} 
  
The successful leptogenesis condition
\be\label{successfullep}
\eta_{B}^{\rm lep}(m_{\rm sol},m_{\rm atm},\theta_{12},\theta_{13};
\theta_{23},\delta,m_1;\rho,\sigma; V_L, \alpha_2) = \eta_B^{\rm exp} 
\ee 
identifies an allowed region in the space of the nine low energy neutrino parameters. In the approximation $V_L = I$ 
this would correspond to an eight-dimensional hypersurface. Turning on the three angles and three phases in $V_L$, this 
hypersurface turns into a hyperlayer with some thickness. 
It is important that it does not fill the whole parameter space and this yields constraints and predictions. 
 In order to visualise this region, we show in Fig.~1 the three-dimensional projection,
\begin{figure}[t]
\centerline{
\psfig{file=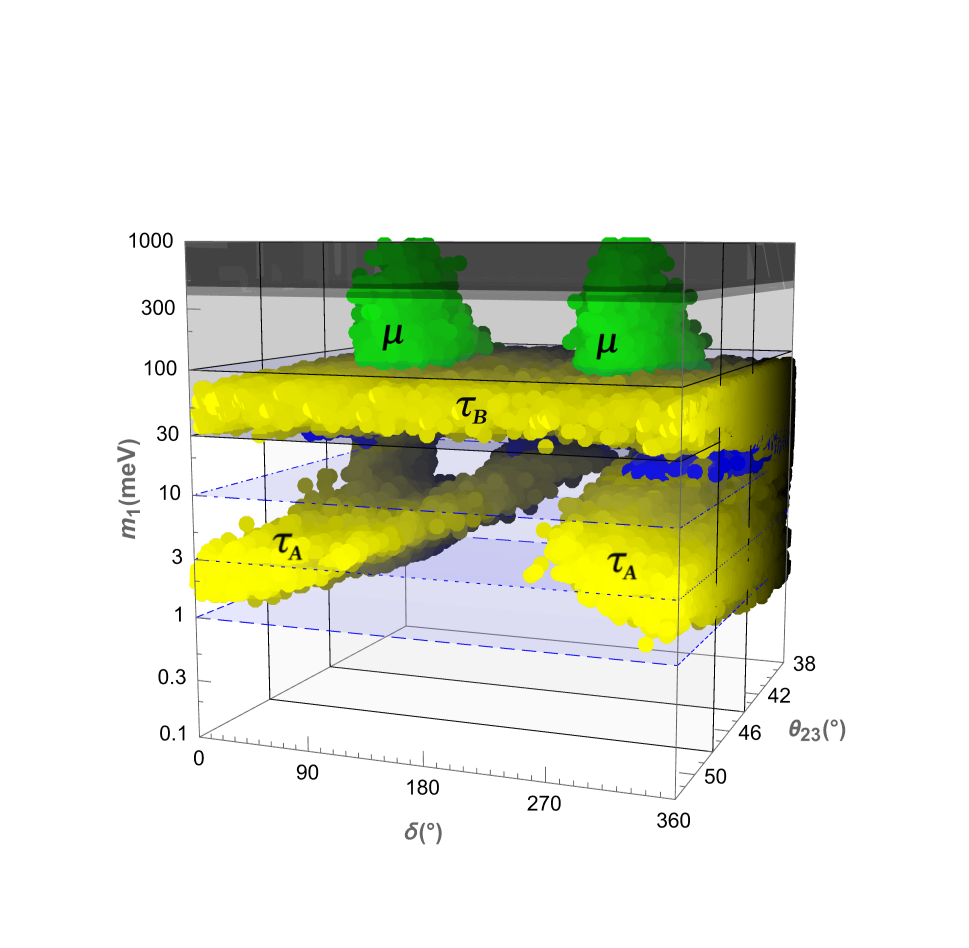,scale=0.6} \hspace{-15mm}
\psfig{file=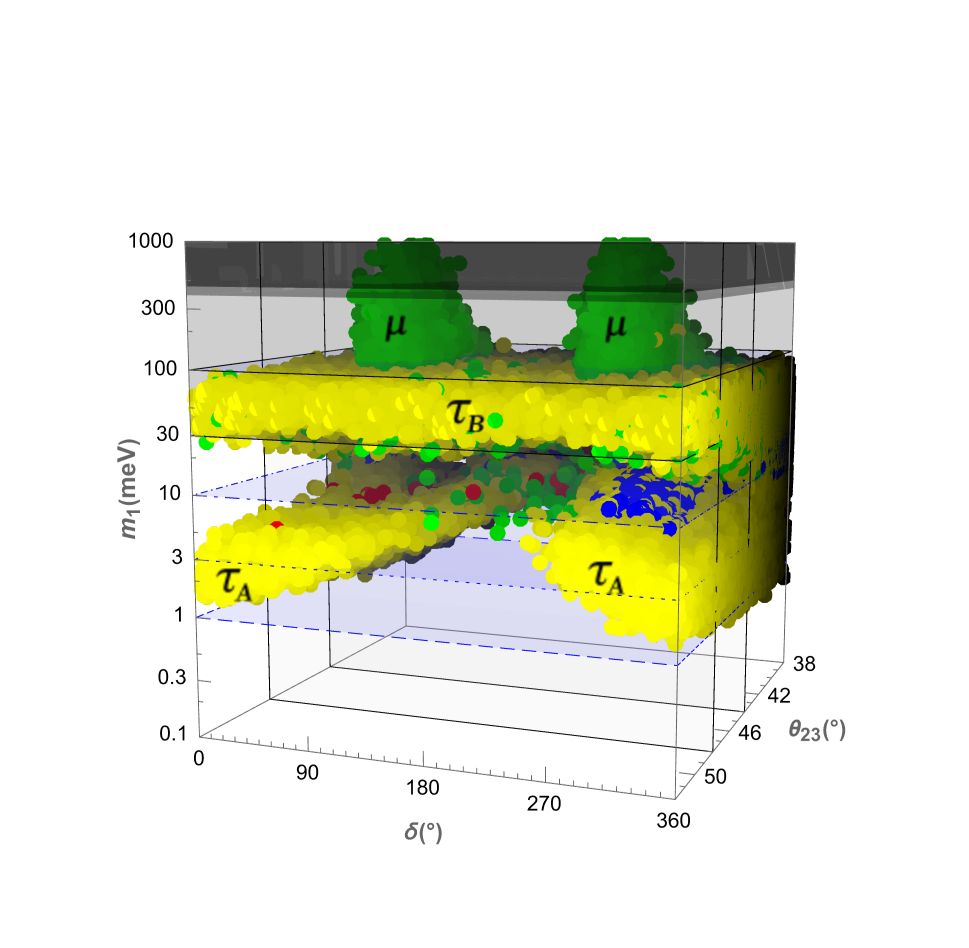,scale=0.6}
}\vspace{-10mm}
\caption{Scatter plot of the solutions obtained  imposing successful SO10INLEP neglecting flavour coupling (left panel)  
and accounting for flavour coupling (right panel). The three grey areas correspond to the
excluded regions by the three upper bounds on the absolute neutrino mass scale: 
Eq.~(\ref{upperbm1}) from cosmological observations, 
Eq.~(\ref{upperbmee}) from $0\nu\b\b$  
and Eq.~(\ref{upperbmnue}) from tritium beta decay.  The three planes in light blue simply help understanding the 3-dim shape.
Colour code: tauonic, muonic and strong thermal solutions are denoted by yellow, green and blue points, respectively.
}
\end{figure}
on the 3-dim space $(\delta, \theta_{23},m_1)$  of the three unknown neutrino parameters \cite{DiBari:2020plh}, of a scatter plot
obtained for $\alpha_2=5$ in the 15-dim space of the parameters in $U$ and $V_L$ by generating about $2\times 10^6$ solutions satisfying
successful $SO(10)$-inspired leptogenesis and all low energy neutrino experimental constraints in Eq.~(\ref{expranges}).\footnote{The success rate
is approximately $0.01\%$ for the parameter ranges we have adopted in the scan.} 
One can immediately notice how solutions are found only for $m_1 \gtrsim 1\,{\rm meV}$,
so that there is a clear lower bound on the absolute neutrino mass scale. 
The three dark grey regions indicate the upper bounds on neutrino masses in
Eq.~(\ref{upperbm1}) from cosmological observations, Eq.~(\ref{upperbmee}) from $0\nu\b\b$  
and Eq.~(\ref{upperbmnue}) from tritium beta decay. 

Each point in the scatter plots is generated imposing the following two conditions:
\begin{itemize}
\item {\small $\chi^2(m_{\rm sol},m_{\rm atm},\theta_{12},\theta_{13}) \equiv 
\left({m_{\rm atm} - \bar{m}_{\rm atm}\over \d m_{\rm atm}}\right)^2 +
\left({m_{\rm sol} - \bar{m}_{\rm sol} \over \d m_{\rm sol}} \right)^2 +
\left({\theta_{12} - \bar{\theta}_{12} \over \d \theta_{12}} \right)^2 +
\left({\theta_{13} - \bar{\theta}_{13} \over \d \theta_{13}} \right)^2
< \chi^2_{\rm max} $;}
\item $\eta_B > \bar{\eta}_B^{\rm exp} - 3\delta \eta_{B}^{\rm exp} = 6.01 \times 10^{-10}$ .
\end{itemize}
We have conservatively used $\chi^2_{\rm max} = 25$. However, the obtained allowed regions are not sensitive to 
a precise value of $\chi^2_{\rm max}$. For example, a more stringent value $\chi^2_{\rm max} = 16$ 
yields basically the same regions. This is because the errors on the four measured neutrino oscillation parameters are sufficiently
 small that fluctuations around the mean value produce a very small change of the asymmetry. 
 We have also tried to combine the two conditions including $\eta_B$ in the calculation of the $\chi^2$. 
 This just simply slows down the search of the solutions without any change, as expected. The reason is quite simple: for any point 
 were the predicted asymmetry is higher than the experimental value, one can always change the values of the other parameters, those not
 involved in the projection, to lower the value of the asymmetry. Therefore, these two conditions provide the most efficient (and sensible) 
 way to produce the allowed regions.

\subsubsection{Three types of solutions}

There are three types of solutions in SO10INLEP \cite{riotto2}. These correspond to three different regions in the parameter space
that can be partly distinguished in the three dimensional projection scatter plot in Fig.~1. 
 The different colours indicate the flavour giving the main contribution to the final asymmetry, specifically: 
\begin{itemize}
\item The yellow points indicate tauon-flavour dominated solutions characterised by $K_{1\tau} \lesssim 1$.
One can distinguish two distinct types of tauonic solutions, $\tau_A$ and $\tau_B$ solutions:\footnote{The two corresponding regions 
in the multi-parameter space are not disjoint but the border line, hybrid, solutions are quite marginal. This means that the 
border between the two regions is quite thin, even though the two-dimensional projections we show might be misleading in that respect.}
\begin{itemize}
\item[-] The $\tau_A$ solutions are characterised
by strong washout at the production ($K_{2\tau} \gtrsim 10$). Moreover, one has  $K_{1\mu} \gg 1$, so that the muon contribution is 
many orders of magnitude below the observed value.
For current allowed experimental  values of $\theta_{23}$ as in Eq.~(\ref{expranges}), they can only be realised for 
$0.9\, {\rm meV} \lesssim m_1 \lesssim 30\,{\rm meV}$.  The blue points indicate the subset of tauon flavour dominated
solutions that realise ST-SO10INLEP and that we will discuss in Section 5.  The $\tau_A$ solutions can realise ST-SO10INLEP while the $\tau_B$ 
cannot in general\footnote{Only a few special ones with large $K_{1\mu} \sim 10$ can be found.}.   
\item[-] The $\tau_B$ solutions are characterised by a mild washout at the production with $2 \lesssim K_{2\tau} \lesssim 10$. 
In this case one can also have simultaneously $K_{1\mu} \lesssim 1$ and in any case $K_{1\mu} \lesssim 10$.
However, one has  $K_{2\tau_2^{\bot}}\gg 10$. The muonic contribution is now only 2-3 orders of magnitudes smaller than the observed value.
They can only be realised for   $m_1 \simeq (20$--$60)\,{\rm meV}$.
\end{itemize}
\item The green points indicate solutions with a sizeable muon-flavour component of the asymmetry. The bulk of these solutions is characterised 
by a dominant muonic contribution. There are some rare solutions where  
the tauonic contribution can be sizeable though sub-dominant, as large as $10\%$.  In general, we will refer to them as muonic solutions.
They are characterised by $K_{1\mu} \lesssim 1$ and $K_{2\tau_2^{\bot}} \lesssim 10$.  Like for $\tau_B$ solutions, one can have simultaneously $K_{1\mu},K_{1\tau} \lesssim 1$
and in any case $K_{1\tau} \lesssim 10$. Therefore, the difference between $\tau_B$ and muonic solutions is mainly given by the value of $K_{2\tau_2^{\bot}}$.
\end{itemize}
One can notice how the cosmological upper bound on neutrino masses is now excluding almost completely muon-dominated and $\tau_B$ solutions. 
However, the $\tau_A$ solutions represent the bulk of the solutions, about $90\%$ of the total. 

It is also interesting to notice that for all $SO(10)$-inspired solutions one has $K_{2e} \ll K_{2\mu} \simeq K_{2\tau_2^{\bot}}$. 
This means that  the  electronic component of the leptons produced from $N_2$-decays is negligible. Essentially, all leptons produced from $N_2$ decays
have a flavour composition that lies quite precisely on the muon-tauon plane. The reason can be easily understood analytically:
from Eq.~(\ref{KialVL}) and Eq.~(\ref{RHnumix}) one easily derives 
$K_{2e}/K_{2\mu} \lesssim |V_{L21}|^2 \sim \theta^2_{\rm c} \sim 0.05$.
One can see from Eq.~(\ref{twofl}) that this result, combined with the fact that also $\ve_{2e}\ll \ve_{2\mu}$, implies that phantom terms are 
always negligible.

In Fig.~2 we plot RH neutrino masses, decay parameters and asymmetries  versus $m_1$ for
three benchmark points in the parameter space and for 
$(\alpha_1,\alpha_2,\alpha_3)=(1,5,1)$ (the values of $\alpha_1$ and $\alpha_3$ only affect 
$M_1$ and $M_3$). For all of them one has $\eta_B^{\rm lep} \geq \eta_B^{\rm exp}$
within some range of $m_1$ values. Therefore, successful SO10INLEP is realised 
for some special values of $m_1$ where $\eta_B^{\rm lep} = \eta_B^{\rm exp}$. 
Each of these three benchmark points provides an example of solution belonging to one of the three types:
$\tau_A$  (left panels), $\tau_B$  (central panels) and $\mu$ (right panels).  At this stage of the discussion,
one should focus only on the the dashed lines, obtained neglecting flavour coupling. 
\begin{figure}[t]
\begin{center}
\psfig{file=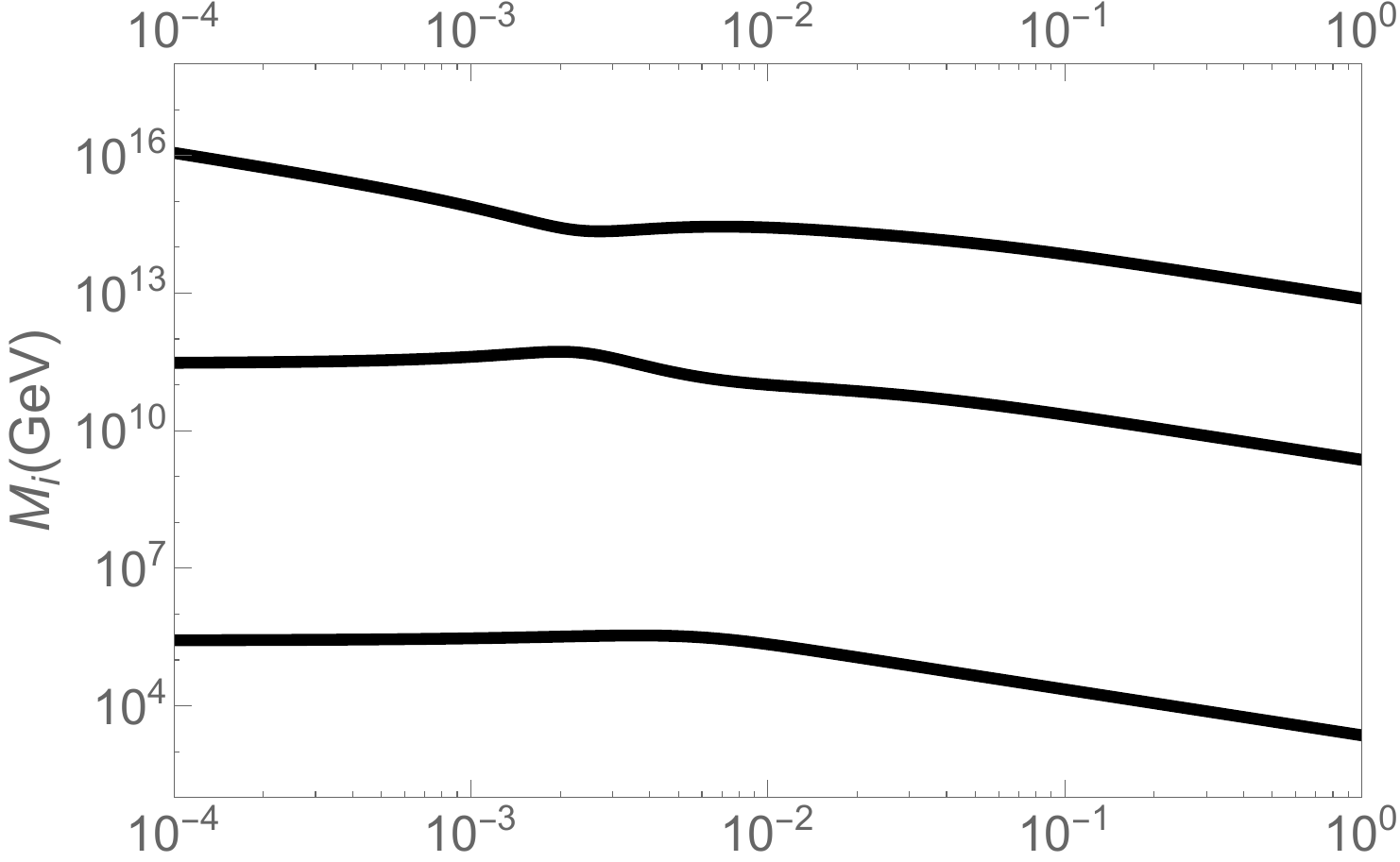,scale=0.18}\hspace{2mm}
\psfig{file=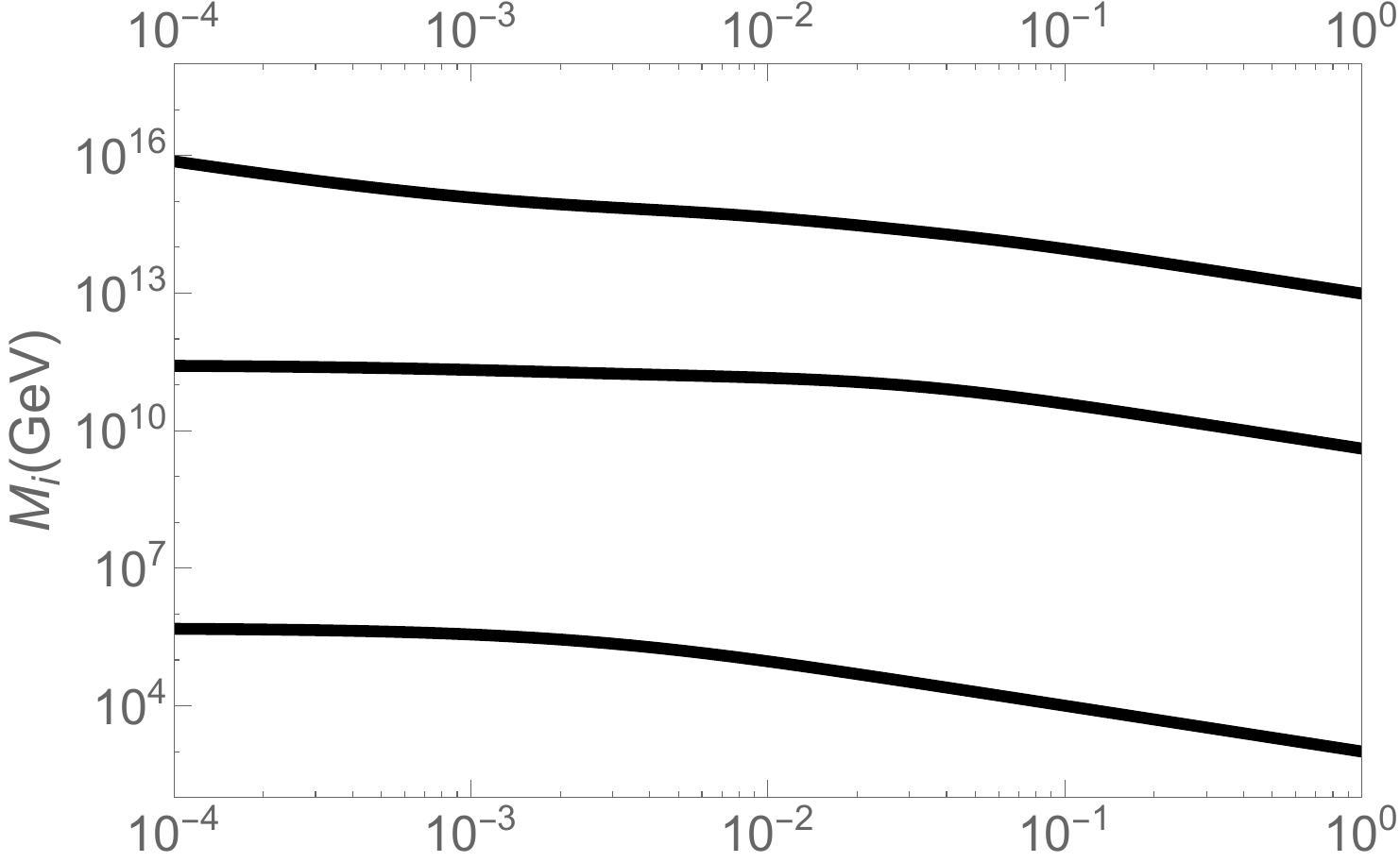,scale=0.18}\hspace{2mm} 
\psfig{file=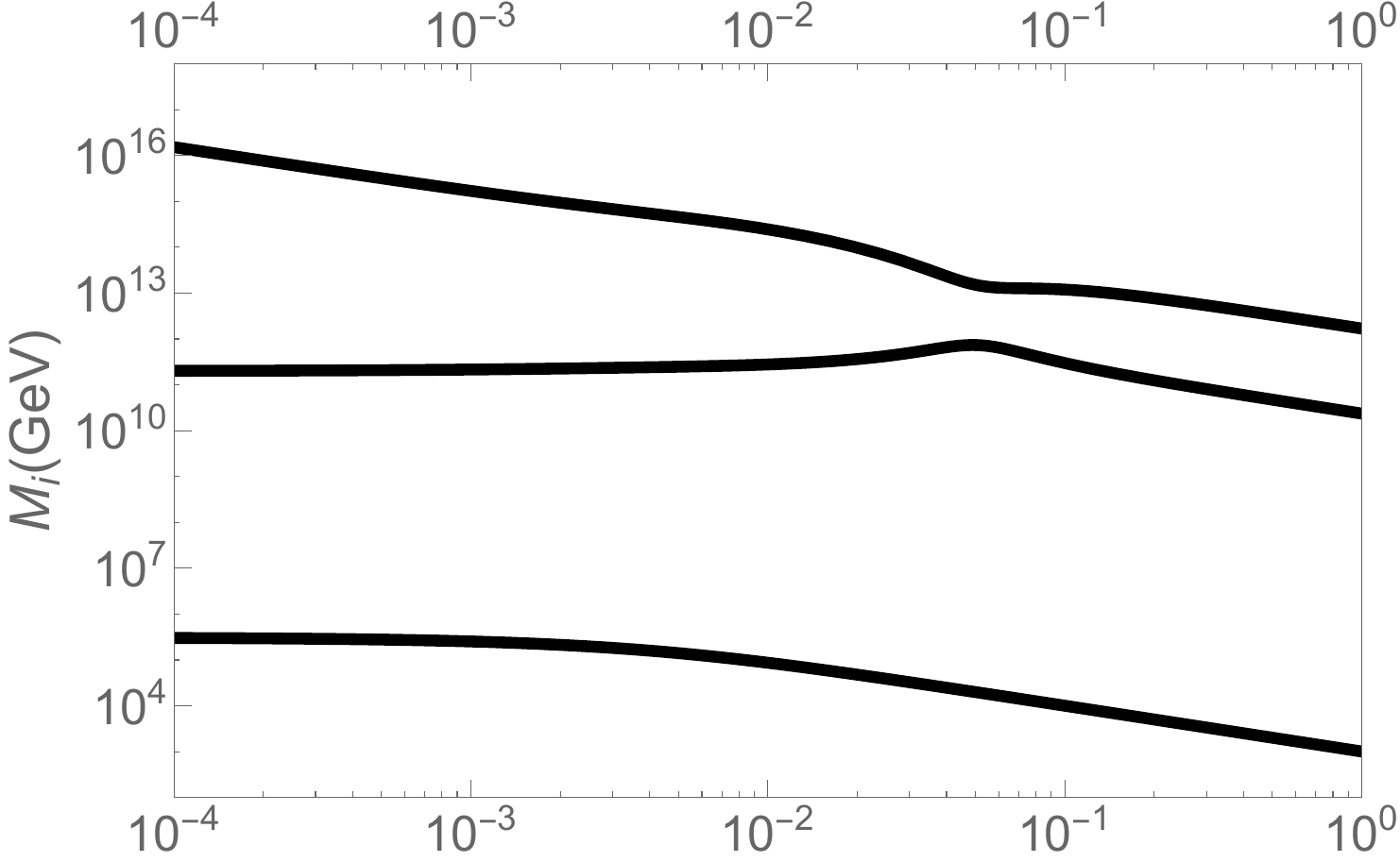,scale=0.18}   \\ \vspace{0mm}
\psfig{file=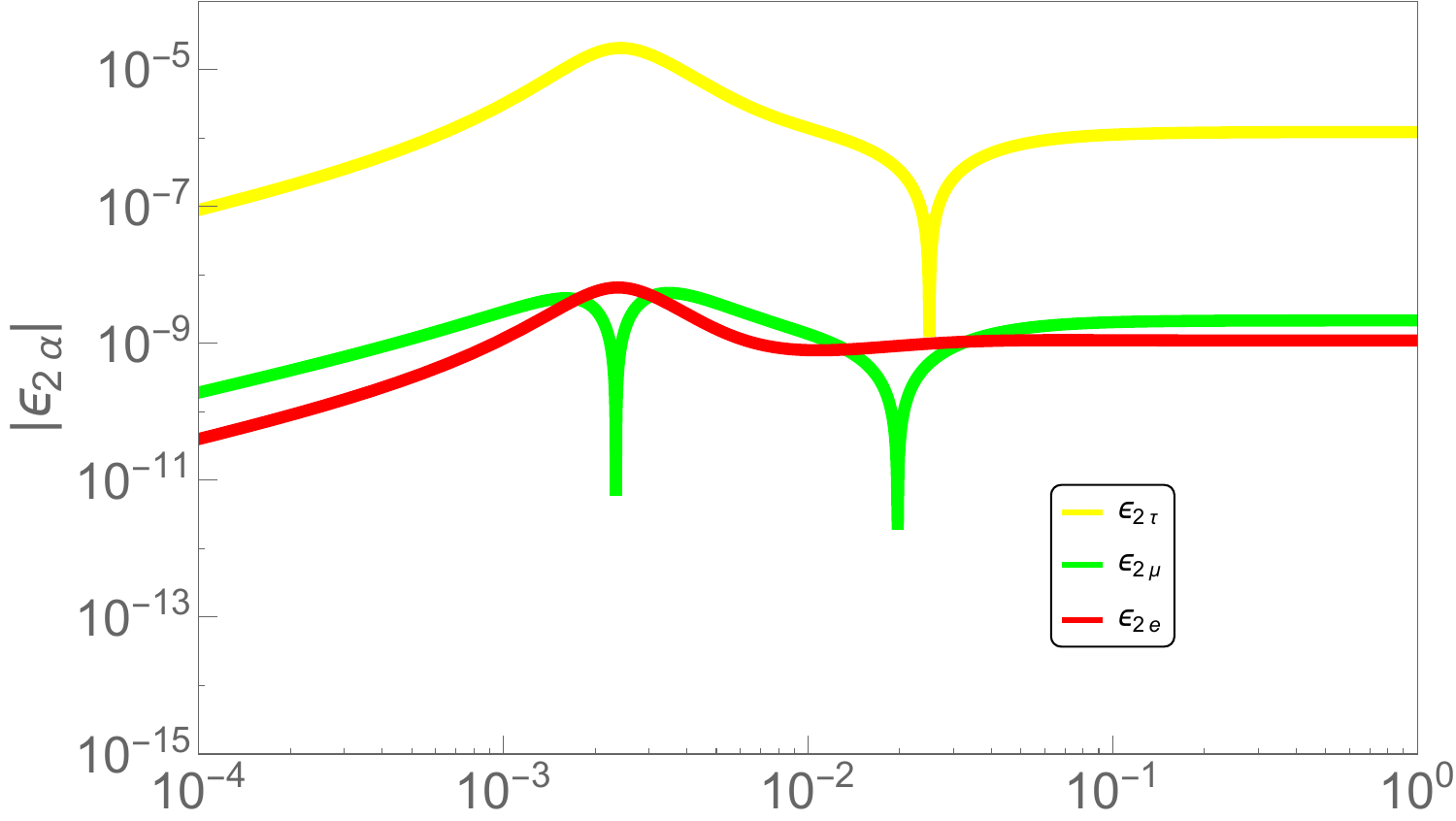,scale=0.18}\hspace{2mm}
\psfig{file=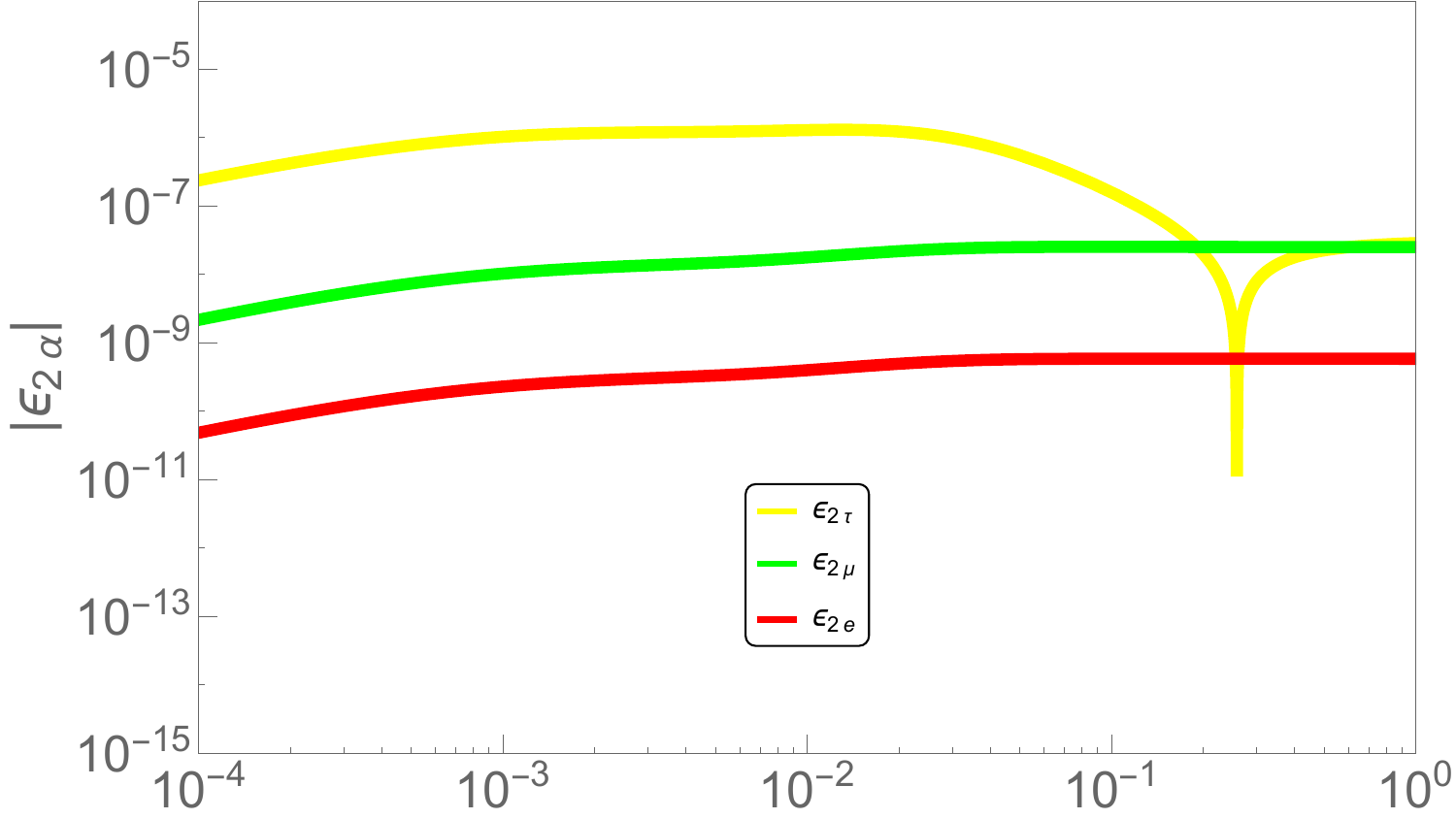,scale=0.18}\hspace{2mm} 
\psfig{file=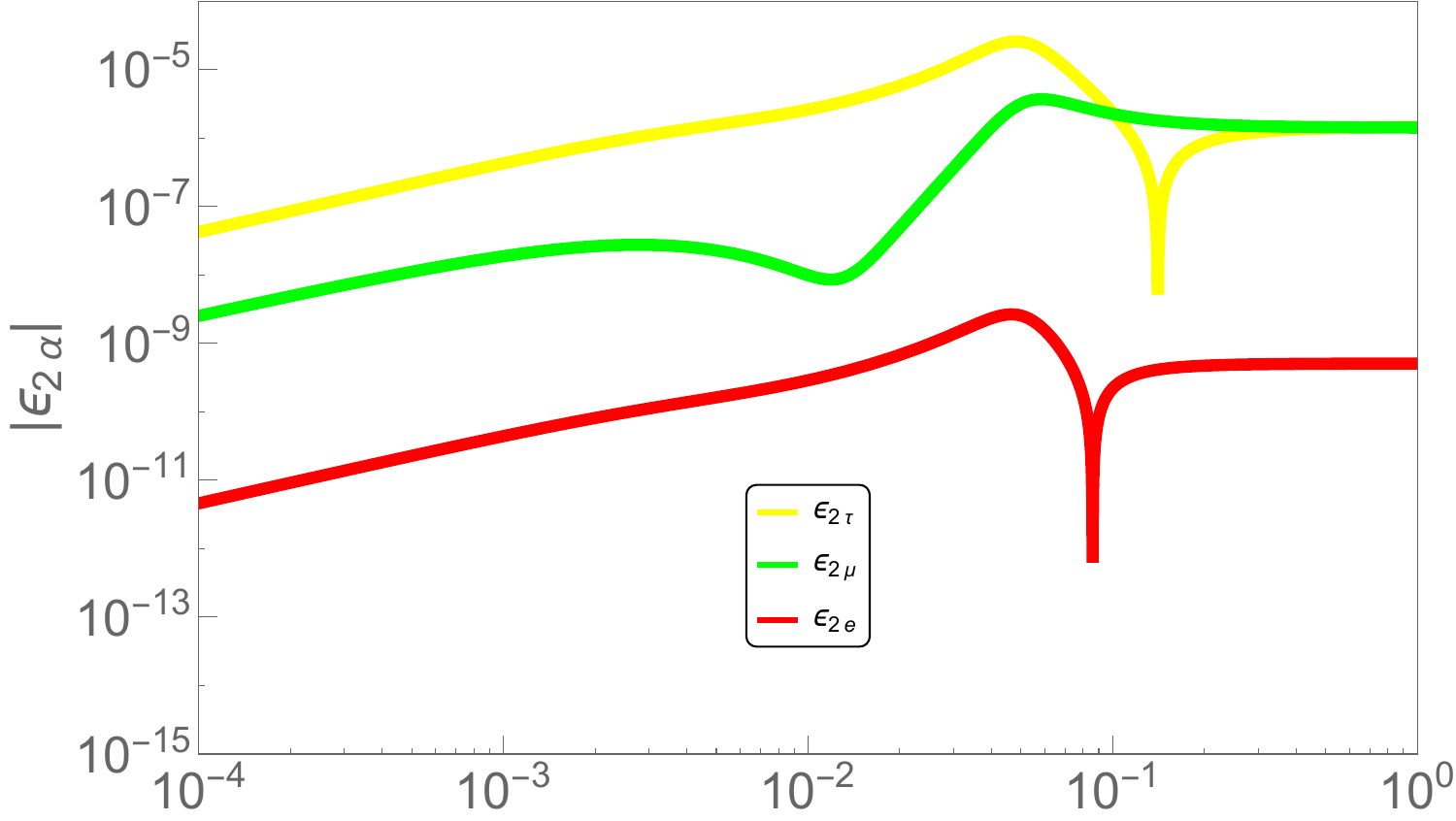,scale=0.18}   \\ \vspace{0mm}
\psfig{file=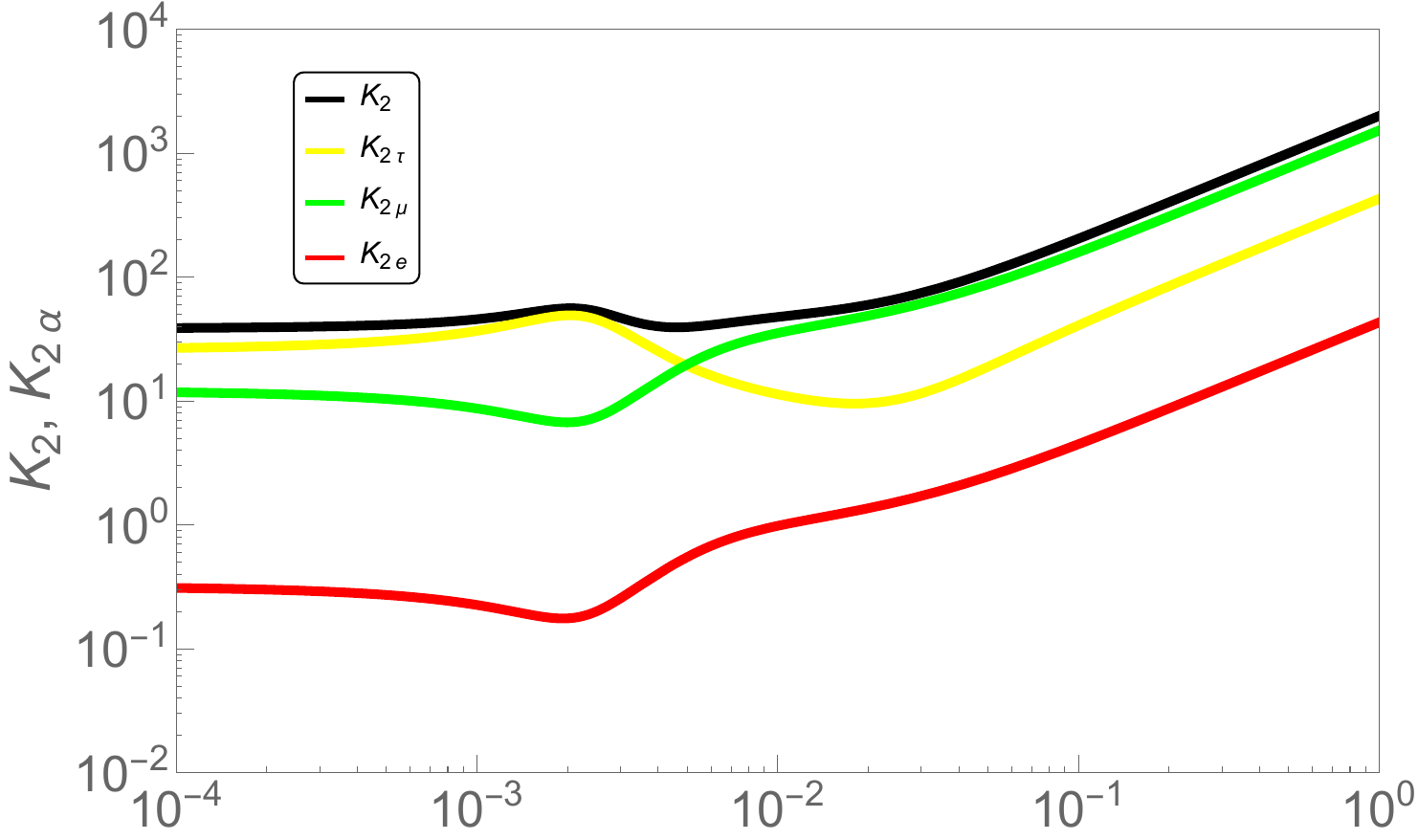,scale=0.18}\hspace{2mm}
\psfig{file=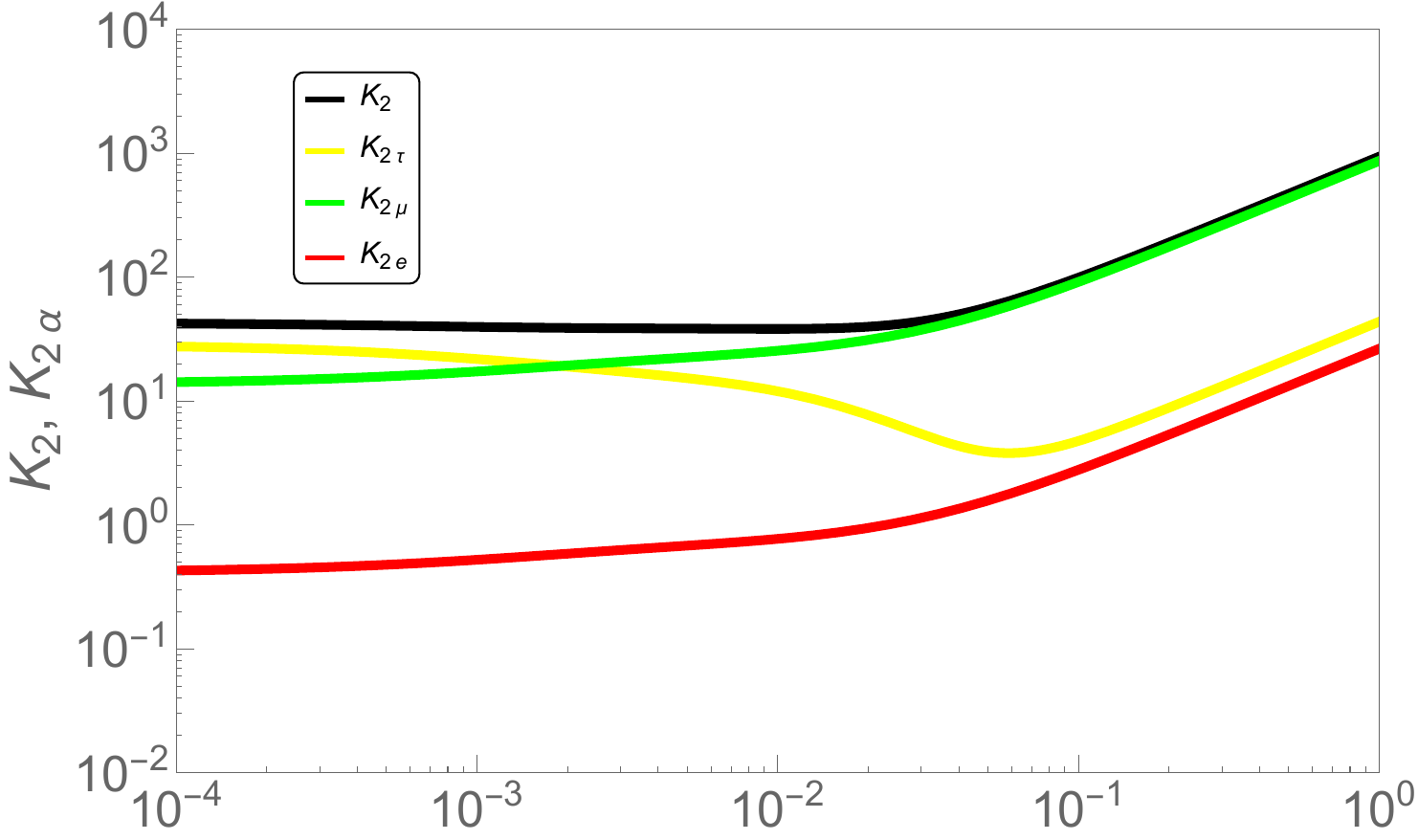,scale=0.18}\hspace{2mm} 
\psfig{file=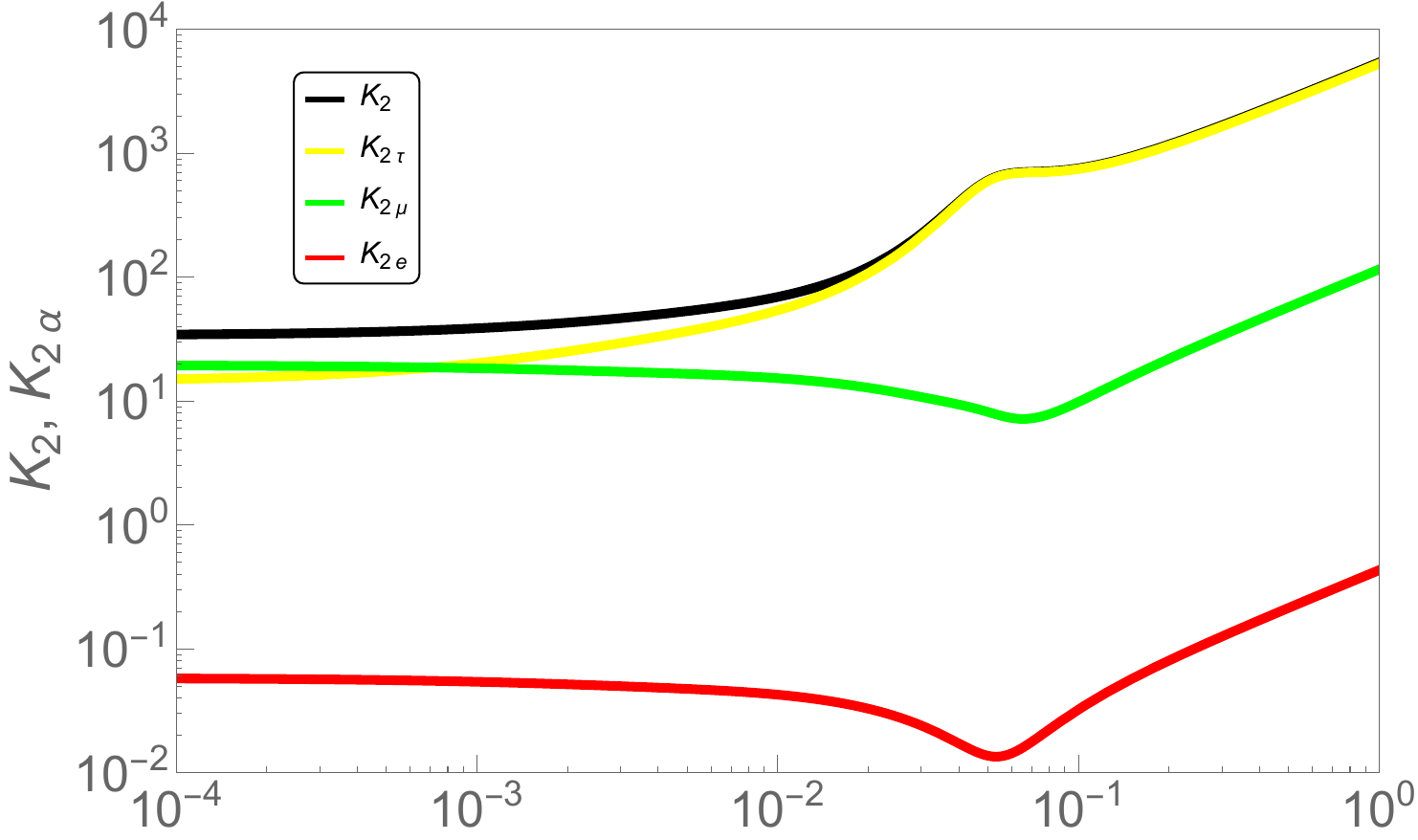,scale=0.18}  \\ \vspace{0mm}
\psfig{file=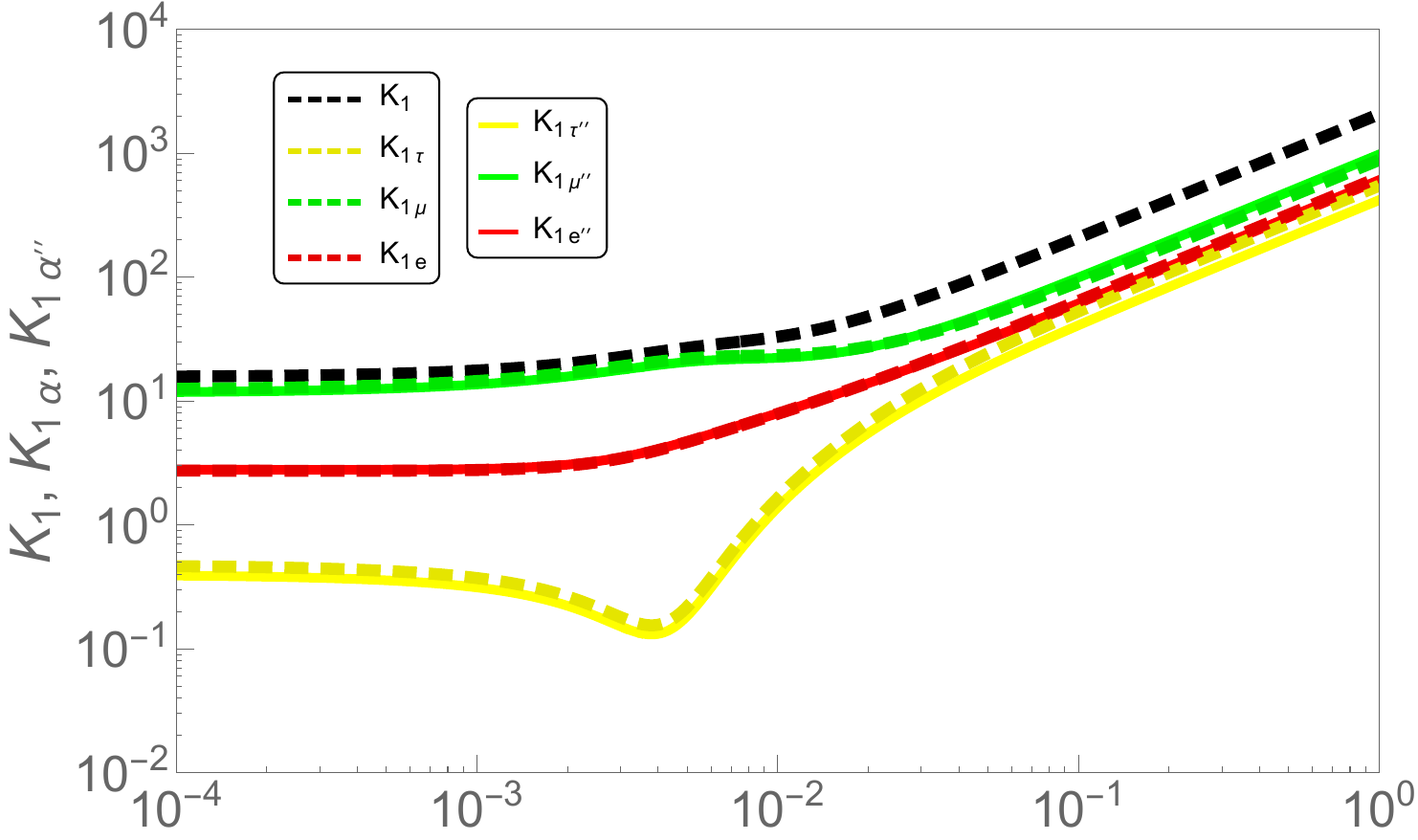,scale=0.18}\hspace{2mm}
\psfig{file=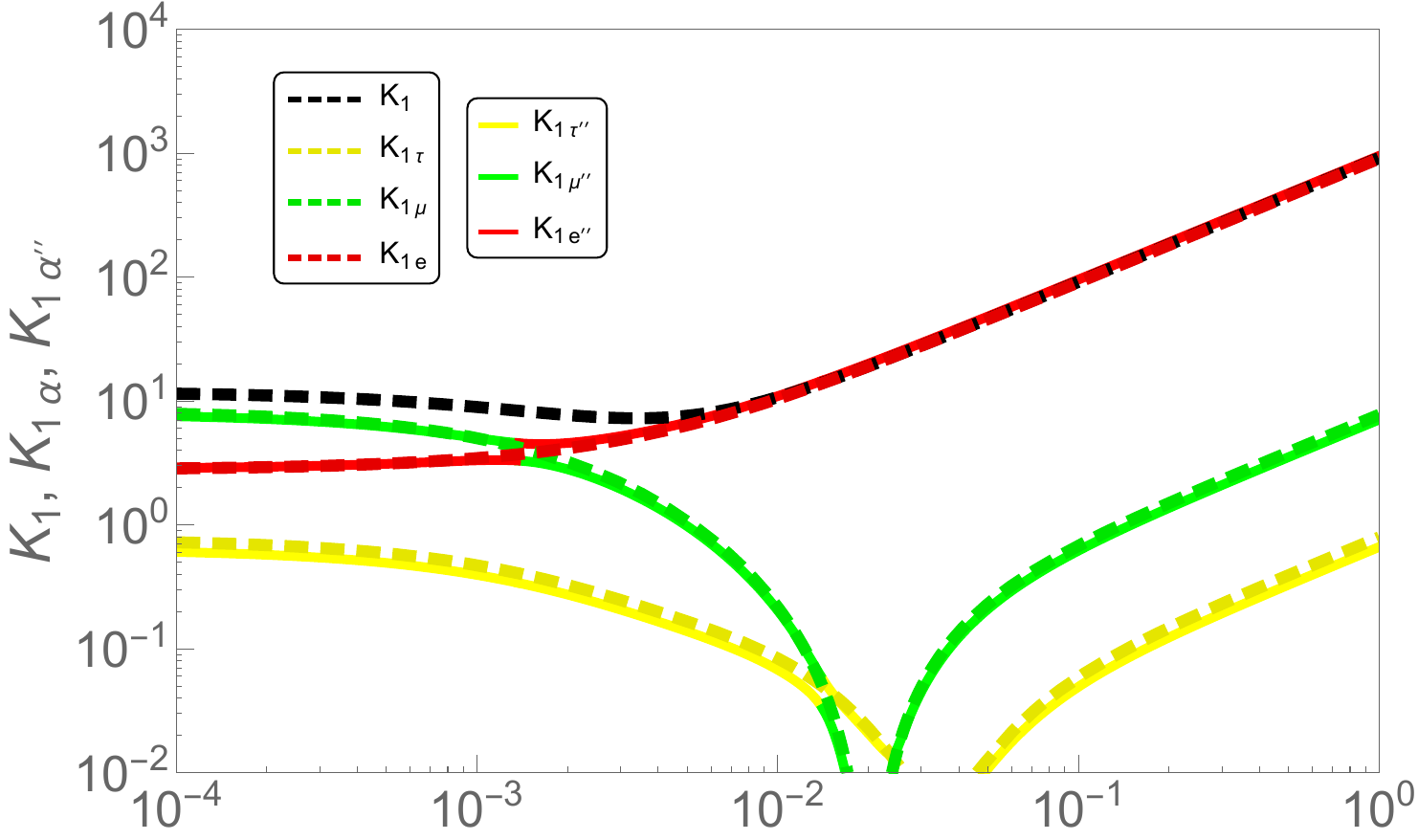,scale=0.18}\hspace{2mm} 
\psfig{file=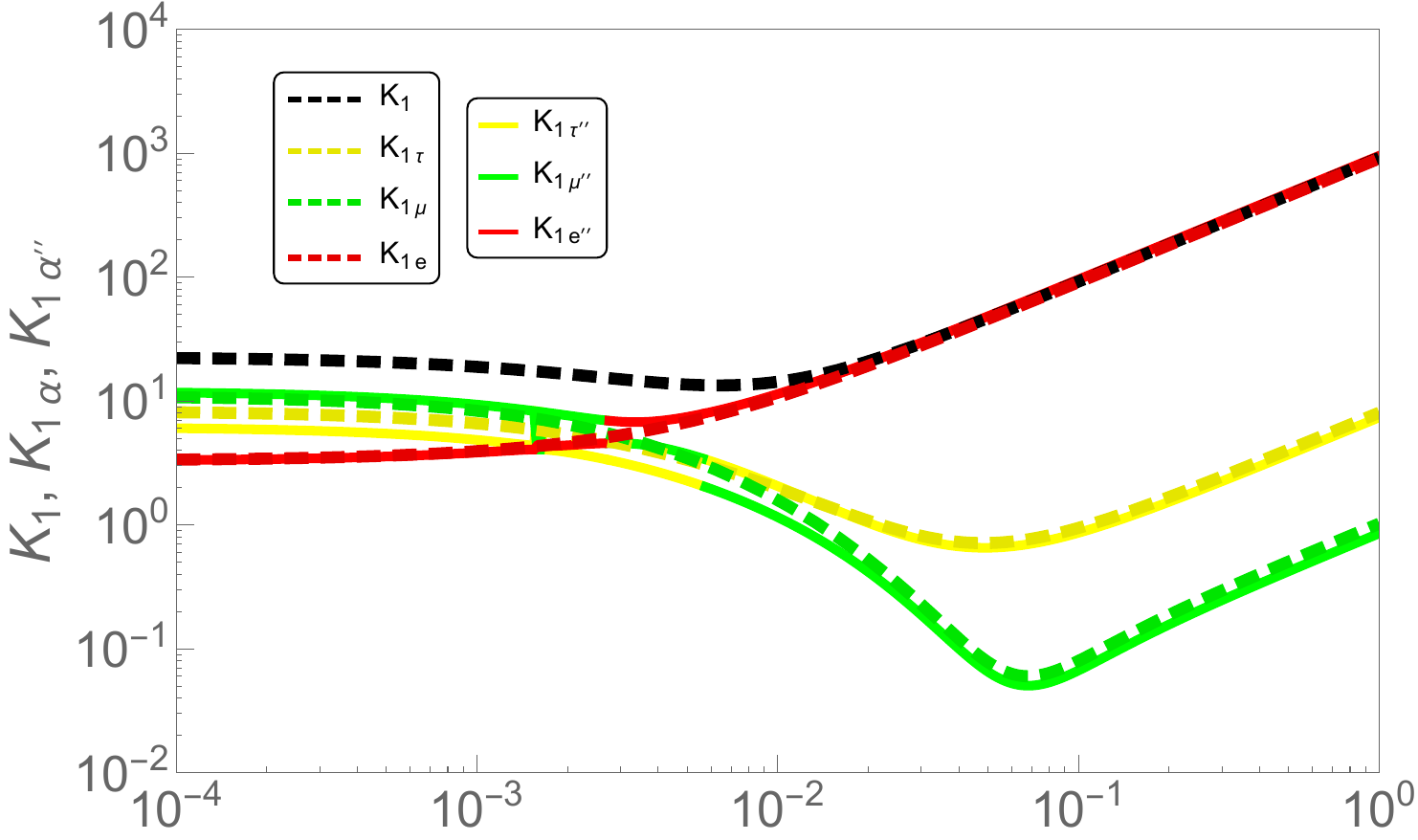,scale=0.18}  \\ \vspace{0mm}
\psfig{file=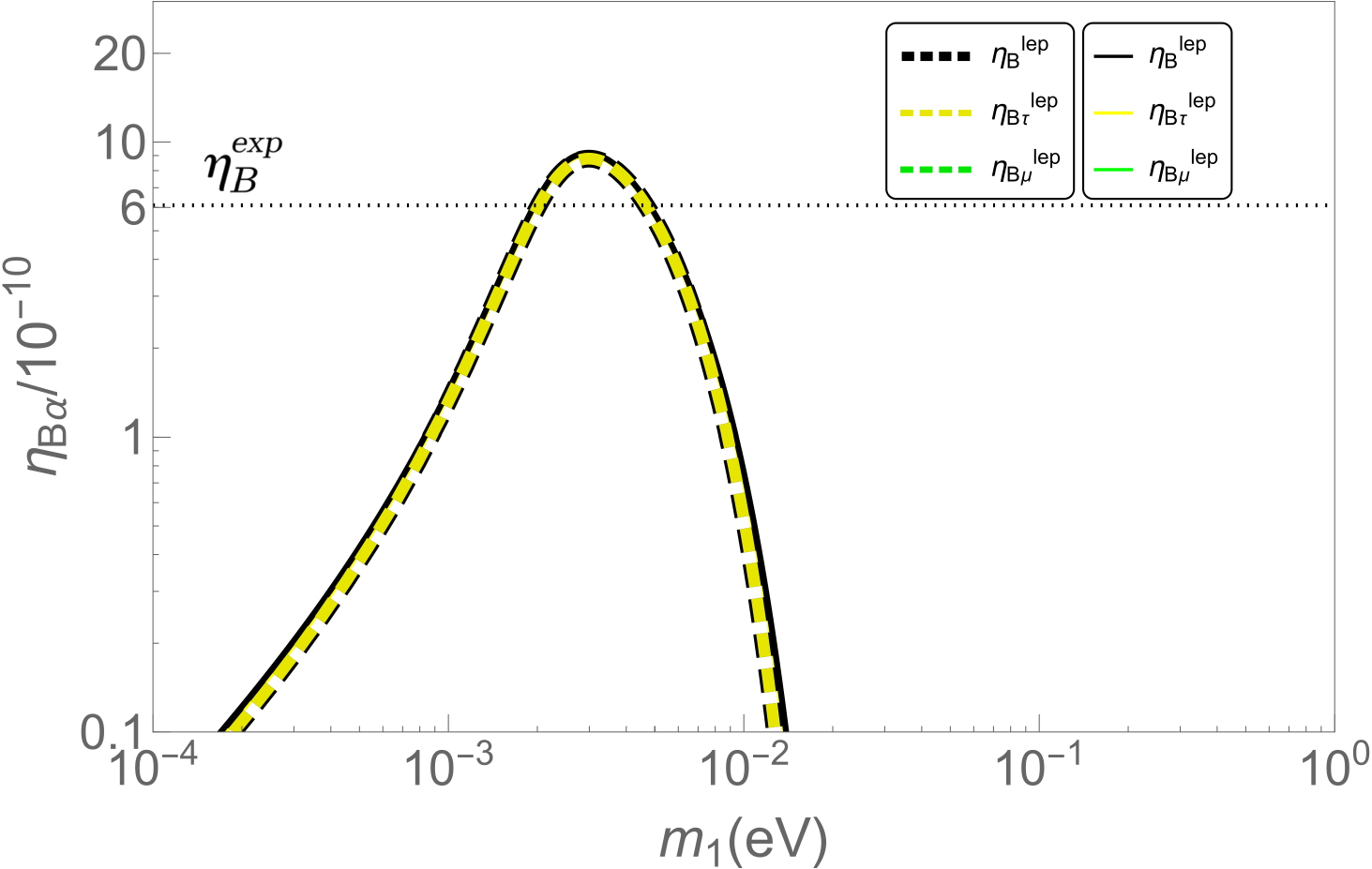,scale=0.18}\hspace{2mm}
\psfig{file=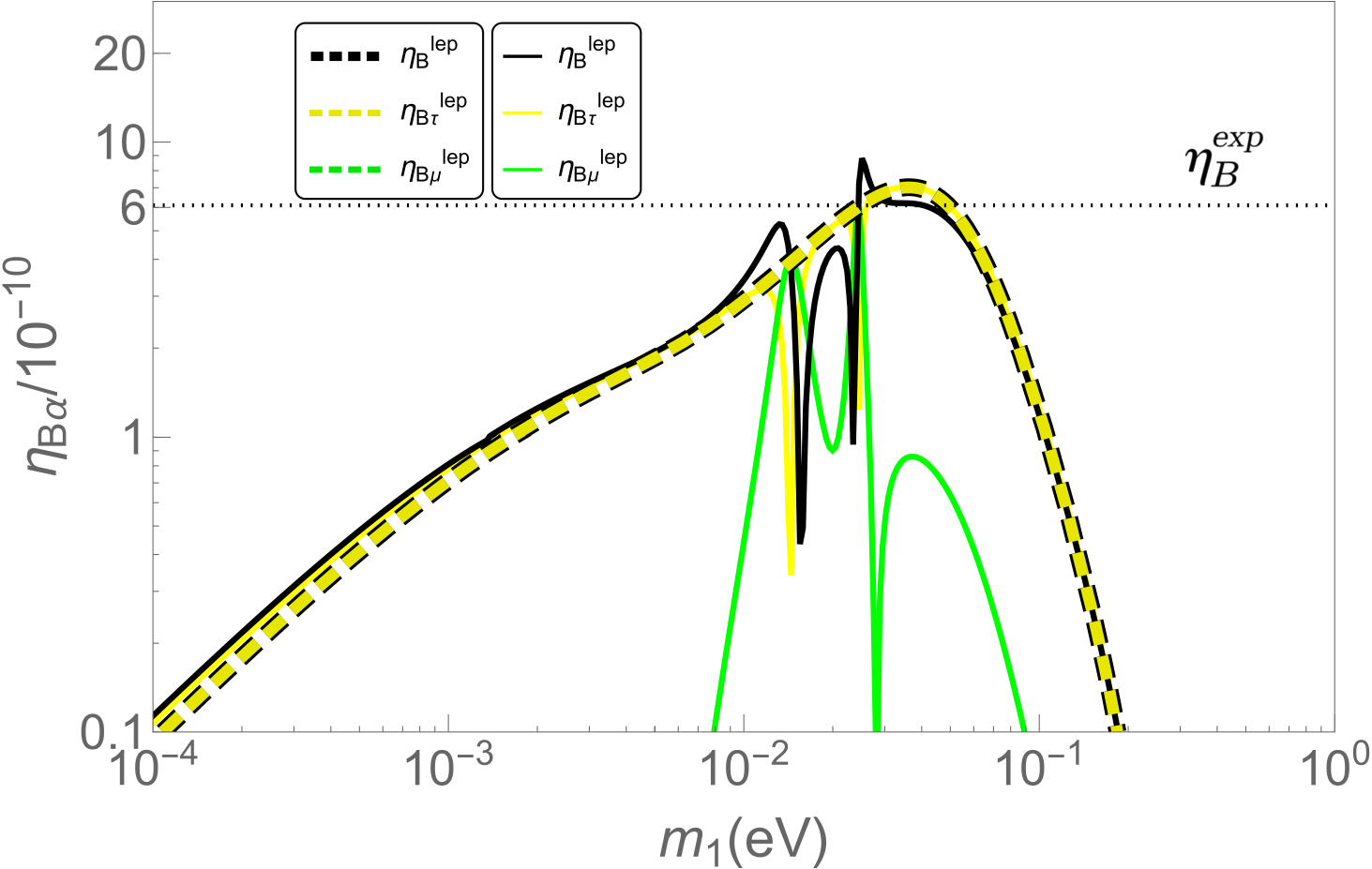,scale=0.18}\hspace{2mm} 
\psfig{file=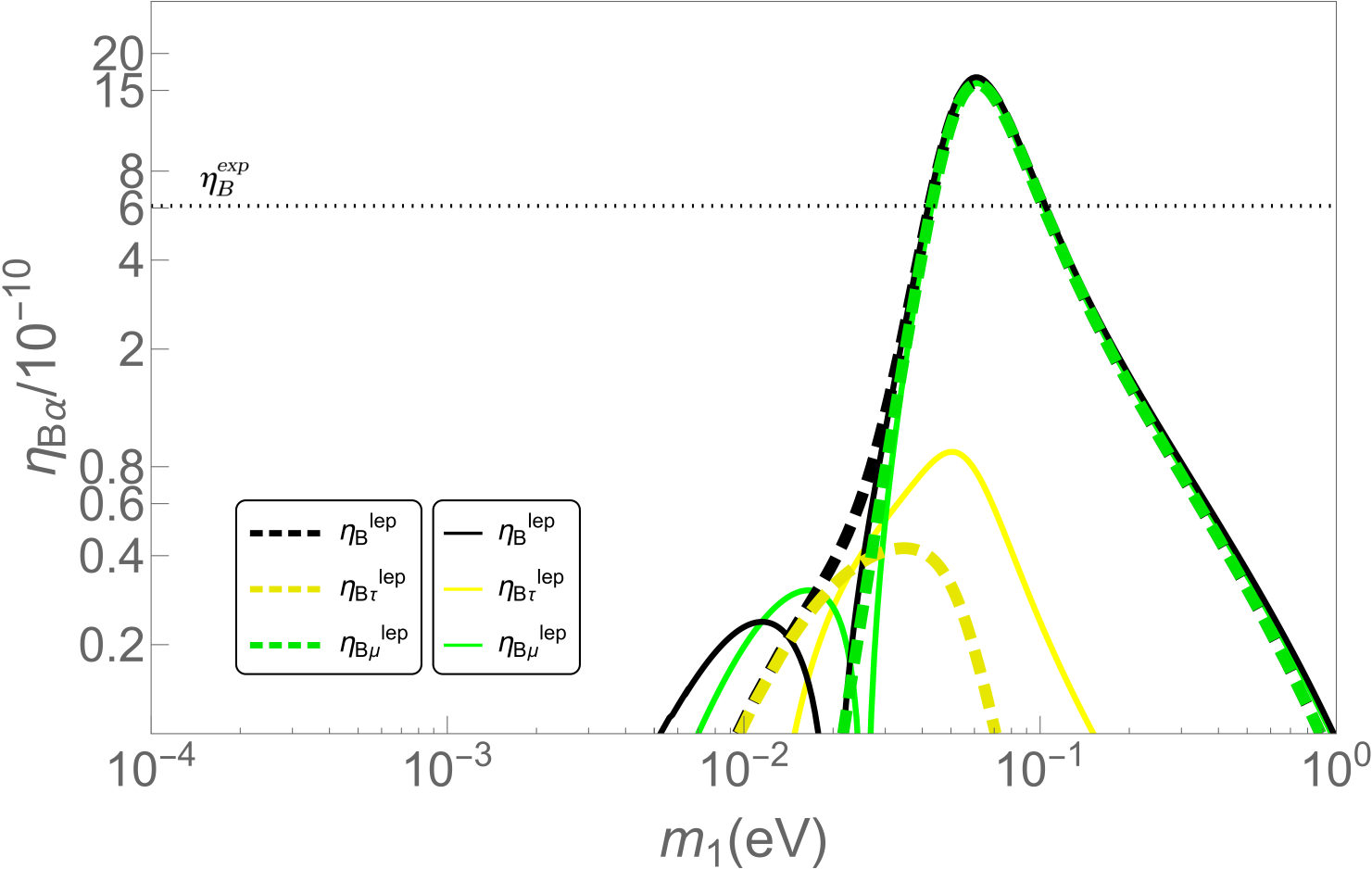,scale=0.18}  
\end{center}
\vspace{-10.9mm}
\caption{Relevant quantities for three benchmark solutions, one for each of the three types discussed in the main text:
$\tau_A$ solution (left panels), $\tau_B$ solution (central panels), $\mu$ solution (right panel).
 The values of the parameters are shown in Table~1. }\label{benchmark}
\end{figure}
The values of the different parameters for each solution are given in Table~1.  The benchmark muonic solution has been selected among those with a sizeable
sub-dominant tauonic contribution to the asymmetry. This can be clearly seen in the bottom right panel (yellow dashed-line), the tauonic 
contribution to the total asymmetry is $\sim 6\%$. However, it requires a value $m_1 = 45\,{\rm meV}$ in tension with the current cosmological upper bound.
\begin{table}
	\centering
	{\tiny
\begin{tabular}{|c|c|c|c|c|c|c|c|c|c|c|c|c|}
\hline
\multicolumn{1}{|c|}{} &  				      
\multicolumn{1}{c|}{$\theta_{12}$} &  
\multicolumn{1}{c|}{$\theta_{13}$} & 
\multicolumn{1}{c|}{$\theta_{23}$} &  
\multicolumn{1}{c|}{$\delta$} &  
\multicolumn{1}{c|}{$\rho/\pi$} & 
\multicolumn{1}{c|}{$\sigma/\pi$} & 
\multicolumn{1}{c|}{$\theta^{\rm L}_{12}$} &  
\multicolumn{1}{c|}{$\theta^{\rm L}_{13}$} & 
\multicolumn{1}{c|}{$\theta^{\rm L}_{23}$} & 
\multicolumn{1}{c|}{$\rho_{\rm L}/\pi$} &  
\multicolumn{1}{c|}{$\sigma_{\rm L}/\pi$} & 
\multicolumn{1}{c|}{$\delta_{\rm L}/\pi$}   \\
\hline 
$\tau_A$ & $33.44^\circ$   &  $8.61^\circ$  & $43.12^\circ$  & $-62.4^\circ$ & 0.31 & 0.92 & $9.46^\circ$ & $0.16^\circ$ & $0.0445^\circ$ & 0.35 & 0.49 & 1.59  \\
\hline
$\tau_B$ & $30.9^\circ$   & $8.2^\circ$  & $41.5^\circ$ & $-34.59^\circ$  &  0.952 & 0.920 &  $9.90^\circ$ & $0.035^\circ$ & $0.86^\circ$ & 0.66 & 0.99 & 1.99    \\
\hline
$\mu$ & $33.03^\circ$ & $8.41^\circ$ & $42.81^\circ$  & $103.4^\circ$ & 0.016 & 0.54 & $3.23^\circ$ & $0.12^\circ$ & $2.29^\circ$ 
& 1.01 & 1.10 & 1.91   \\
\hline
\end{tabular}}
			\caption{Values of the six low energy neutrino parameters in $U$ and six parameters in the unitary matrix $V_L$ for the
			 three benchmark solutions in Fig.~\ref{benchmark}, as indicated.  Best fit values of $m_{\rm sol}$ and $m_{\rm atm}$ are assumed.
			For each solution, the observed asymmetry is reproduced for two values of $m_1$. These three benchmark points are indicated with black stars in all scatter plots, where we choose $m_1 = 2.6\,{\rm meV}, 39\,{\rm meV}$ and $45\,{\rm meV}$ for $\tau_A$, $\tau_B$ and $\mu$ solutions, respectively.}
		\label{table}
\end{table}
Indeed, as we mentioned and one can see from Fig.~1, the current cosmological upper bound Eq.~(\ref{upperbm1}) tends to rule out
$\tau_B$ and $\mu$ solutions, so that  only the bulk of $\tau_A$ solutions is still fully allowed. 

\subsubsection{Lower bound on $m_1$}

The major feature that can be noticed in the three-dimensional scatter plot in Fig.~1 is the existence of a lower bound on $m_1$.  
It is also clear that this lower bound depends both on $\theta_{23}$ and even more strongly on $\delta$. In the light of an 
accurate measurement of these two neutrino mixing unknowns at DUNE+T2HK, it is then important to 
derive the value of the lower bound on $m_1$ for each point in the plane $\theta_{23}$-$\delta$.
In Fig.~3 we show iso-contour lines of the lower bound on $m_1$ in the plane $\delta$ versus $\theta_{23}$.
\begin{figure}[t] \vspace{-1mm}
\psfig{file=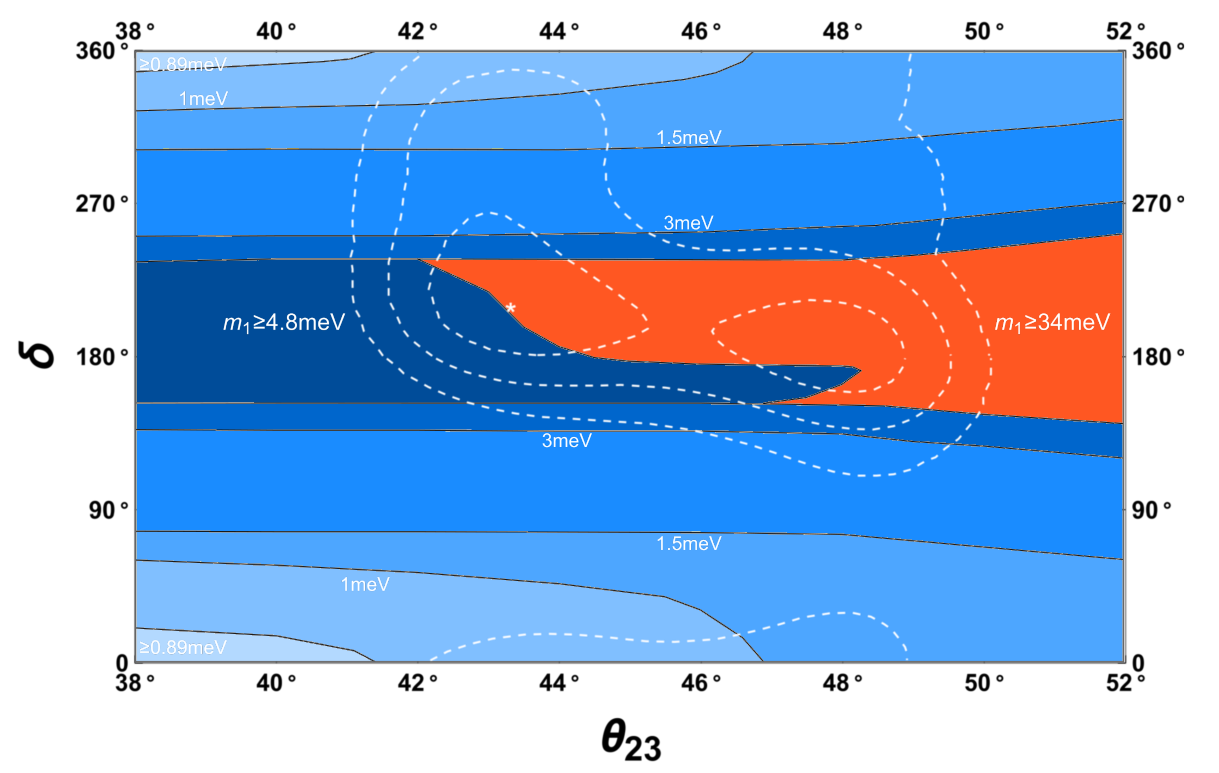,scale=0.4}
\hspace{1mm}
\psfig{file=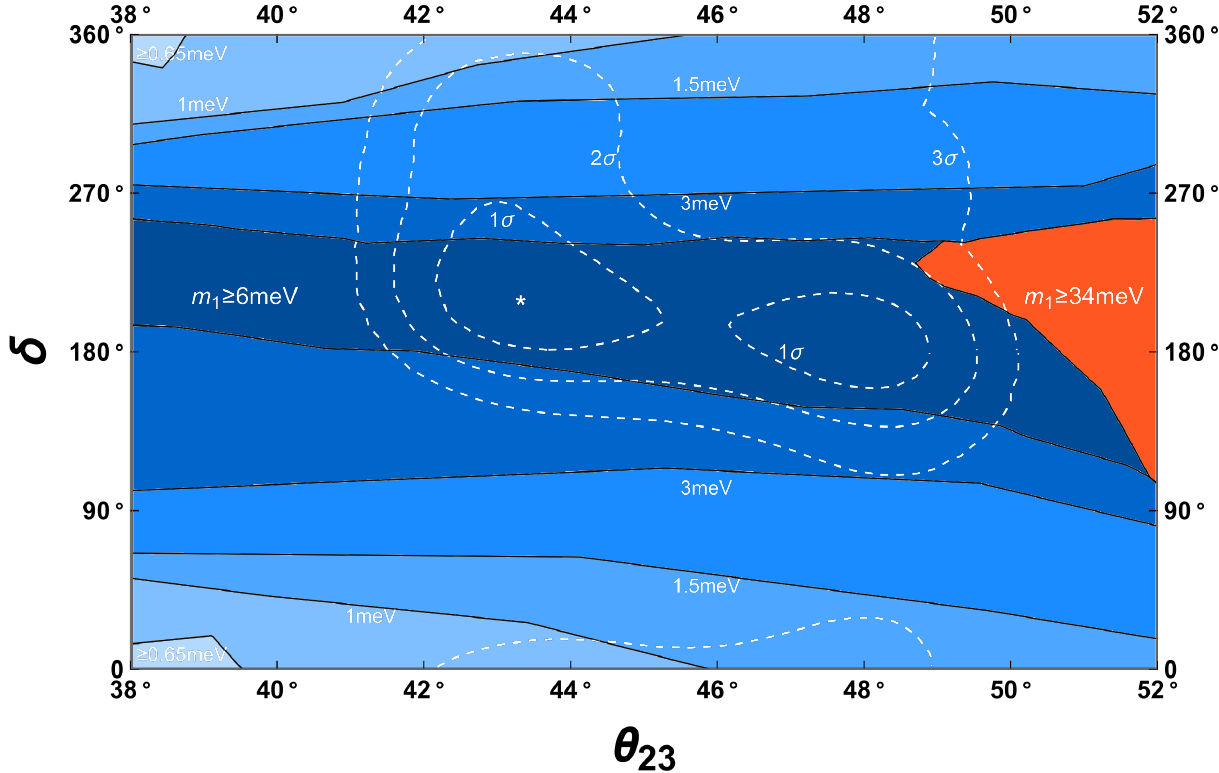,scale=0.4}
\caption{Values of the lower bound on $m_1$ in the $\theta_{23}-\delta$ plane (isocontour lines) 
without flavour coupling (left) and with flavour coupling (right). The white dashed lines are the $1\sigma$, $2\sigma$and  $3\sigma$ experimental
constraints from \cite{nufit24}. The orange area is the area excluded by the cosmological upper bound (\ref{upperbm1}).}
\end{figure}
We also superimpose the  two-dimensional experimental constraints from
latest global analyses \cite{nufit24}. It can be seen how at $3\sigma$ the global lower bound is $m_1 \gtrsim 0.9\,{\rm meV}$.
An interesting feature is that there is a large region of the plane $\theta_{23}$-$\delta$, in red colour, that is incompatible with the
cosmological upper bound Eq.~(\ref{upperbm}) and that the $1\sigma$ experimental regions fall almost entirely just within this
region. This clearly shows how SO10INLEP is being strongly tested by low energy neutrino experimental results. Therefore, it is 
very important to have a clear understanding of theoretical uncertainties to draw firm conclusions from a comparison
between theoretical calculations and experimental constraints. As we will see in the next section, flavour coupling 
plays an important role in this respect.

\subsubsection{Upper bound on $\theta_{23}$}

We can further project these solutions from the 3-dim scatter plot on two-dimensional planes, 
to show better the constraints on the different parameters.
In the left panel of Fig~4 we show the projection on the $m_1-\theta_{23}$ plane. 
\begin{figure}[t]
\centerline{
\psfig{file=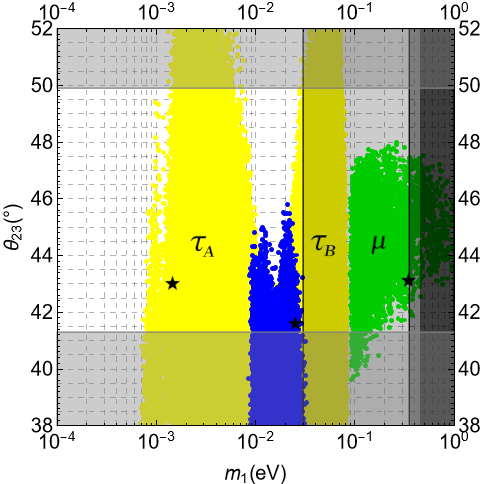,scale=0.5} \hspace{10mm}
\psfig{file=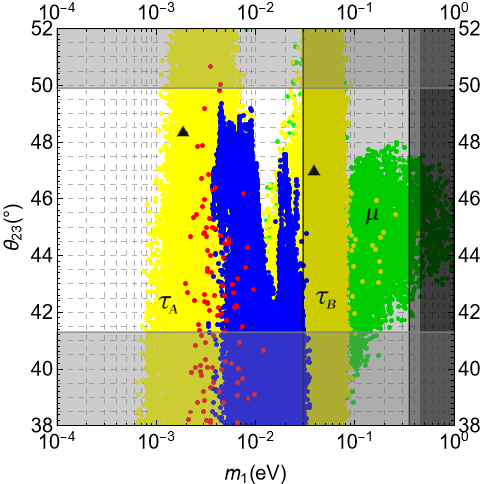,scale=0.5}
}
\caption{Two-dimensional projection on the plane $m_1$-$\theta_{23}$ of the 
scatter plot of the solutions obtained imposing successful SO10INLEP, neglecting flavour coupling
effects (left panel)  and accounting for flavour coupling effects (right panel).  Colour code as in Fig.~1.
In the left (right) panel the 3 (2) stars (triangles) denote the 3 (2) benchmark solutions in Table 1 (2).
}
\end{figure}
The most interesting feature is that, in the range $m_1 \sim (10$--$30)\,{\rm meV}$, the second octant
is not allowed. This is another important clear test for SO10INLEP and one should notice that it is independent of the value of $\delta$. 

\subsubsection{Excluded regions for $\delta$}

In the left panel of Fig~5 we also show the projection on the $m_1-\delta$ plane. 
\begin{figure}[t]
\centerline{
\psfig{file=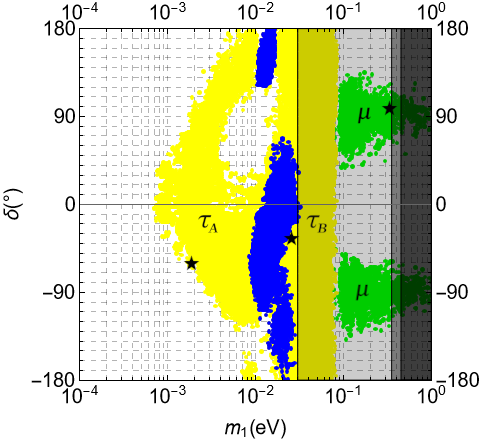,scale=0.5} \hspace{10mm}
\psfig{file=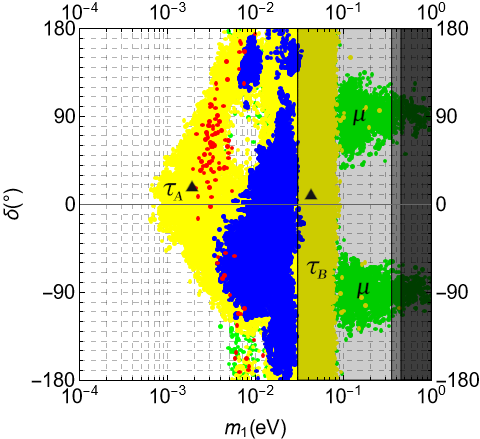,scale=0.5}
}
\caption{Two-dimensional projection on the plane $m_1$-$\theta_{23}$ of the 
scatter plot of the solutions obtained imposing successful SO10INLEP, neglecting flavour coupling (left)  
and accounting for flavour coupling (right). As in Fig.~1, the vertical grey areas
denote the excluded regions by the three upper bounds on $m_1$:
Eq.~(\ref{upperbm1}) from cosmological observations, 
Eq.~(\ref{upperbmee}) from $0\nu\b\b$  
and Eq.~(\ref{upperbmnue}) from tritium beta decay. 
The horizontal grey area is the $3\sigma$ excluded $\theta_{23}$ range of values from Eq.~(\ref{expranges}).
Same conventions as in Fig.~4.}
\end{figure}
One can notice how  the lower bound on $m_1$ is much more stringent in the half $\delta = 180^\circ \pm 90^{\circ}$ 
compared to the other half range $\delta \sim 0 \pm 90^{\circ}$. In particular, the current $1 \sigma$ experimental range of
values for $\delta$ falls in the first half that is allowed only for $m_1 \gtrsim 10\,{\rm meV}$.
It is also curious how there is a kind of hole for $\delta \sim 30^\circ-120^\circ$ and $m_1 \sim (5$--$15)\,{\rm meV}$.
This shows again how a future precise determination of $\delta$ will represent a crucial test for SO10INLEP.

\subsubsection{Majorana phases}
 
In the left panel of Fig~6 we show the projection of the scatter plot solutions on the $\sigma-\rho$ plane.  One can notice
how large part of the plane is excluded. This is because, as well understood analytically \cite{decrypting,full}, the Majorana
phases need to be within certain ranges in order for either $K_{1\tau} \lesssim 1$ or $K_{1\mu} \lesssim 1$ (or both). 
\begin{figure}[t]
\centerline{
\psfig{file=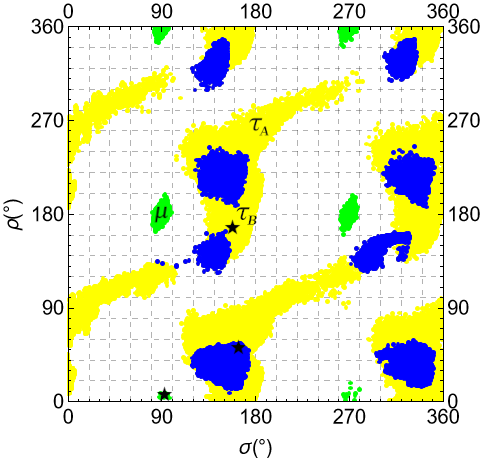,scale=0.5} \hspace{10mm}
\psfig{file=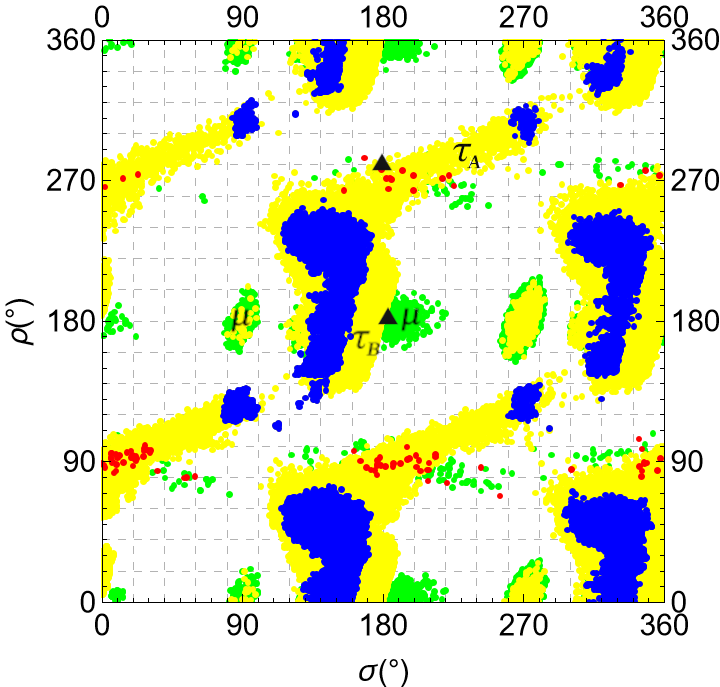,scale=0.5}
}
\caption{Two-dimensional projection on the plane $\sigma$-$\rho$ of the 
scatter plot of the solutions obtained imposing successful SO10INLEP, neglecting flavour coupling
effects (left panel)  and accounting for flavour coupling effects (right panel). Same conventions as in Fig.~4.
}
\end{figure}
Notice how in this plane the muon solution allowed  regions (green points) 
are well disconnected by the tauon solution allowed region (yellow points) 
and lay around the points $(\sigma,\rho) = ( (2n+1)\pi/2, m\pi)$, with $m,n \in {\mathbb Z}$.  
These are indeed the values that make possible to have $K_{1\mu} \lesssim 1$ \cite{full}.
  
In Fig.~7 we also show the projection on the plane $m_1-\sigma$.  It can be clearly seen how the values of $\sigma$
for the muon solutions are different from those for the $\tau$ solutions.
\begin{figure}[t]
\centerline{
\psfig{file=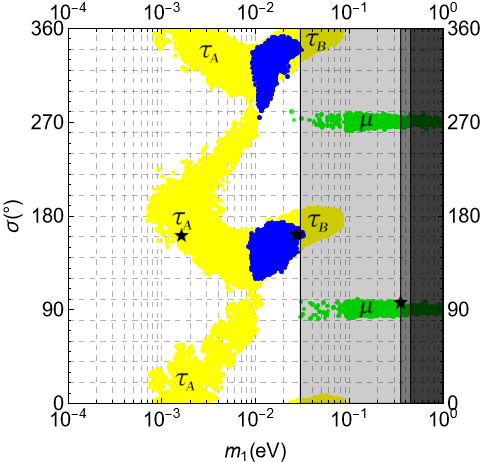,scale=0.5} \hspace{10mm}
\psfig{file=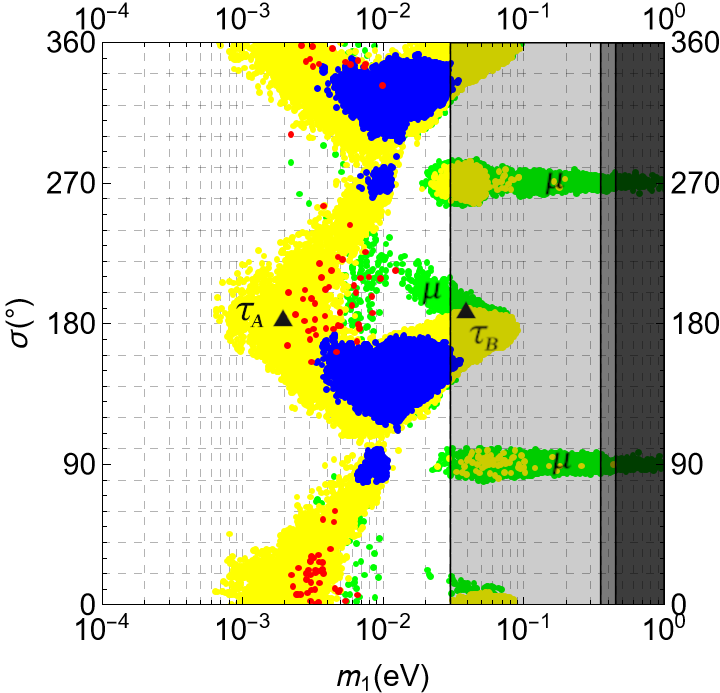,scale=0.5}
}
\caption{Two-dimensional projection on the plane $m_1$-$\theta_{23}$ of the 
scatter plot of the solutions obtained imposing successful SO10INLEP, neglecting flavour coupling
effects (left panel)  and accounting for flavour coupling effects (right panel). Same conventions as in Fig.~4.
}
\end{figure}

\subsubsection{$0\nu\b\b$ effective neutrino mass}

The observation of a $0\nu\b\b$ signal would be a crucial discovery supporting, in general, leptogenesis since it would
be a clear sign that lepton number is violated at tree level.  Specifically, for SO10INLEP, it would provide the measurement 
of an additional low energy neutrino observable depending on all nine low energy neutrino parameters. This would provide
another important experimental constraint to be satisfied. In Fig.~8 we show the allowed region in the usual plane $m_{ee}$ versus $m_1$.    
\begin{figure}[t]
\centerline{
\psfig{file=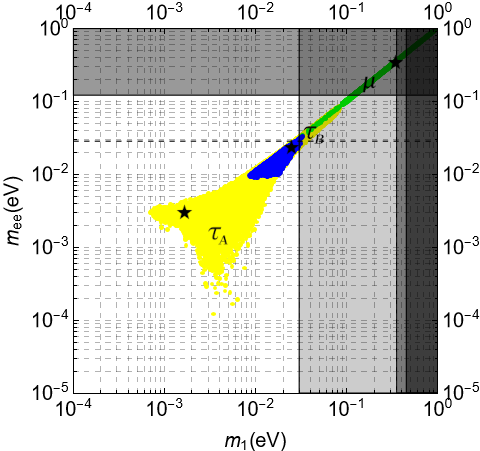,scale=0.5} \hspace{3mm}
\psfig{file=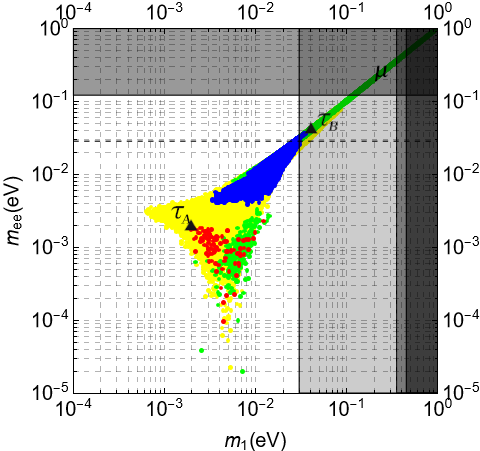,scale=0.5}
}
\caption{Projection in the plane $m_1-m_{ee}$ of the scatter plot of the solutions obtained imposing successful SO10INLEP, neglecting flavour coupling
effects (left panel)  and accounting for flavour coupling (right panel). Same conventions as in Fig.~4. The grey lines delimit the general allowed region
obtained from the experimental ranges in Eq.~(\ref{expranges}).
}
\end{figure}
The vertical grey area is the conservative excluded region from the KamLAND-Zen upper bound in Eq.~(\ref{upperbmee}), while the dashed line indicates
the most stringent upper bound also given in Eq.~(\ref{upperbmee}).   
The grey lines indicate the general allowed region compatible with the experimental constraints (\ref{expranges})
(i.e., no successful SO10INLEP condition is imposed). One can see how SO10INLEP strongly restricts this region and, importantly,
now there is a lower bound on $m_{ee}$ also in the range $2\,{\rm meV} \lesssim m_1 \lesssim 7\,{\rm meV}$,  
even though we are considering NO.  Therefore, the simultaneous experimental determination of $m_1$ and $m_{ee}$, 
realistically for both quantities $\gtrsim 1\,{\rm meV}$, would provide a very strong test in favour of SO10INLEP. This is true especially  
if $m_1$ is found in that particular range, also considering that the bulk of solutions, of $\tau_A$-type, do fall in that range.\footnote{This argument
contains some subjectivity depending whether one considers $m_1$ linearly or logarithmically: in the latter case, without SO10INLEP, one would expect
not to measure $m_{ee}$ for $2\,{\rm meV} \lesssim m_1 \lesssim 7\,{\rm meV}$. This expectation neglects that
arbitrarily small values of $m_{ee}$ are obtained for fine-tuned phase cancellations. This also contributes to the level of subjectiveness.
In any case, it is true, that with SO10INLEP such fine-tuned phase cancellations are not possible, since Majorana phases, as we have seen, 
are constrained within regions allowing $K_{1\tau}\lesssim 1$ and this justifies our statement 
that a simultaneous determination of $m_1$ and $m_{ee}$ would be a very strong test in favour of SO10INLEP.}

\section{Including flavour coupling}

The expression for the final $B-L$ asymmetry in Eq.~(\ref{twofl})  assumes that the evolution of the three flavoured asymmetries 
$\D_\alpha \equiv B/3 - L_{\alpha}$ ($\alpha=e,\mu,\tau$), where the $L_{\alpha}$'s 
are the  lepton asymmetries in lepton doublets, proceeds independently of each other, in an uncoupled way. 
Sphaleron process conserve each $\D_\alpha$ and the total $B-L$ asymmetry is the sum of the three. 
In this way one can solve the set of uncoupled differential equations for the $\D_\alpha$'s.
At the end $N_{B-L}^{\rm f} = \sum_\alpha N_{\D_\a}^{\rm f}$ and the result is the expression in Eq.~(\ref{twofl}) \cite{vives,bounds,fuller,density}.
The assumption of uncoupled evolution of the flavoured asymmetries relies on a simple picture where:  $N_2$-decays source initially the $B-L$
asymmetry in a specific direction in flavour space (a ray, in space vector language) 
denoted by  ${\ell}_2$;  decoherence of leptons states results into an independent washout from $N_2$-inverse processes 
for the two flavoured asymmetries  $\Delta_{\tau}$ and $\Delta_e + \Delta_{\mu}$ at $T\sim M_2$ (two flavour regime) 
and into an independent washout from $N_1$-inverse processes for all three $\D_{\alpha}$ at $T\sim M_1$;
finally, it assumes that the washout is just depending on each $\D_\alpha$ separately. Within this simple picture, though decoherence
splits the evolution of the asymmetry first in two and then in three flavoured asymmetries, these still develop independently of each other because the
 the washout in a certain flavour is always depending, linearly in very good approximation, just on the asymmetry in that flavour. 
 However, this picture is incomplete because it neglects spectator processes of different nature 
 \cite{Barbieri:1999ma,Buchmuller:2001sr,flavoureffects,bounds}:
(i) the generation of an asymmetry from $N_2$-decays into lepton doublets is also accompanied by the generation of 
a hypercharge asymmetry into the Higgs bosons and 
(ii) by a baryonic asymmetry into quarks via sphaleron processes;
(iii) the lepton asymmetry from lepton doublets is also redistributed to RH charged particles. 
In this way the washout of a $\D_{\alpha}$ asymmetry also depends on the asymmetries stored into the
Higgs doublets (this is the primary source of flavour coupling) and quarks (this a secondary source of flavour coupling).  
To be more specific, since the Higgs doublet asymmetry is flavour blind,  if a $\Delta_\alpha$ asymmetry has been generated in a certain 
flavour, necessarily at the washout the asymmetry into the Higgs doublets will induce the generation of a $\Delta_{\beta\neq\alpha}$ 
asymmetry in the other flavour(s) $\beta$:  therefore, inverse processes do in this case generate a flavoured asymmetry rather than wash it out.   
This means that the washout applies to the Higgs asymmetry and the $N_2$-decay produced $\D_{\alpha}$ asymmetry, 
not to  $\Delta_{\beta\neq\alpha}$.  This is the primary source of flavour coupling. 
Secondarily, the redistribution of the asymmetries into quarks and RH charged particles also 
couples the $\D_\alpha$ asymmetries. 

\subsection{A more tangled expression for the final asymmetry}

Solving the system of coupled differential equations at the production, for $T\sim M_2$, and at the $N_1$-washout, for $T \sim M_1$,  
leads to the following analytical expression for each final $\D_{\alpha}$ asymmetry \cite{fuller}:
\be\label{NfDalpha}
N^{\rm f}_{\D_{\a}}  =  \sum_{\a''}\,V^{-1}_{\a\a''}\,
\left[N^{T\sim T_L}_{\a''}\,e^{-{3\pi\over 8}\,K_{1\a''}}\right] \,  ,
\ee
where the three $N^{T\sim T_L}_{\a''}$ ($\alpha'' = e'', \mu'', \tau''$) are the asymmetries produced by $N_2$-decays
at $T= T_L \sim M_2$ in the (rotated) flavours $\alpha''$ and are given by
\be\label{NfDalpha2}
N^{T\sim T_L}_{\a''} = \sum_{\b=e,\mu,\tau}\,V_{\a''\b}\,N_{\D_{\b}}^{T\sim T_L} \,  .
\ee
Before explaining in detail the meaning and definition of all involved quantities in this general expression, notice that
now the final $B-L$ asymmetry, though still given by $N^{\rm f}_{B-L} = \sum_{\a =e,\mu,\tau} N^{\rm f}_{\D_{\a}}$,
is the sum of 27 terms instead of just 3. The matrix
\be
V\equiv \left(\begin{array}{ccc}
V_{ e'' e} & V_{e''\m} & V_{e''\t} \\
V_{\m'' e} & V_{\m''\m} & V_{\m''\t} \\
V_{\t'' e} & V_{\t''\m} & V_{\t''\t}
\end{array}\right)
\ee
is the matrix that diagonalizes $P^0_1$, i.e. $V\,P^0_{1}\,V^{-1} = P^0_{1''} \equiv {\rm diag}(P^0_{1 e''},P^0_{1\m''},P^0_{1\t''}) $,
where we defined $P^0_{1\alpha} \equiv K_{1\alpha}/K_1$ and the rotated flavoured decay parameters 
as $K_{1\alpha''} \equiv P^0_{1 \alpha''} \, K_1$.  The matrix $P^0_1$ is given by
\be
P_1^0 \equiv
\left(\begin{array}{ccc}
P^0_{1e}\,C_{ee}^{(3)} & P^0_{1e}\,C_{e\m}^{(3)} & P^0_{1e}\,C_{e\t}^{(3)} \\
P^0_{1\m}\,C_{\m e}^{(3)} & P^0_{1\m}\,C_{\m\m}^{(3)} & P^0_{1\m}\,C_{\m \t}^{(3)} \\
P^0_{1\t}\,C_{\t e}^{(3)} & P^0_{1\t}\,C_{\t\m}^{(3)} & P^0_{1\t}\,C_{\t\t}^{(3)} 	
\end{array}\right) \, ,
\ee
and the three-flavour coupling matrix \cite{Davidson:2008bu}
\be
C^{(3)} \equiv
\left(\begin{array}{ccc}
C_{ee}^{(3)} & C_{e\m}^{(3)} & C_{e\t}^{(3)} \\
C_{\m e}^{(3)} & C_{\m\m}^{(3)} & C_{\m \t}^{(3)} \\
C_{\t e}^{(3)} & C_{\t\m}^{(3)} & C_{\t\t}^{(3)}	
\end{array}\right) =
\left(\begin{array}{ccc}
188/179 & 32/179 & 32/179 \\ 49/358 & 500/537 & 142/537 \\ 49/358 & 142/537 & 500/537	
\end{array}\right) \,  .
\ee
The three flavour asymmetries $N_{\D_{\b}}^{T\sim T_L}$ in the standard flavours $\beta = e,\mu,\tau$, at the production, are given by
\bea \nonumber
N_{\D_{\t}}^{T\sim T_L}  & = & 
U^{-1}_{\t\tau_2^{\bot'}}\left[U_{\tau_2^{\bot'}\tau_2^\bot}\,\ve_{2\tau_2^{\bot}}+U_{\tau_2^{\bot'}\t}\,\ve_{2\t}\right]\,\kappa(K_{2\tau_2^{\bot}})
+U^{-1}_{\t\t'}\left[U_{\t'\tau_2^{\bot}}\,\ve_{2\tau_2^{\bot}}+U_{\t'\t}\,\ve_{2\t}\right]\,\kappa(K_{2\t})  \,   ,  \\ \nonumber
N_{\D_{e}}^{T\sim T_L}  & = &  \left[{K_{2e}\over K_{2\tau_2^{\bot}}}\,N_{\D_{\tau_2^{\bot}}}^{T\sim T_L} 
+ \left(\ve_{2e} - {K_{2e}\over K_{2\tau_2^{\bot}}}\, \ve_{2 \tau_2^{\bot}} \right)\,\kappa(K_{2 \tau_2^{\bot}}/2)\right] \,   ,\\
N_{\D_{\mu}}^{T\sim T_L}  & = &  \left[{K_{2\mu}\over K_{2 \tau_2^{\bot}}}\,N_{\D_{\tau_2^{\bot}}}^{T\sim T_L} +
\left(\ve_{2\mu} - {K_{2\mu}\over K_{2\tau_2^{\bot}}}\, \ve_{2 \tau_2^{\bot}} \right)\,
\kappa(K_{2 \tau_2^{\bot}}/2) \right] \,  ,
\eea
where
\bea \label{Dg} \nonumber
N_{\D_{\tau_2^{\bot}}}^{T\sim T_L} & = &
 U^{-1}_{\tau_2^{\bot}\tau_2^{\bot '}}\left[U_{\tau_2^{\bot '}\tau_2^{\bot}}\,\ve_{2\tau_2^{\bot}}+U_{\tau_2^{\bot '}\t}\,\ve_{2\t}\right]\,\kappa(K_{2\tau_2^{\bot}})
+U^{-1}_{\tau_2^{\bot}\t'}\left[U_{\t'\tau_2^{\bot}}\,\ve_{2\tau_2^{\bot}}+U_{\t'\t}\,\ve_{2\t}\right]\,\kappa(K_{2\t}) \,  .
\eea
The matrix 
\be\label{Umatrix}
U\equiv \left(\begin{array}{cc}
U_{\tau^{\bot '}_2\tau^\bot_2} & U_{\tau^{\bot '}_2\t}  \\
U_{\t'\tau^\bot_2} & U_{\t'\t}
\end{array}\right)
\ee
is the analogous of $V$, accounting for flavour coupling in the two-flavour regime. It is defined as the matrix that diagonalises
\be
P^0_2 \equiv
\left(\begin{array}{cc}
P^0_{2\tau^\bot}\,C_{\tau^\bot_2\tau^\bot_2}^{(2)} & P^0_{2\tau^\bot_2}\,C_{\tau^\bot_2\t}^{(2)}  \\
P^0_{2\tau}\,C_{\t\tau^\bot_2}^{(2)} & P^0_{2\tau}\,C_{\t\t}^{(2)}
\end{array}\right) \, ,
\ee
i.e. $U\,P^0_{2}\,U^{-1} ={\rm diag}(P^0_{2\tau^{\bot '}_2},P^0_{2\t'})$, where  the
two-flavour coupling matrix is given by \cite{Davidson:2008bu}
\be
C^{(2)} \equiv
\left(\begin{array}{ccc}
C^{(2)}_{\tau^\bot_2\tau^\bot_2} & C^{(2)}_{\tau^\bot_2\t}  \\  C^{(2)}_{\t\tau^\bot_2} & C^{(2)}_{\t\t}
\end{array}\right) =
\left(\begin{array}{ccc}
581/589 & 104/589 \\ 194/589 & 614/589 \end{array}\right) 
\ee
and we defined $P^0_{2\alpha} \equiv K_{2\alpha}/K_2$. We have now fully defined all quantities entering the expression for the
flavoured asymmetries in Eq.~(\ref{NfDalpha}). As we said, the final $B-L$ asymmetry is still obtained as the sum of the three flavoured asymmetries 
and it can be checked that, for $U=V=I$, one recovers the expression Eq.~(\ref{twofl}) obtained neglecting flavour coupling.
We can also unpack the first sum in  Eq.~(\ref{NfDalpha}) and write explicitly:
\bea\label{NfDa2}
N^{\rm f}_{\D_{\a}}
 & = &  V^{-1}_{\a e''}\,
\left[\sum_{\b}\,V_{e''\b}\,N_{\D_{\b}}^{T\sim T_L}\right]
\,e^{-{3\pi\over 8}\,K_{1 e''}} \\  \nonumber
& + &  V^{-1}_{\a \m''}\,
\left[\sum_{\b}\,V_{\m''\b}\,N_{\D_{\b}}^{T\sim T_L}\right]
\,e^{-{3\pi\over 8}\,K_{1 \m''}} \\ \nonumber
& + &  V^{-1}_{\a \t''}\,
\left[\sum_{\b}\,V_{\t''\b}\,N_{\D_{\b}}^{T\sim T_L}\right]
\,e^{-{3\pi\over 8}\,K_{1 \t''}} \, .
\eea
This expression shows how now each flavoured asymmetry $N^{\rm f}_{\D_{\a}}$
is not simply given by one term containing the $N_1$ washout exponential
suppression term  described by $e^{-3\pi\,K_{1\a}/8}$ but it also contains terms
that are washed out by exponentials $e^{-3\pi\,K_{1\d''}/8}$ with $\delta\neq \alpha$. 
In this way, even though $K_{1\a}\gg 1$, there can still be unsuppressed contributions
to $N^{\rm f}_{\D_{\a}}$ from terms with $K_{1\d}\ll 1$. Even though
these terms are weighted by factors $V^{-1}_{\a\d}$,
containing off-diagonal terms of the $C^{(3)}$ matrix that are ${\cal O}(0.1)$, they
might be dominant in some cases and therefore, in general, they have to
be accounted for.

\subsection{Successful leptogenesis solutions} 

Despite the much higher intricacy of the expression Eq.~(\ref{NfDalpha}), obtained accounting for flavour coupling, compared
to Eq.~(\ref{twofl}), obtained neglecting flavour coupling, the discussion made in 2.3 on the parameter dependence
and the consequent successful leptogenesis condition Eq.~(\ref{successfullep}) still holds. Flavour coupling
does not introduce any new parameter. It only introduces many new terms in the expression for the final asymmetry
that are proportional to the off-diagonal terms in the flavour coupling matrix. These are usually sub-dominant and give a small
correction but for particular values of the parameters they might yield new solutions in  regions of parameter space 
that would otherwise be unaccessible when flavour coupling is neglected.

The search of these solutions is the main objective of including flavour coupling. The result  is shown in the right panel of Fig.~1. 
Here we show again the three-dimensional projection of a scatter plot containing about $2\times 10^6$ solutions
obtained for $\alpha_2 = 5$ and scanning over the 15 parameters in $m_\nu$ and $V_L$, 
respecting the experimental constraints in Eq.~(\ref{expranges}). The solutions have been again obtained imposing
$\chi^2 < \chi^2_{\rm max} =25$ and  $\eta_B > 6.01 \times 10^{-10}$. 
The first thing one can notice is that the main structure of the allowed region is not changed by flavour coupling and, in particular,
one still has a lower bound on $m_1$. On the other hand, it can be noticed how there are new muonic solution appearing in a region that 
was before excluded, at values of $m_1 \sim 10\,{\rm meV}$. Moreover, electronic solutions, indicated by red points, now appear in the region of $\tau_A$ solutions.  
 
\subsubsection{Two new types of solutions}

Let us have a closer look a these new solutions. As we did in the last section for the three types of old solutions, in 
Fig.~9 we plot different quantities, $M_I, \varepsilon_\alpha, K_{1\alpha}, K_{2\tau}$ and $K_{2\tau^{\bot}_2}$, versus $m_1$ for 
two benchmark new solutions, one muonic and one electronic, with values of the parameters given in Table 2. 
\begin{figure}[t]
\begin{center}
\psfig{file=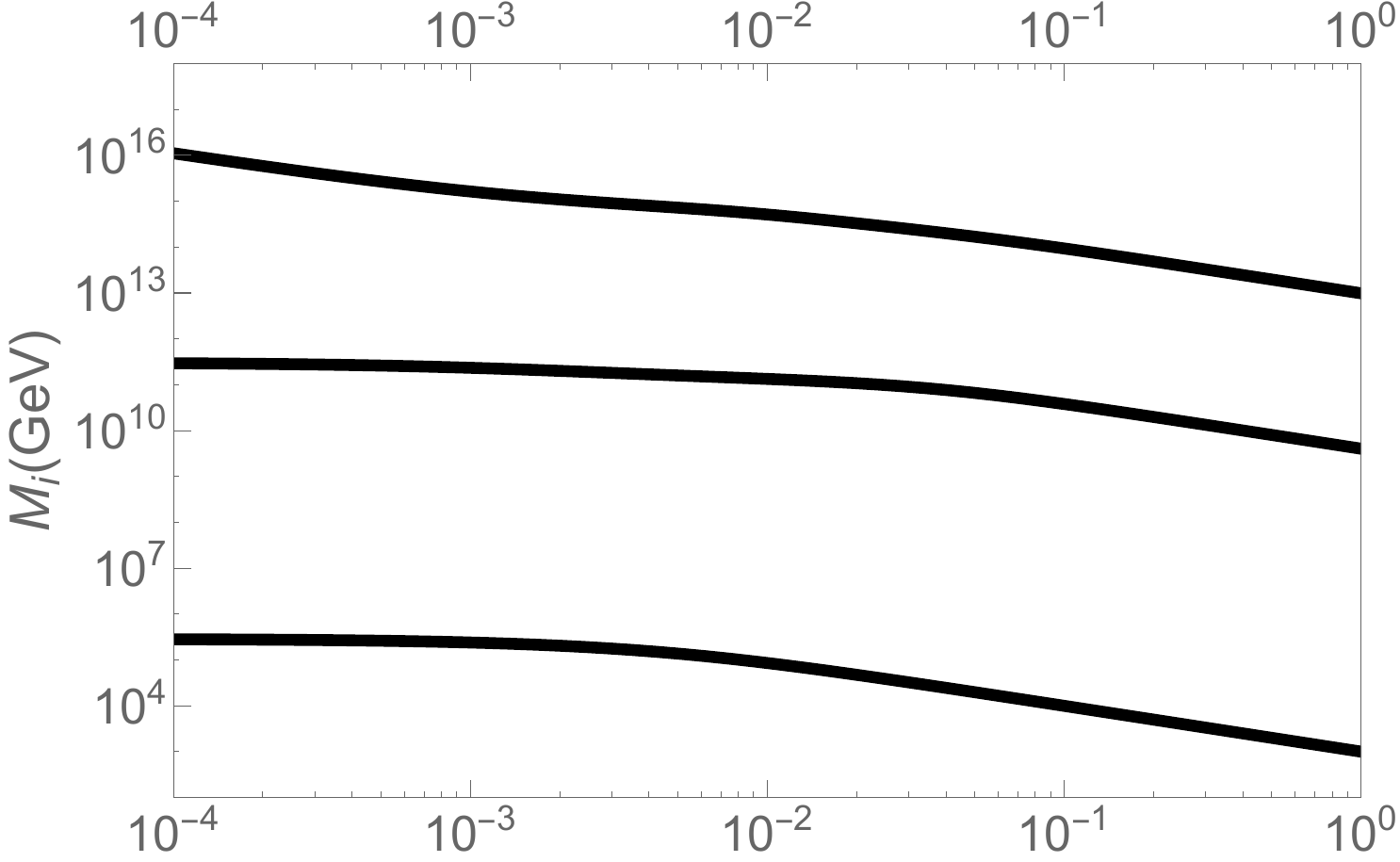,scale=0.18}\hspace{2mm}
\psfig{file=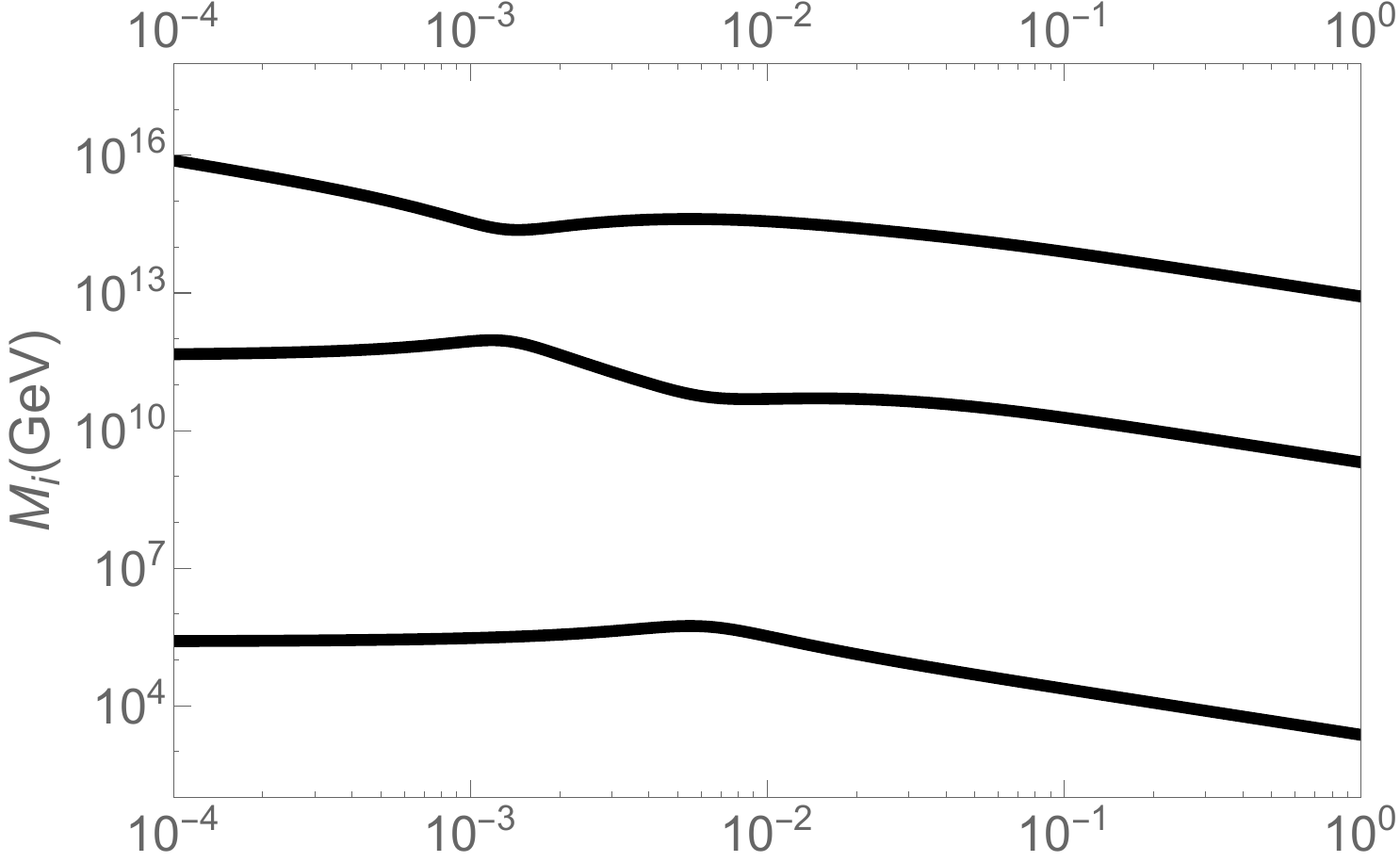,scale=0.18}\hspace{2mm}   \\ \vspace{0mm}
\psfig{file=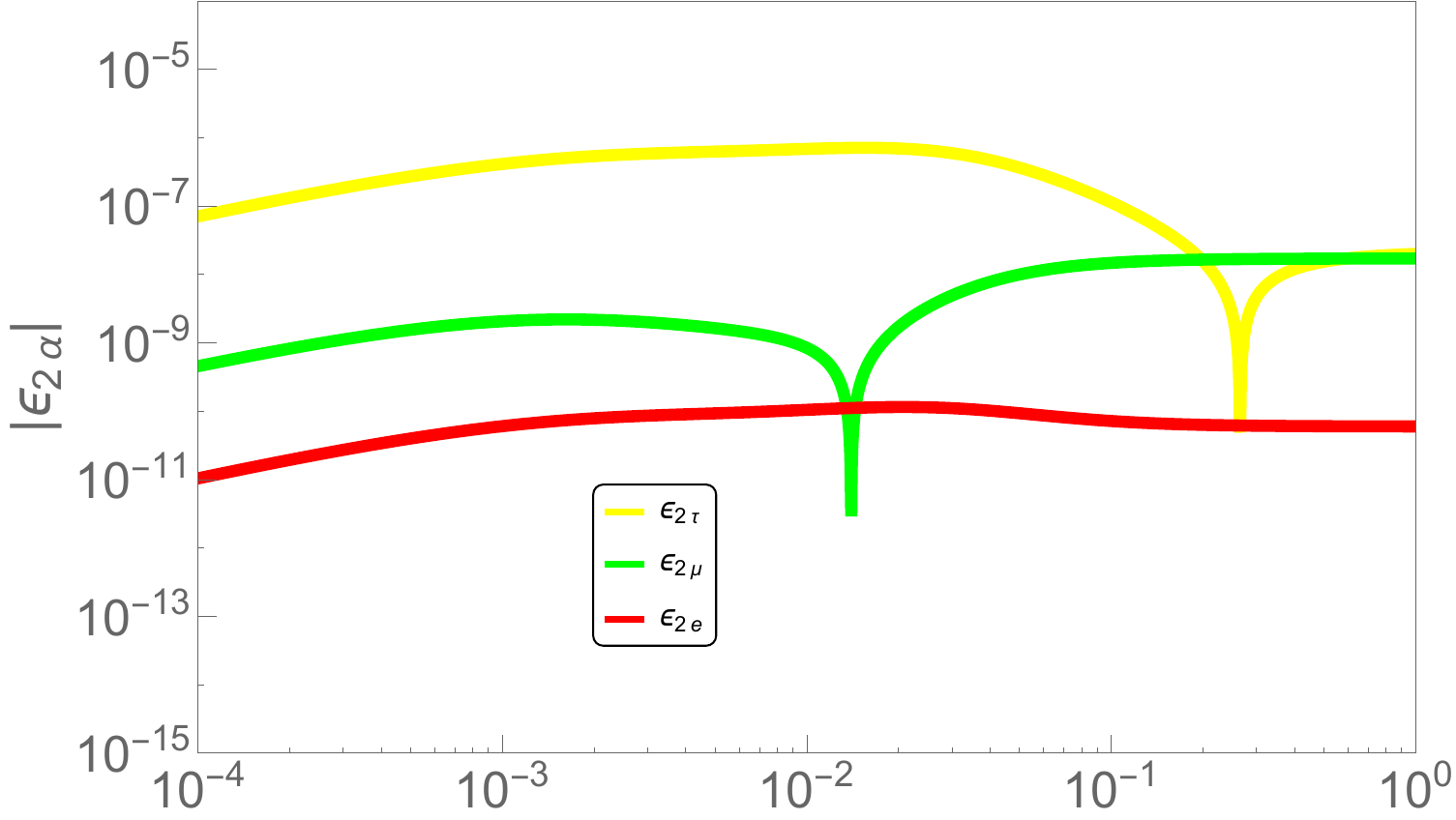,scale=0.18}\hspace{2mm}
\psfig{file=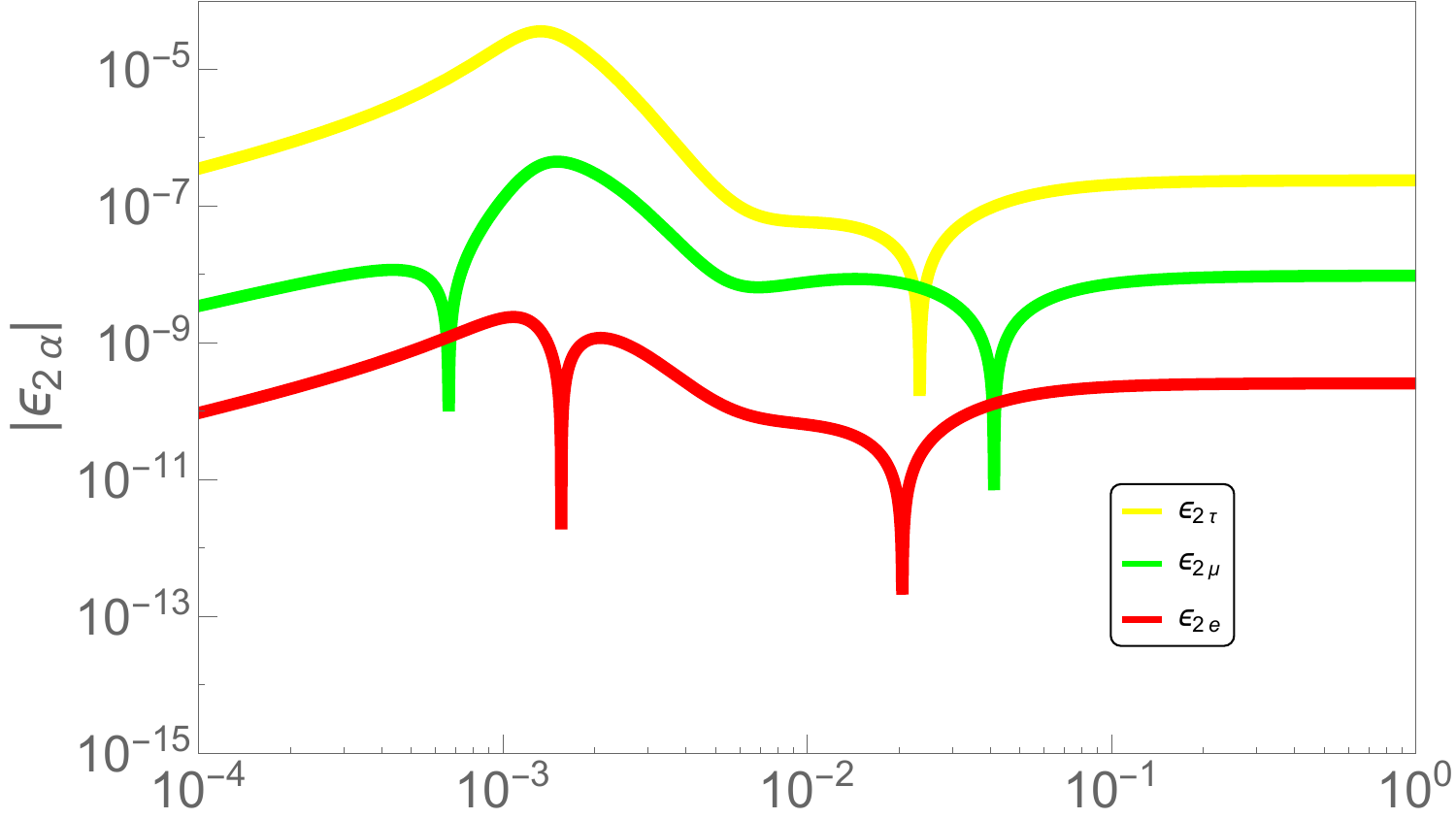,scale=0.18}\hspace{2mm} \\ \vspace{0mm}
\psfig{file=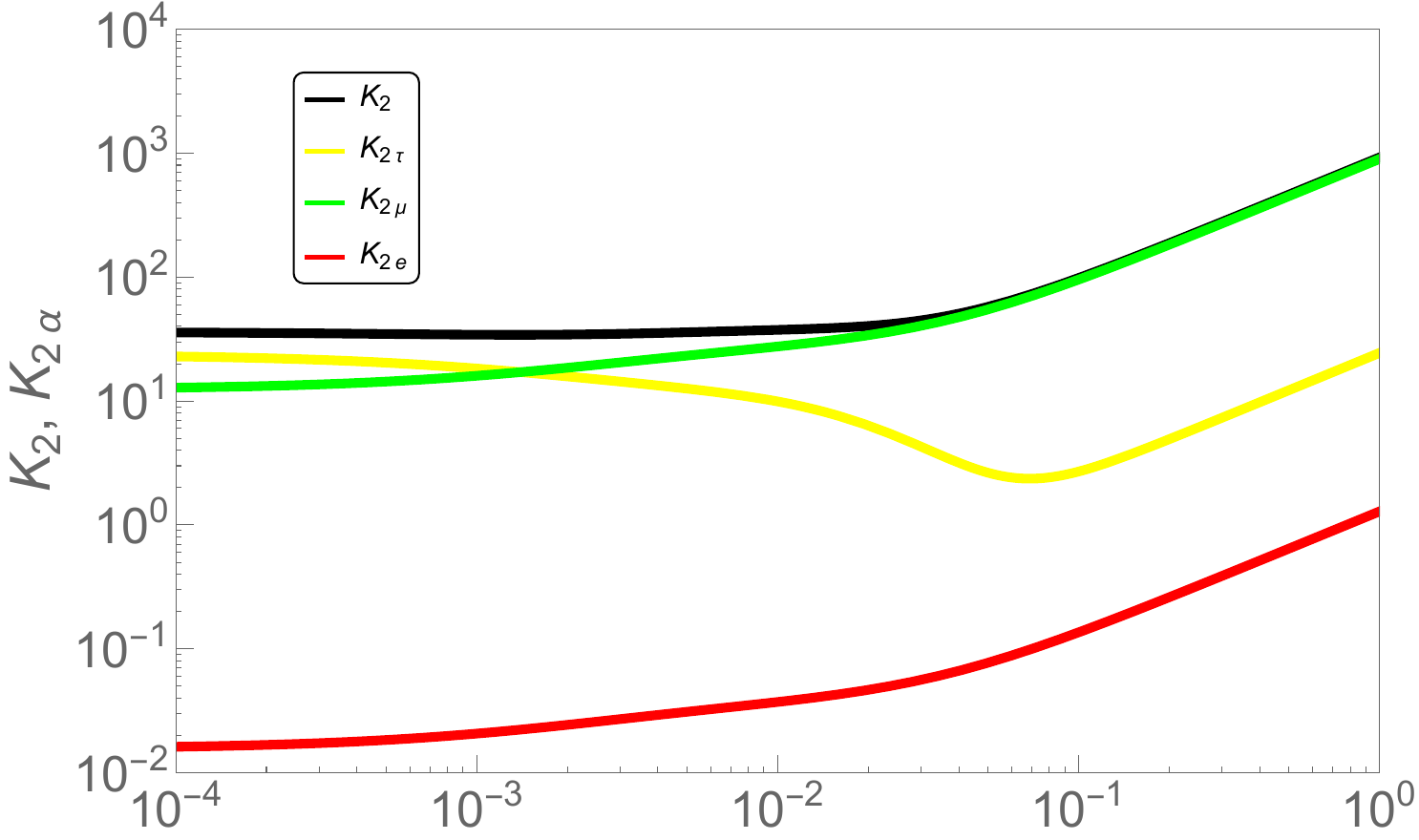,scale=0.18}\hspace{2mm}
\psfig{file=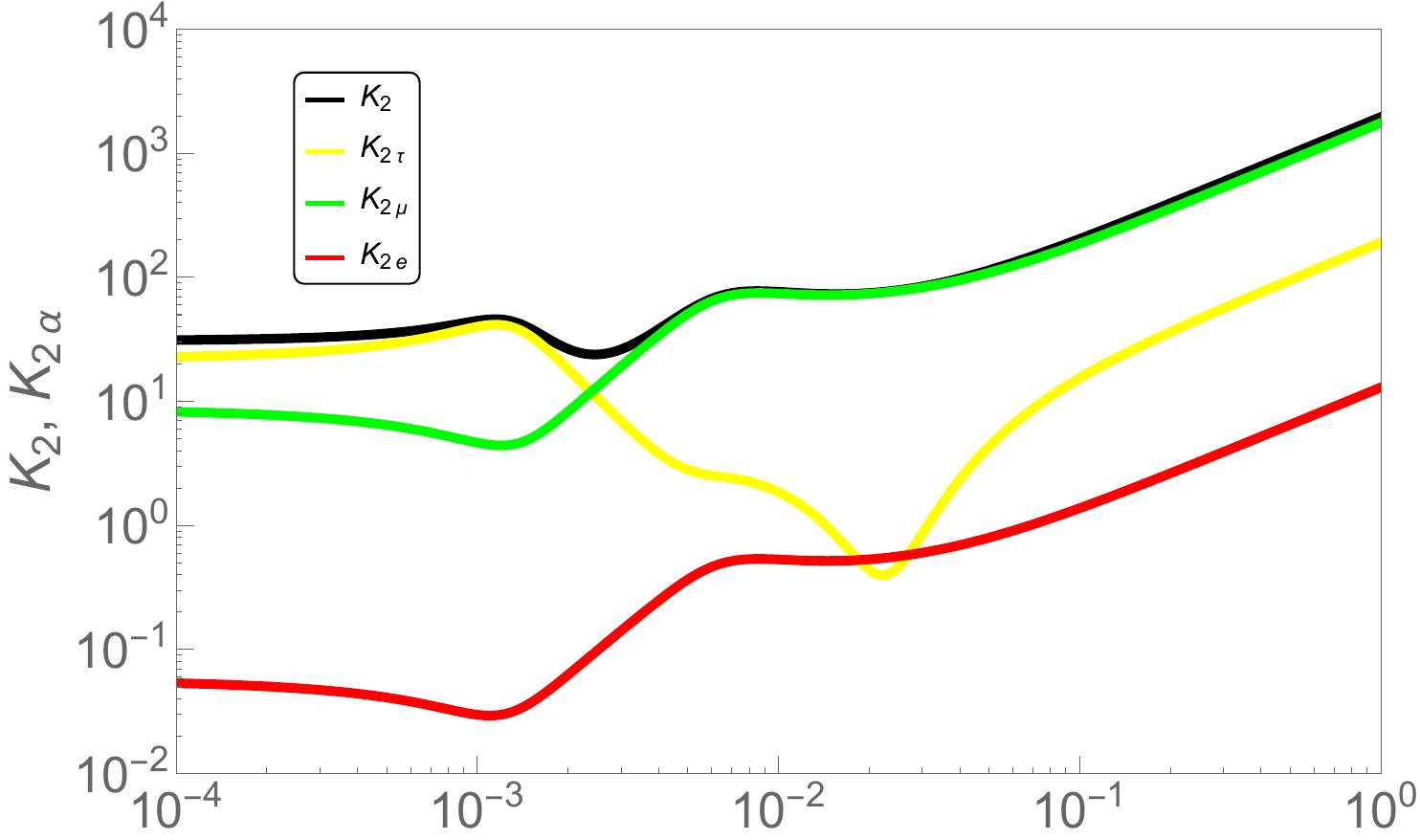,scale=0.18}\hspace{2mm}  \\ \vspace{0mm}
\psfig{file=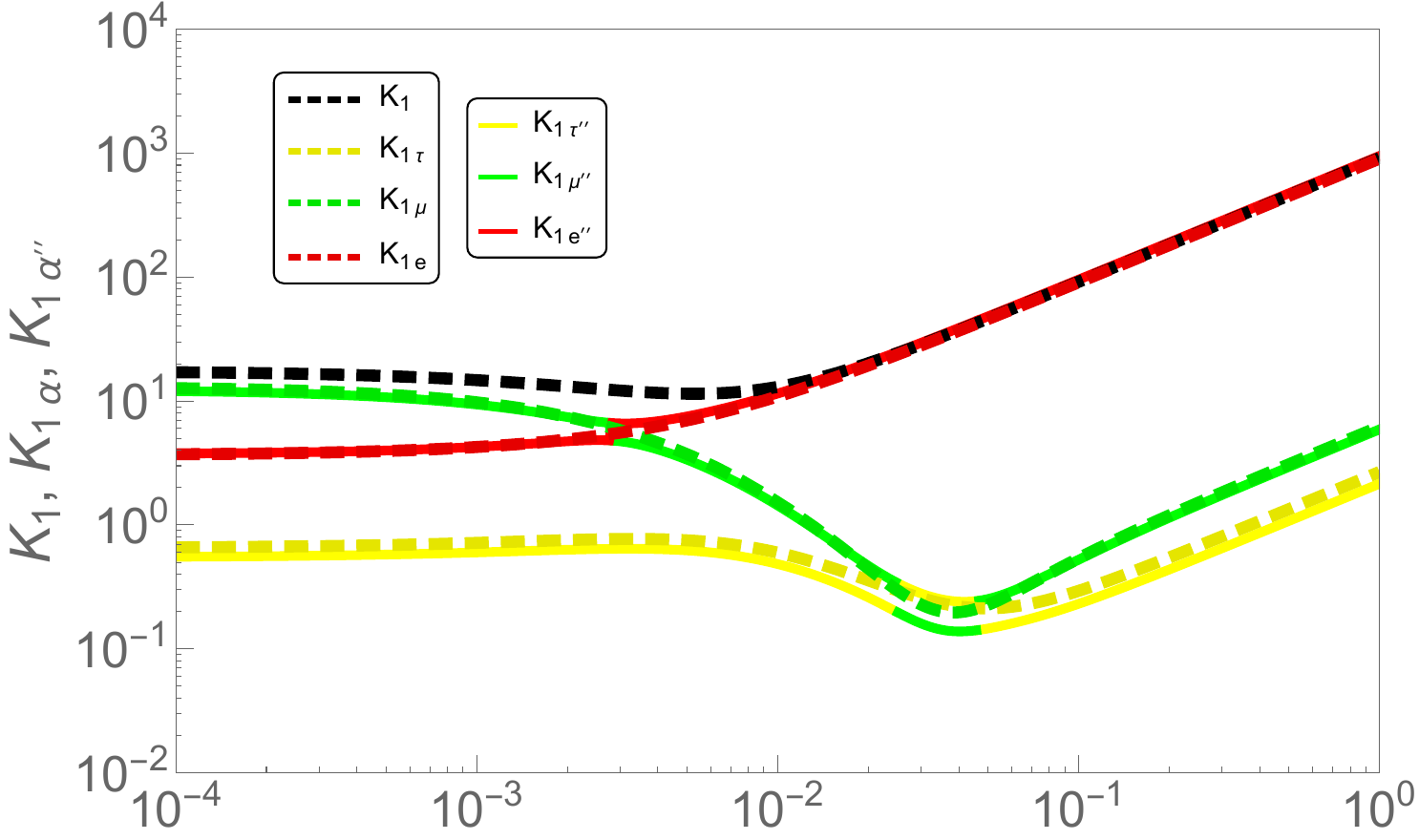,scale=0.18}\hspace{2mm}
\psfig{file=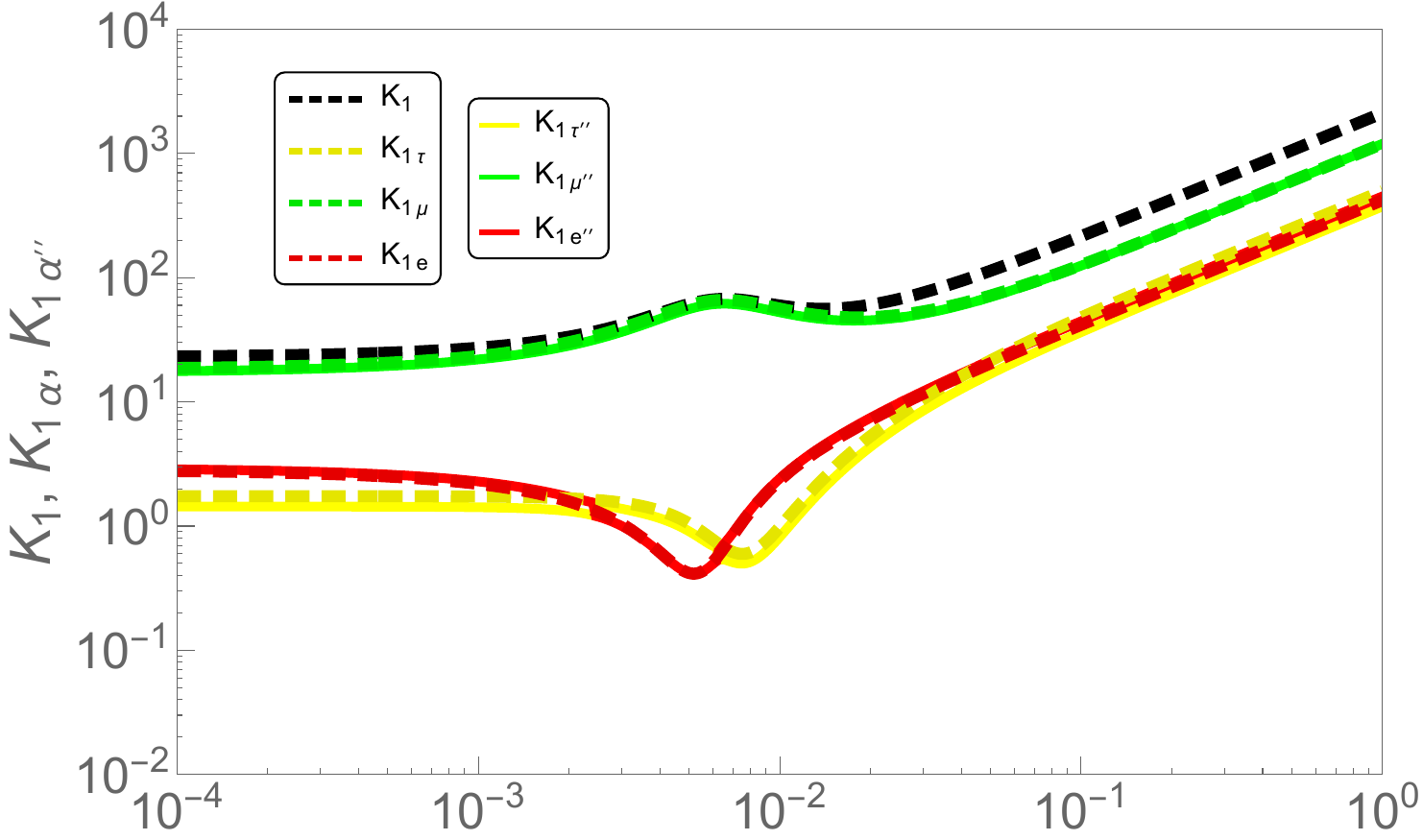,scale=0.18}\hspace{2mm}  \\ \vspace{0mm}
\psfig{file=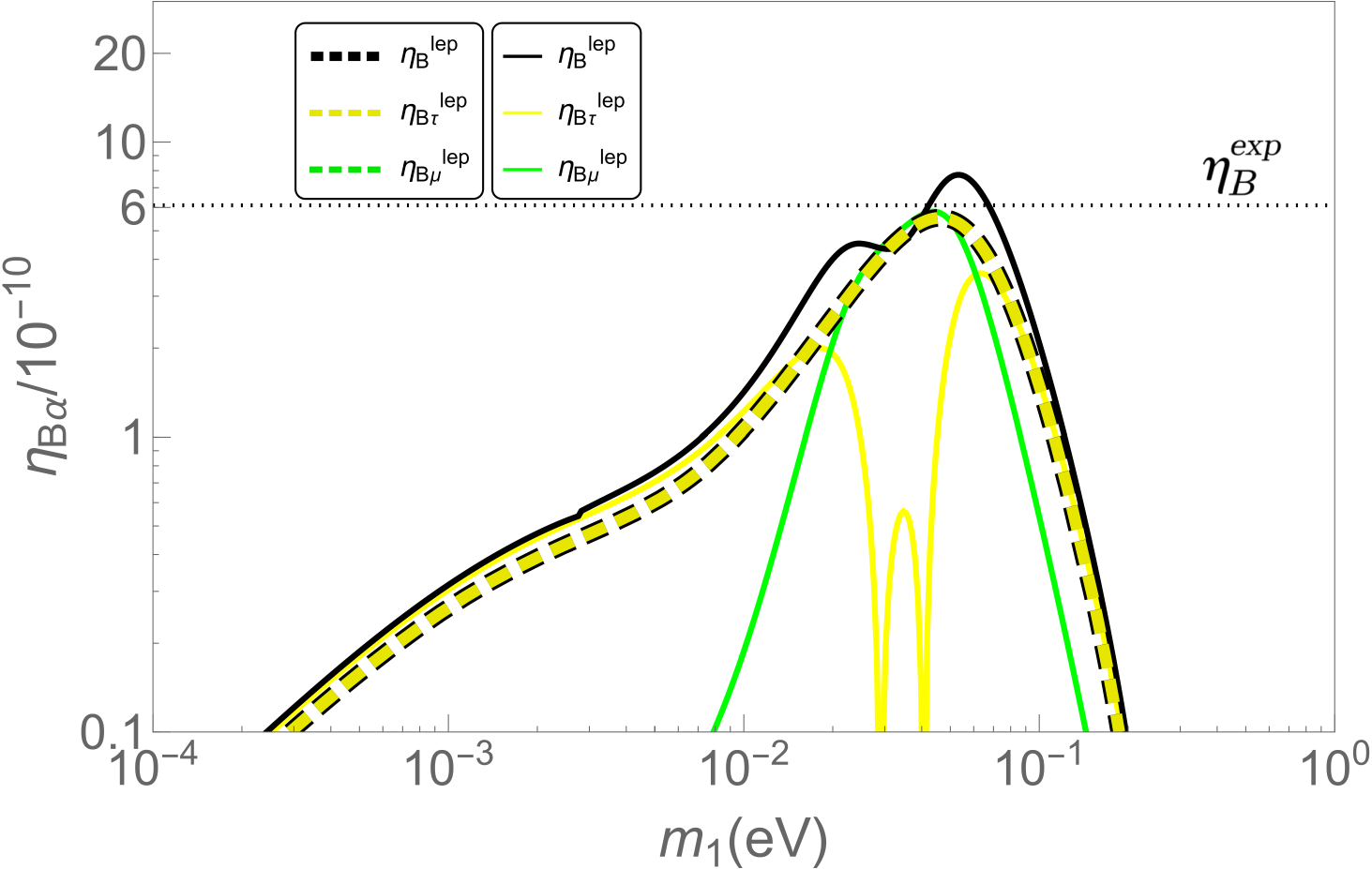,scale=0.18}\hspace{2mm} 
\psfig{file=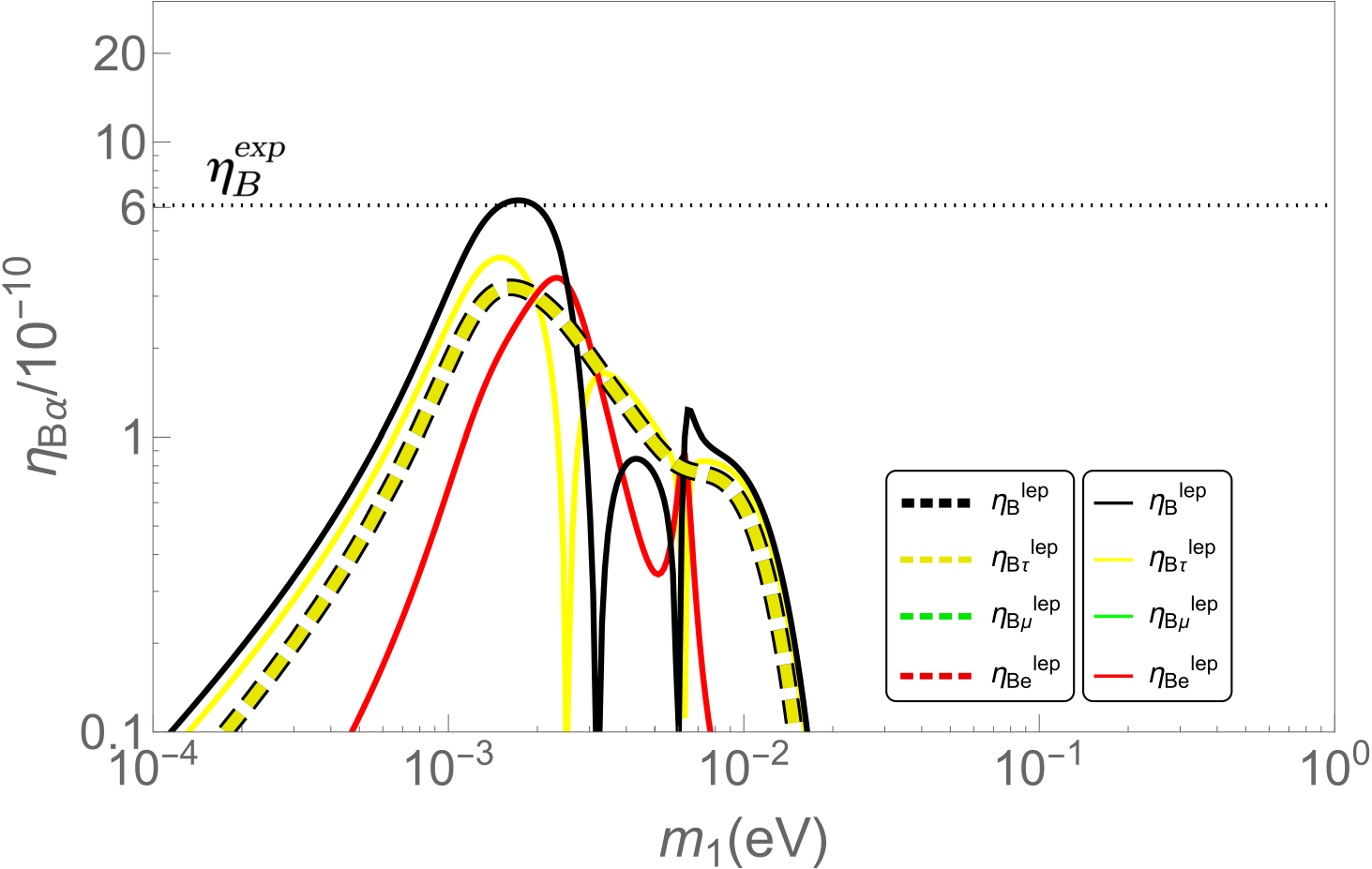,scale=0.18}  
\end{center}
\vspace{-10mm}
\caption{Relevant quantities for two benchmark new type of solutions: one new $\mu$-solution (left panels), one new $e$-solution (right panels).
 The values of the parameters for these solutions are shown in Table~2.}\label{benchmark2}
\end{figure}
\begin{table}
	\centering
	{\tiny
\begin{tabular}{|c|c|c|c|c|c|c|c|c|c|c|c|c|}
\hline
\multicolumn{1}{|c|}{} &  				      
\multicolumn{1}{c|}{$\theta_{12}$} &  
\multicolumn{1}{c|}{$\theta_{13}$} & 
\multicolumn{1}{c|}{$\theta_{23}$} &  
\multicolumn{1}{c|}{$\delta$} &  
\multicolumn{1}{c|}{$\rho/\pi$} & 
\multicolumn{1}{c|}{$\sigma/\pi$} & 
\multicolumn{1}{c|}{$\theta^{\rm L}_{12}$} &  
\multicolumn{1}{c|}{$\theta^{\rm L}_{13}$} & 
\multicolumn{1}{c|}{$\theta^{\rm L}_{23}$} & 
\multicolumn{1}{c|}{$\rho_{\rm L}/\pi$} &  
\multicolumn{1}{c|}{$\sigma_{\rm L}/\pi$} & 
\multicolumn{1}{c|}{$\delta_{\rm L}/\pi$}   \\
\hline 
$\tau\ra\mu$ & $35.18^\circ$   & $9.11^\circ$  & $46.73^\circ$ & $9.97^\circ$  &  1.033 & 1.064 &  $2.172^\circ$ & $0.109^\circ$ & $0.842^\circ$ & 1.05 & 1.13 & 0.23    \\
\hline
$\tau\ra e$ & $30.04^\circ$ & $8.325^\circ$ & $48.24^\circ$  & $16.56^\circ$ & 1.47 & 1.021 & $4.859^\circ$ & $0.155^\circ$ & $1.238^\circ$ 
& 1.72 & 1.25 & 0.31   \\
\hline
\end{tabular}}
			\caption{Values of the six low energy neutrino parameters in $U$ and six parameters in the unitary matrix $V_L$ for the
			 two benchmark solutions shown in Fig.~\ref{benchmark2}.  Best fit values of $m_{\rm sol}$ and $m_{\rm atm}$ are assumed.
			For each solution, the observed asymmetry is reproduced for two values of $m_1$. 
			For $m_1 = 42\,{\rm meV}$ (left) and $m_1 = 2\,{\rm meV}$ (right) one has a muon-dominated and electron-dominated solution, respectively.
			These two solution are indicated with triangles in the scatter plots.}
		\label{table2}
\end{table}
As one can see, successful leptogenesis is satisfied for $m_1 = 42\,{\rm meV}$ and $m_1 = 2.1\,{\rm meV}$ for the muon-dominated 
and the electron-dominated solution, respectively. One can also see how
the dominant $C\!P$ asymmetry is the tauonic one, while both the muonic and the electronic $C\!P$ asymmetries would be too small, in the 
absence of flavour coupling, to yield the observed baryon asymmetry. Therefore, the existence of these two new types of solutions 
crucially relies on the account of flavour coupling (without flavour coupling the muon and electron asymmetries, respectively, would be orders of magnitude
below the observed value). We should add, however, that these new electronic solutions are very marginal ones. 
This can be somehow understood by the fact that the peak of the asymmetry in the shown benchmark case is just above the observed value. 
In the $2\times 10^6$ point scatter plot we 
generated, we obtained only less than $200$ electronic solution (i.e., less than $0.1\%$). 
The new muonic solutions are less special and we obtained about $2 \times 10^3$ of them (i.e.,  $\sim 1\%$). 

It is useful to give some analytical insight.  Starting from Eq.~(\ref{NfDa2}), we can select in the second sum only terms with $\beta=\tau$
and focus only on the muonic and tauon asymmetry, if we consider the muonic benchmark solution in Fig.~9, obtaining:
\bea\label{NfDmu}
N^{\rm f}_{\D\mu} & = & N_{\D\t}^{T\sim T_L}\left[ V^{-1}_{\mu\mu''}V_{\mu''\tau} e^{-{3\pi\over 8}K_{1\mu''}} + 
V^{-1}_{\mu\tau''}V_{\tau''\tau} e^{-{3\pi\over 8}K_{1\tau''}}\right ] \\ \label{NDftau}
N^{\rm f}_{\D\tau} & = & N_{\D\t}^{T\sim T_L}\left[ V^{-1}_{\tau\mu''}V_{\mu''\tau} e^{-{3\pi\over 8}K_{1\mu''}} + 
V^{-1}_{\tau\tau''}V_{\tau''\tau} e^{-{3\pi\over 8}K_{1\tau''}}\right ] \,  .
\eea
In the absence of flavour coupling, for $V=I$, one recovers $N^{\rm f}_{\D\mu} = 0$, since we are neglecting terms $\propto N_{\D\m}^{T\sim T_L}$, and 
$N^{\rm f}_{\D\tau} \simeq N_{\D\t}^{T\sim T_L}e^{-{3\pi\over 8}K_{1\tau}}$. When flavour coupling is taken into account
the muon asymmetry can get filled while the tauonic asymmetry can get depleted since, though the matrix V and its inverse 
contain negative terms. Notice that both in Eq.~(\ref{NfDmu}) and in Eq.~(\ref{NDftau}) there is apparently no role played by the
processes involving the electron flavour, the only one undergoing a strong wash-out ($K_{1e}'' \gg 1)$ while the muon and tauon flavours seem
to be decoupled ($K_{1\mu}''$, $K_{1\tau}'' \lesssim 1$). One can wonder then why the $V$ and $V^{-1}$ entries coupling muon and tauon flavours
should not be vanishing. The answer is that these entries are indeed sensitive to the large value of $K_{1e}$ resulting in large
entries of the first row in the matrix $P^0_1$. Physically, this is because of the strong modification of the Higgs asymmetry due to electron
washout that yields a strong feedback producing contribution to the the asymmetries in the muon and tauon flavour in a way that
the muon asymmetry gets enhanced and the tauon asymmetry gets suppressed.\footnote{As an example, we give here the matrices
$P^0_1$, $V$ and $V^{-1}$ for the benchmark muon solution with values of the parameters in Table 2 well illustrating the role played by the large
washout in the electron flavour:
{\tiny \be
P^0_{1} =
\begin{pmatrix}
1.039 & 0.177 & 0.177 \\
0.0007 & 0.0048 & 0.0014 \\
0.00076 & 0.0015 & 0.0052 
\end{pmatrix}
\, ,  \;
V =
\begin{pmatrix}
0.99999 & 0.00068 & 0.00073 \\
-0.000054 & 0.77 & -0.64 \\
-0.001 & 0.64 & 0.77 
\end{pmatrix}
\, , \;
V^{-1} =
\begin{pmatrix}
0.99999 & -0.00005 & -0.000998 \\
0.00068 & 0.77 & 0.638 \\
0.00073& -0.64 & 0.77
\end{pmatrix} \,  .
\ee 
}}

\subsubsection{Lower bound on $m_1$}

In the right panel of Fig.~3 we again show the isocontour lines for the $m_1$ lower bound in the $\delta$ versus $\theta_{23}$ plane.
One can see how they are just slightly modified by the inclusion of flavour coupling and, for example, the absolute minimum value gets 
slightly relaxed from $m_1 \geq 0.89\,{\rm meV}$ to $m_1 \geq 0.65\,{\rm meV}$. The most remarkable difference is that now the
excluded region by the cosmological upper bound Eq.~(\ref{upperbm1}), in red in Fig.~3, gets very stronly reduced in a way that there are solutions compatible with the cosmological upper bound (blue region) basically for all allowed experimental values of $\delta$ and $\theta_{23}$, even at $3\s$. 

\subsubsection{Upper bound on $\theta_{23}$}

In the right panel of Fig.~4, we show how flavour coupling affects the upper bound on $\theta_{23}$ versus $m_1$. 
One can see that the new muonic solutions tend to cover, very marginally, 
the region for $m_1 \simeq 15\,{\rm meV}$ and $\theta_{23} \simeq 44^\circ$--$45^\circ$
that would be excluded without an account of flavour coupling. However, all second octant values of $\theta_{23}$ still remain excluded for $m_1=(10$--$30)\,{\rm meV}$.

\subsubsection{Excluded regions for $\delta$}

In the right panel of Fig.~5 we show the constraints in the plane $\delta$ versus $m_1$. One can see how the new muonic solutions
tend to fill some existing gaps for $m_1\simeq 10\,{\rm meV}$ and $\delta$ in the range between $\pi/2$ and $3\pi/2$. However,
the hole at $\delta \simeq \pi/2$, though shrinking, still survives. For $m_1$ much below 10 meV the constraints are practically unchanged.  

\subsubsection{Majorana phases}

In the right panel of Fig.~6 we show the constraints in the plane $\rho$ versus $\sigma$.  This plot is quite interesting because it clearly shows 
how the new solutions, both muonic and electronic, stem from the tauonic $C\!P$ asymmetry. The regions they fill are 
completely disconnected from the primary muonic solutions that are present also in the absence of flavour coupling and for which the 
asymmetry is proportional to the muon $C\!P$ asymmetry.  One can also notice how the new solutions extend to higher values of $\sigma$ for $\rho \sim \pi$. 
The same conclusions can be also drawn looking at the right panel of Fig.~7 where we show the constraints in the plane $\sigma$ versus $m_1$.

\subsubsection{$0\nu\b\b$ effective neutrino mass}

Finally, one can see from the right panel of Fig.~8 that flavour coupling does not change much the constraints in the plane $m_{ee}$ versus $m_1$.
The new solutions are confined within the region that was anyway accessible also without accounting for flavour coupling. The most remarkable difference
is that the lower bound on $m_{ee}$ gets relaxed by about one order-of-magnitude but it is still exists.


\section{Strong thermal $SO(10)$-inspired leptogenesis}

In this section, we finally discuss ST-SO10INLEP \cite{ST}. It is intriguing  that the rather special conditions
that are needed to realise strong thermal leptogenesis \cite{problem,sophie}, 
are naturally satisfied by a subset of the solutions realising successful SO10INLEP. 
In this section we first re-derive, and update, the predictions on low energy neutrino parameters 
from ST-SO10INLEP neglecting flavour coupling, and then we discuss the impact of flavour coupling. 

\subsection{Neglecting flavour coupling}

Let us assume that some external mechanism has created an initial pre-existing B-L asymmetry 
$N_{B-L}^{\rm p,i}$ prior to leptogenesis and prior to any washout from RH neutrinos. We also
assume that the $N_3$ mass is larger than the reheat temperature, so that the $N_3$-washout is absent.\footnote{This assumption
is conservative but ultimately it is irrelevant whether $N_3$-washout is included or not, the predictions on the low 
energy neutrino parameters change just slightly \cite{ST}.} Because of the RH neutrino inverse processes, the pre-existing asymmetry will 
undergo a dynamical evolution that has to take into account also the different flavour regimes. Finally,
the relic value of the pre-existing asymmetry can be also expressed as the sum of three contributions from each flavour 
i.e., $N_{B-L}^{\rm p, f} = \sum_{\a} \,  N_{\D_\a}^{\rm p,f}$.
Neglecting flavour coupling, the expressions for the $N_{\D_\a}^{\rm p,f}$ are given by \cite{ST,chianese}
\bea\label{finalpas}
N_{\D_\t}^{\rm p,f} & = & 
(p^0_{{\rm p}\t}+\D p_{{\rm p}\tau})\,  e^{-{3\pi\over 8}\,(K_{1\t}+K_{2\t})} \, N_{B-L}^{\rm p,i} \,  , \\  \nonumber
N_{\D_\m}^{\rm p,f} & = & 
\left\{(1-p^0_{{\rm p}\t})\,\left[
p^0_{\mu\t_2^{\bot}}\, p^0_{{\rm p}\t^\bot_2}\,
e^{-{3\pi\over 8}\,(K_{2e}+K_{2\m})} + (1-p^0_{\m\t_2^{\bot}})\,(1-p^0_{{\rm p}\t^\bot_2}) \right]  + 
\D p_{{\rm p}\mu}\right\}
\,e^{-{3\pi\over 8}\,K_{1\m}}\, N_{B-L}^{\rm p,i}
 ,  \\  \nonumber
N_{\D_e}^{\rm p,f}& = & 
\left\{(1-p^0_{{\rm p}\t})\,\left[ 
p^0_{e\t_2^{\bot}}\,p^0_{{\rm p}\t^\bot_2}\,
e^{-{3\pi\over 8}\,(K_{2e}+K_{2\m})} + (1-p^0_{e \t_2^{\bot}})\,(1-p^0_{{\rm p}\t^\bot_2}) \right]  + \D p_{{\rm p} e}\right\}
 \,e^{-{3\pi\over 8}\,K_{1e}} \,\, N_{B-L}^{\rm p,i} \,   .
\eea
In this expression the quantities $p^0_{{\rm p} \tau}$ and $p^0_{{\rm p}\tau_2^{\bot}}$ denote
the fractions of the initial pre-existing asymmetry in the tauon flavour and 
in the flavour $\tau_2^{\bot}$, where, as already defined, $\tau_2^{\bot}$  is the  electron and muon flavour  
superposition component in the leptons  
produced by $N_2$-decays (or equivalently the flavour component that is 
washed-out in the inverse processes producing $N_2$).\footnote{Notice that in general $p^0_{{\rm p} \tau}+
p^0_{{\rm p}\tau_2^{\bot}} \neq 1$ since in general the projection of the pre-existing asymmetry flavour direction on the
$\tau^\bot$ plane, is not parallel to $\tau^\bot_2$.}
We also introduced the two quantities $p^0_{\alpha \t_2^{\bot}} \equiv K_{2\alpha}/(K_{2e}+K_{2\mu}) \; (\a=e,\mu)$. These give
the fractions of the $\a$ flavour  in the  $\tau_2^{\bot}$ component of the lepton state $|\ell_2\rangle$ produced by $N_2$-decays.
The terms $\D p_{{\rm p} \t},\D p_{{\rm p} e}$ and $\D p_{{\rm p}\mu}$ are the so-called phantom terms and 
such that $\D p_{{\rm p} e}+\D p_{{\rm p}\mu} +  \D p_{{\rm p}\tau} = 0$. They are
phantom terms  taking into account the possibility that the mechanism
that created the pre-existing asymmetry produced lepton states that  are not exactly the $C\!P$ conjugated of the anti-lepton states. 
It can be crossed-checked that in Eq.~(\ref{finalpas}), if one switches-off the washout terms, 
then the sum of the three final flavoured pre-existing asymmetries
is equal to the total initial pre-existing asymmetry, as it should be.

The strong thermal leptogenesis condition requires that the relic value of the pre-existing asymmetry is negligible with respect
to the final value of the asymmetry produced by $N_2$-decays, for generic initial values of the flavoured pre-existing asymmetries. 
For definiteness, we will impose a condition $N_{B-L}^{\rm p,f} < 0.1 \, N_{B-L}^{\rm lep,f}$. 

The only possibility to have successful strong thermal leptogenesis is just the $N_2$ tauon-dominated leptogenesis scenario \cite{problem} 
that is also representing the bulk of solutions in SO10INLEP, as we discussed. This is quite an intriguing coincidence. 
The reason why only tauon-dominates solutions can realise strong thermal leptogenesis 
can be understood directly from Eq.~(\ref{finalpas}). One can indeed have in this case $K_{2\tau}, K_{1\mu}, K_{1e} \gg 1$ and $K_{1\tau} \lesssim 1$
in a way to wash out at $T \sim M_2$ the tauonic component of the pre-existing asymmetry and at $T \sim M_1$ also
the electronic and muonic components. At the same time, having $K_{1\tau} \lesssim 1$, the tauonic asymmetry produced from leptogenesis
is the only that survives the $N_1$-washout. 

There is only a very special caveat \cite{chianese}. If one imposes $(1-p^0_{\m\t_2^{\bot}})=K_{2e}/(K_{2e}+K_{2\mu}) \lesssim 
0.1 N_{B-L}^{\rm lep,f}/N_{B-L}^{\rm p,i} \lll 1$ and $K_{1\mu} \lesssim 1$, instead of $K_{1\mu}\gg 1$, one can have a strong thermal
muon-dominated solution. We noticed that $K_{2e}/(K_{2e}+K_{2\mu})\sim \theta_{\rm c}^2 \sim 0.05$ is indeed small
but in this case, for a meaningfully large value of pre-existing asymmetry $\sim 10^{-3}$, as we will assume,  one needs this ratio
at least four orders of magnitude smaller. These are incredibly special solutions, unless they can be justified within a particular model.
In \cite{chianese} these solutions were found with a tiny $\sim 10^{-9}$ rate. Therefore, they have to be considered  more like oddities,
rather than real ST-SO10INLEP solutions. 

By imposing a strong thermal leptogenesis condition in addition to successful SO10INLEP, one singles out a special subset of solutions
within those we obtained just imposing successful SO10INLEP.  We have assumed as benchmark value $N^{\rm p,i}_{B-L} = 10^{-3}$.
In the scatter plots, in Fig.~1 and Figs.~3--9, these solutions are denoted by blue points. The success rate for these solutions in our scan
is $10^{-6}$. Our results fully confirm previous findings \cite{ST,decrypting,full,chianese}.  It is still useful to notice some important aspects:
\begin{itemize}
\item As one can see they are all tauon-dominated solutions obtained with a $10^{-6}$ success rate, 
except for a couple of special muon-dominated solutions (success rate $10^{-9}$) we mentioned.
\item From Fig.~6 one can see how the requested values of the Majorana phases for ST-SO10INLEP become
so constrained that  ST-SO10INLEP would be a way to determine their values.
\item The absolute neutrino mass scale range is quite restricted. For the benchmark value we
assumed, $N^{\rm p,i}_{B-L} = 10^{-3}$, one has that successful ST-SO10INLEP requires $m_1 = (10$--$30)\,${\rm meV}
and, from Fig.~8, one can see a corresponding similar range for $m_{ee}$.
It is interesting that both cosmological observations and $0\nu\beta\beta$ experiments are now starting to test this range.
\item The bulk of solutions realising ST-SO10INLEP are of $\tau_A$ type. The $\tau_B$ type solutions 
have the problem that $K_{1\mu}$  tends to be too small to washout the muonic component of the pre-existing asymmetry.
For example for $V_L = I$ one finds $K_{1\mu}\lesssim 4$ \cite{decrypting} and in this case they would be absolutely unable to
yield strong thermal solutions. However, turning on angles in $V_L$ values as large as $K_{1\mu} \sim 10$ are possible and for this reason
some $\tau_B$ strong thermal solutions do exist, though they contribute only $\sim 1\%$ to the total number of solutions. 
\end{itemize}
Finally, we have focused on an accurate determination of the allowed region in the plane $\delta$ versus $\theta_{23}$ for strong thermal leptogenesis.
\begin{figure}[t]
\centerline{
\psfig{file=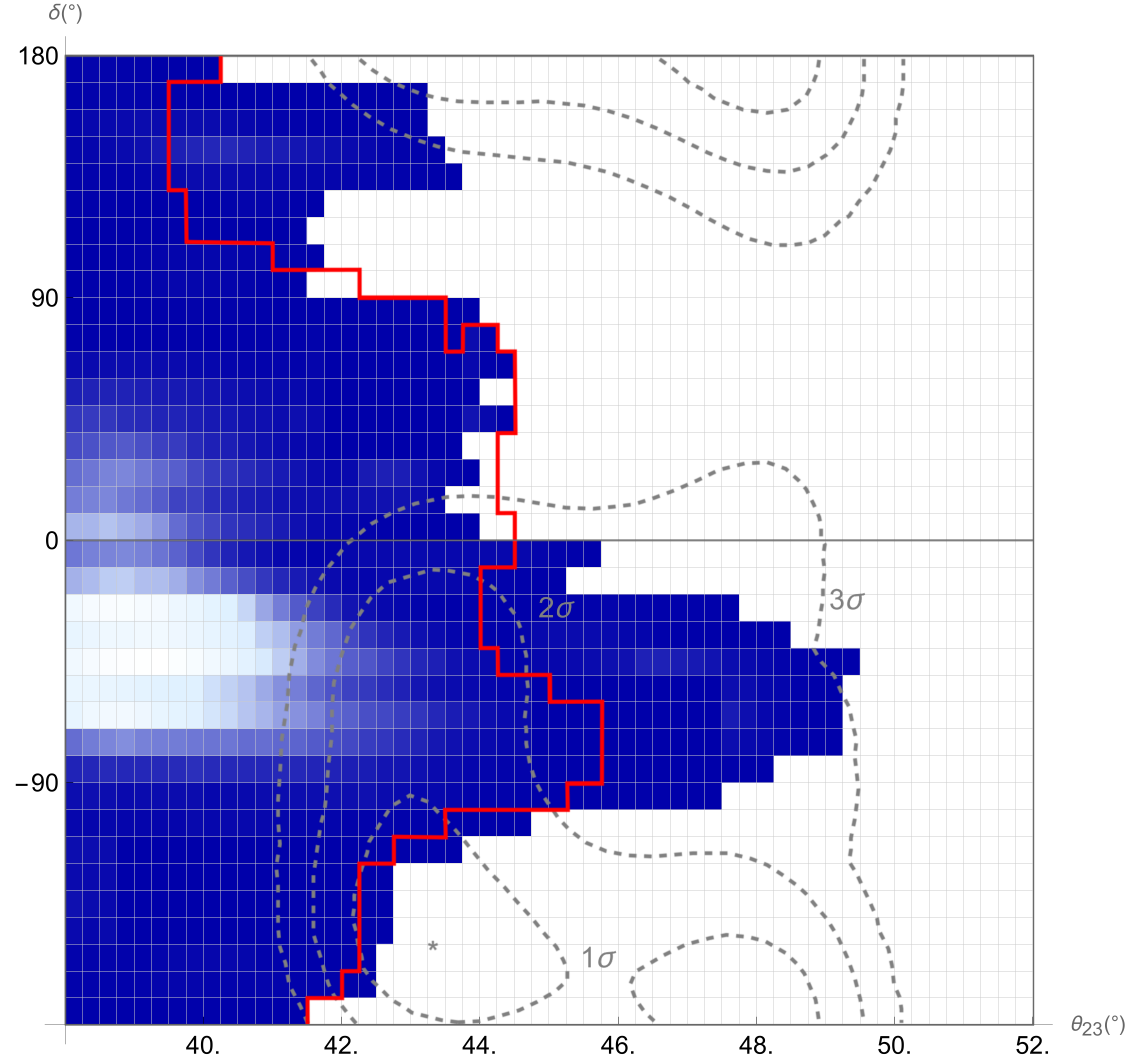,scale=0.4}
}\vspace{-1mm}
\caption{ST-SO10INLEP allowed region in the plane $\delta$ versus $\theta_{23}$. The red line  delimits the region when 
flavour coupling is neglected, the coloured region is the result in the case when flavour coupling is included. The colour code
denotes point density:  lighter blue means higher density of solutions. The thin dashed lines are
the $1\sigma, 2\sigma$ and $3\sigma$ experimental constraints from \cite{nufit24}.}
\end{figure}
The result is shown in Fig.~10 and indicated by a red line. This well reproduces the result found in \cite{chianese}. The small differences can be ascribed 
to the updated experimental ranges for neutrino oscillation parameters in Eq.~(\ref{expranges}) we are adopting. The thin dashed lines are 
the the $1\sigma, 2\sigma$ and $3\sigma$ experimental constraints from \cite{nufit24}. One can see how  ST-SO10INLEP is in
nice agreement with latest experimental results within $1\sigma$. The fact that the current best fit falls in the first octant releases the tension that
was found with old experimental data in \cite{chianese}. The best fit $\delta$ value is actually too low but this is very sensitive to the
employed data sets in the analysis, mainly because of an existing tension between T2K and NOVA data.   In the left panel of Fig.~11 we show
\begin{figure}[t]
\centerline{
\psfig{file=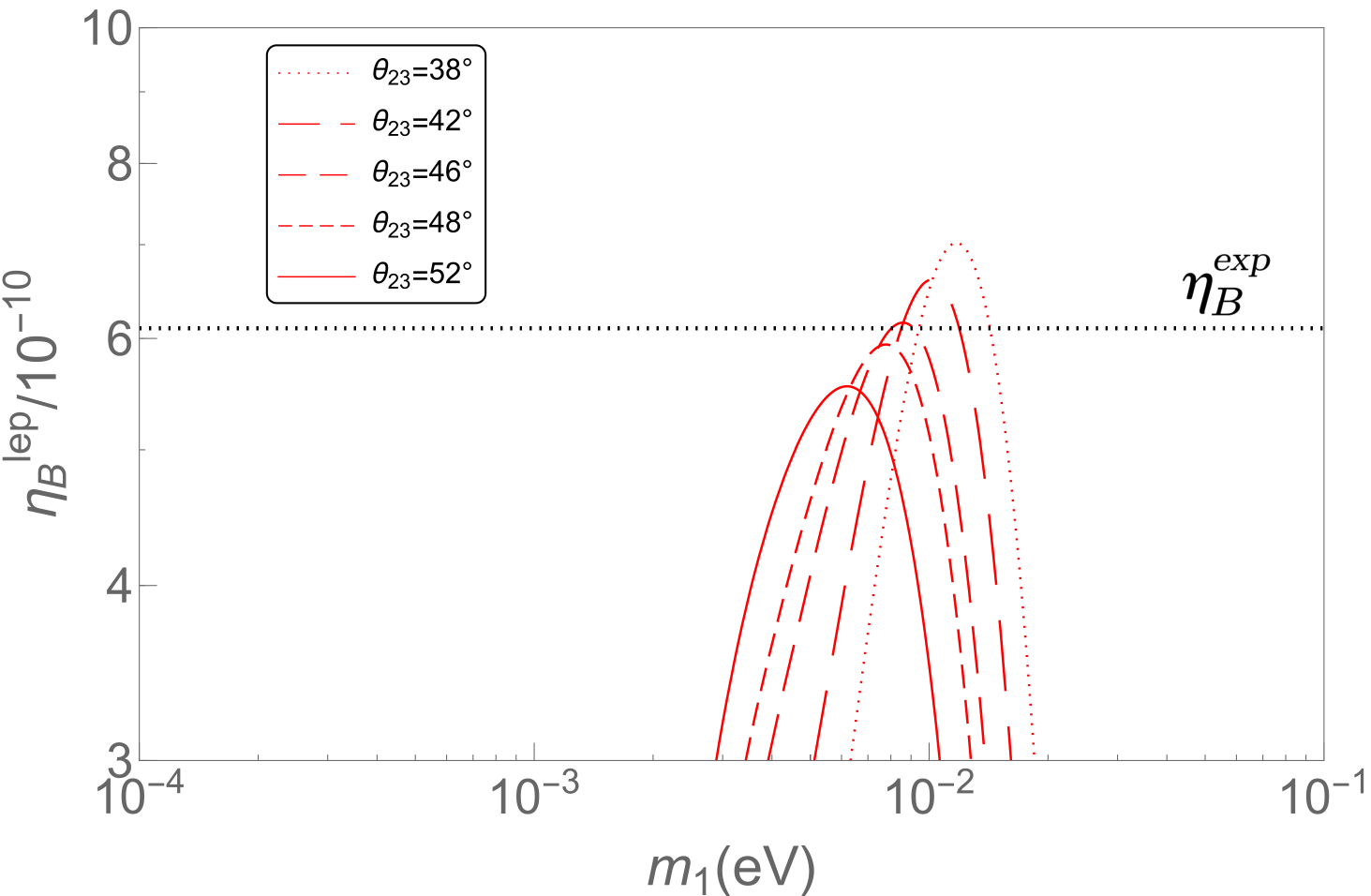,scale=0.3} \hspace{5mm}
\psfig{file=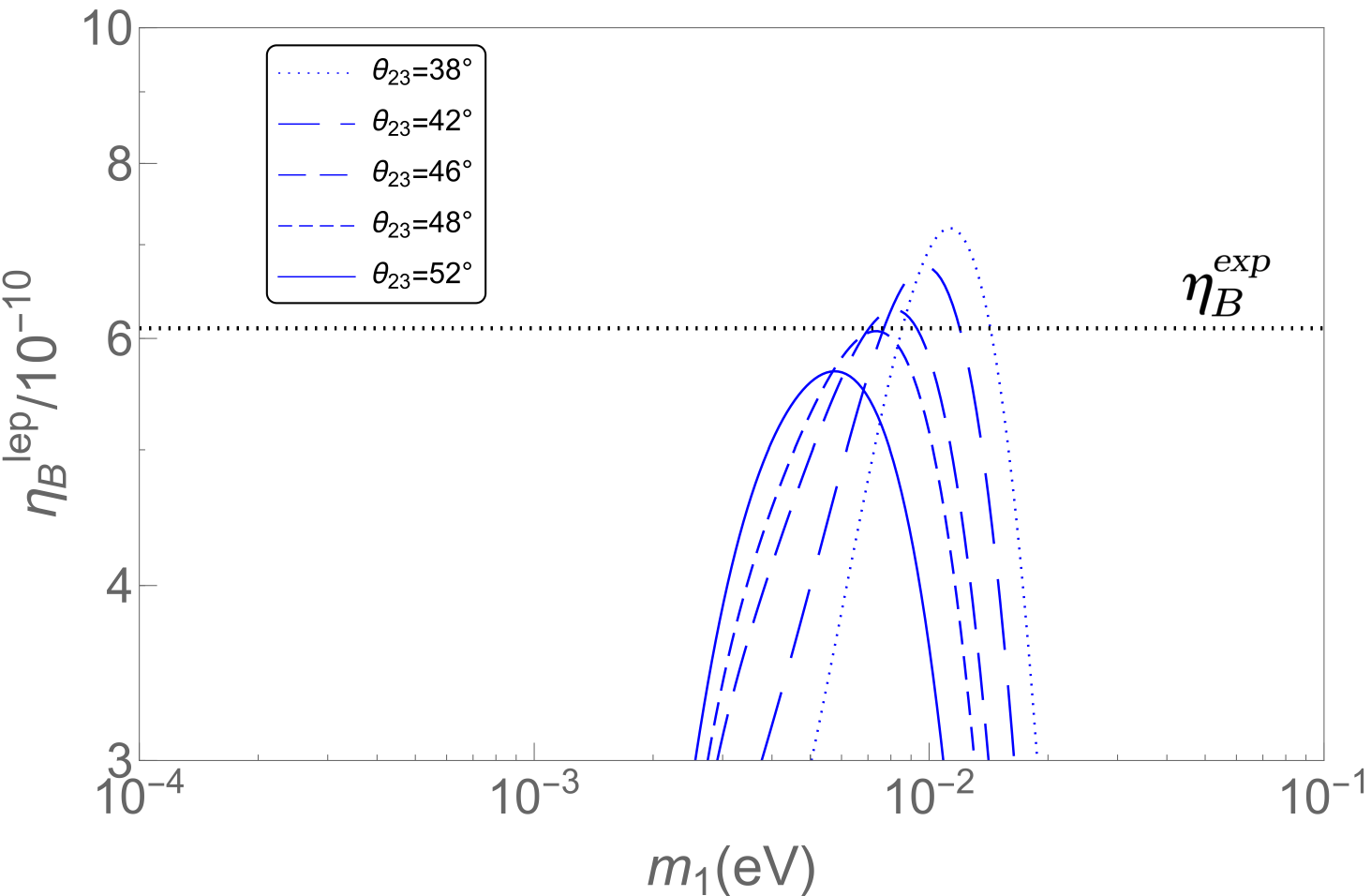,scale=0.3}
}\vspace{-1mm}
\caption{Plots of the final value of $\eta_B^{\rm lep}$ versus $m_1$ for different values of $\theta_{23}$, 
for fixed $\delta = -50^\circ$ and for values of the other parameters maximising the asymmetry shown in Table 3. 
The right (left) panel is for the case with (no) flavour coupling.}
\end{figure}
plots of $\eta_B^{\rm lep}$ versus $m_1$ for the indicated values of $\theta_{23}$ and for fixed $\delta = -50^\circ$, corresponding
to the value allowing the maximum value of $\theta_{23}$. All other parameters are fixed to values that maximise the asymmetry and are given in Table 3.
One can see how for increasing values of $\theta_{23}$ the peak value of $\eta_B^{\rm lep}$ decreases. This is well understood
since the asymmetry is approximately $\propto 1/\sin^4 \theta_{23}$ \cite{decrypting,full}. One can see how the
maximum value of $\theta_{23}$ we obtain is $46^\circ$. This is consistent with the value obtained in \cite{chianese} and it clearly shows
how the second octant is incompatible with ST-SO10INLEP.

\begin{table}
	\centering
	{\tiny
\begin{tabular}{|c|c|c|c|c|c|c|c|c|c|c|c|c|}
\hline
\multicolumn{1}{|c|}{} &  				      
\multicolumn{1}{c|}{$\theta_{12}$} &  
\multicolumn{1}{c|}{$\theta_{13}$} & 
\multicolumn{1}{c|}{$\theta_{23}$} &  
\multicolumn{1}{c|}{$\delta$} &  
\multicolumn{1}{c|}{$\rho/\pi$} & 
\multicolumn{1}{c|}{$\sigma/\pi$} & 
\multicolumn{1}{c|}{$\theta^{\rm L}_{12}$} &  
\multicolumn{1}{c|}{$\theta^{\rm L}_{13}$} & 
\multicolumn{1}{c|}{$\theta^{\rm L}_{23}$} & 
\multicolumn{1}{c|}{$\rho_{\rm L}/\pi$} &  
\multicolumn{1}{c|}{$\sigma_{\rm L}/\pi$} & 
\multicolumn{1}{c|}{$\delta_{\rm L}/\pi$}   \\
\hline 
No F.C. & $33.62^\circ$   & $9.09^\circ$  & See Figure & $-49^\circ$  &  0.312 & 0.834 &  $4.11^\circ$ & $0.183^\circ$ & $2.41^\circ$ & 1.92 & 1.24 & 0.18    \\
\hline
With F.C. & $33.62^\circ$   & $9.09^\circ$  & See Figure & $-49^\circ$  &  0.312 & 0.834 &  $4.11^\circ$ & $0.183^\circ$ & $2.41^\circ$ & 1.92 & 1.24 & 0.18   \\
\hline
\end{tabular}}
			\caption{Values of the low energy neutrino parameters in $U$ (except $\theta_{23}$ given in Fig.~11) and six parameters in the unitary matrix $V_L$ for the plots shown in Fig.~11.}
		\label{table3}
\end{table}

\subsection{Including flavour coupling}

We have now finally to discuss the impact of flavour coupling on ST-SO10INLEP.  This is one of the main motivations of the paper. 
A simplistic, but legitimate, expectation is that flavour coupling might completely disrupt ST-SO10INLEP. The reason is that 
a large pre-existing asymmetry, thanks to flavour coupling, can find its way to survive into the tauon flavour where the washout is weak. 
As we are going to explain, this expectation is not met.  

First of all, we have to generalise the expression (\ref{finalpas}) for the relic value of the pre-existing asymmetry. If we start with an initial
pre-existing asymmetry $N^{\rm p,i}_{B-L}$, that we assume to be generated at some temperature above $\sim 10^{12}\,{\rm GeV}$,
below $10^{12}\,{\rm GeV}$ the tauon component of lepton states is measured by tauon charged lepton interactions and the pre-existing asymmetry will
be the sum of a tauon component, given by $p^0_{{\rm p}\t}\,N^{\rm p,i}_{B-L} $, and a component along a flavour direction 
given by the projection of the  flavour composition of pre-existing leptons  
on the $e$-$\mu$ flavour plane given by $(1-p^0_{{\rm p}\t})\, N^{\rm p,i}_{B-L}$. 
In addition, in general, there should be also the phantom terms that however we neglect just to shorten the notation
(in any case results are not sensitive to them).   Similarly to the solution for the asymmetry generated by leptogenesis,
at $T\sim M_2$ the $N_2$-washout, in the presence of flavour coupling, uncouples in a rotated flavour basis $\tau'$ and $\tau^{\bot '}_2$. 
The difference is that now there is no source term and, therefore, at the end of the washout processes the washout is described by exponential factors
and, rotating back to the standard flavour basis, one finds
\bea \nonumber
N_{\D_{\t}}^{{\rm p},T\lesssim T_L}  & = & 
U^{-1}_{\t\tau_2^{\bot'}}\left[U_{\tau_2^{\bot'}\tau_2^\bot}\,
(1-p^0_{{\rm p}\t}) + U_{\tau_2^{\bot'}\t}\,p^0_{{\rm p}\t} \right]\,N^{\rm p,i}_{B-L} \,e^{-{3\pi\over 8}\,(K_{2e}+K_{2\m})} \\ \nonumber
&  & + U^{-1}_{\t\t'}\left[U_{\t'\tau_2^{\bot}}\,(1-p^0_{{\rm p}\t}) +U_{\t'\t}\,p^0_{{\rm p}\t} \right] 
\,N^{\rm p,i}_{B-L} \,e^{-{3\pi\over 8}\,K_{2\t}}  \,   ,  \\ 
N_{\D_{\tau_2^{\bot}}}^{{\rm p},T\lesssim T_L} & = &
 U^{-1}_{\tau_2^{\bot}\tau_2^{\bot '}}\left[U_{\tau_2^{\bot '}\tau_2^{\bot}}\,(1-p^0_{{\rm p}\t}) 
 +U_{\tau_2^{\bot '}\t}\,p^0_{{\rm p}\t} \right]\,N^{\rm p,i}_{B-L}\,e^{-{3\pi\over 8}\,(K_{2e}+K_{2\m})}  \\
&  & +U^{-1}_{\tau_2^{\bot}\t'}\left[U_{\t'\tau_2^{\bot}}\,(1-p^0_{{\rm p}\t}) +U_{\t'\t}\,p^0_{{\rm p}\t} \right]\,N^{\rm p,i}_{B-L}\,e^{-{3\pi\over 8}\,K_{2\t}} \nonumber   \\
N_{\D_{\tau_{2^\bot}^{\bot}}}^{{\rm p},T\lesssim T_L} & = &
 (1-p^0_{{\rm p}\t})\,(1-p^0_{{\rm p}\t^\bot_2})\, N^{\rm p,i}_{B-L}\,  .
\eea
where the matrix $U$ is the same we introduced in Eq.~(\ref{Umatrix}) when we introduced flavour coupling for the asymmetry produced by $N_2$-decays.
It can be noticed that, for $U=I$, one recovers the result in the absence of flavour coupling and $N_1$-washout in Eq.~(\ref{finalpas}).
Also notice that the component of the pre-existing asymmetry along the flavour $\tau_{2^\bot}^{\bot}$, orthogonal to $\tau$ and $\tau_2^\bot$,
escapes the washout at $T\sim M_2$ \cite{Barbieri:1999ma,Engelhard:2006yg}. 

We have now still to evolve the asymmetry down to temperatures below $M_1$. First of all, at temperatures $T\sim 10^9\,{\rm GeV}$
also RH muon interactions become effective and break the coherence of $\tau^\bot$-flavour leptons. The asymmetries in the electron
and muon flavour, prior to $N_1$-washout, can then be written as
\bea
N_{\D_{e}}^{{\rm p},T\gtrsim M_1} & = & p^0_{e\t_2^{\bot}}\, N_{\D_{\tau_2^{\bot}}}^{{\rm p},T\lesssim T_L} + (1-p^0_{e\t_2^{\bot}}) \, N_{\D_{\tau_{2^\bot}^{\bot}}}^{{\rm p},T\lesssim T_L} \,  , \\
N_{\D_{\mu}}^{{\rm p},T\gtrsim M_1} & = & p^0_{\mu\t_2^{\bot}}\, N_{\D_{\tau_2^{\bot}}}^{{\rm p},T\lesssim T_L}
+ (1-p^0_{\mu\t_2^{\bot}}) \, N_{\D_{\tau_{2^\bot}^{\bot}}}^{{\rm p},T\lesssim T_L} \,  .
\eea
Finally, the $N_1$-washout stage can be taken into account exactly as for the asymmetry produced from leptogenesis, using an equation similar to
Eq.~(\ref{NfDalpha}) and Eq.~(\ref{NfDalpha2}), obtaining
\be
N^{\rm p,f}_{\D_{\a}}  =  \sum_{\a''}\,V^{-1}_{\a\a''}\,
\left[\sum_{\b=e,\mu,\tau}\,V_{\a''\b}\,N_{\D_{\b}}^{{\rm p},T\gtrsim M_1} \,e^{-{3\pi\over 8}\,K_{1\a''}}\right] \,  .
\ee

Using now this much more intricate expression, we have obtained ST-SO10INLEP solutions shown, again with blue points,
in the right panels of Fig.~1 and Figs. 2--8. The first thing to notice is that the region survives and in fact it even slightly 
increases. In particular, the most noticeable change is that the lower bound on $m_1$ gets relaxed and from
$m_1 \gtrsim 8\,{\rm meV}$, obtained neglecting flavour coupling, we have now found $m_1 \gtrsim 3\,{\rm meV}$.  
One can also see from Fig.~8 that, corresponding, the lower bound on $m_{ee}$ gets relaxed from
$m_{ee} \gtrsim 9\,{\rm meV}$ to $m_{ee}\gtrsim 4\,{\rm meV}$. One can also see from Fig.~6 that now the two disconnected subregions
get linked by the appearance of the new solutions. 

This implies, that in fact the washout of a pre-existing asymmetry, gets even easier when flavour coupling is taken into account,
contrarily to naive expectations. The reasons why that happens were already partially envisaged in \cite{full}.
Even though flavour coupling induces part of the pre-existing asymmetry in the electron and muon flavours (that survived the $N_2$-washout)
to be transferred in the tauon flavour, where the washout is 
absent ($K_{1\tau}\ll 1$) to let the asymmetry produced from leptogenesis to survive, the off-diagonal term in the V and $V^{-1}$ matrices responsible for the
coupling of $e$ ($\mu$) flavour to the $\tau$ flavour is proportional to $K_{1\tau}/K_{1e} (K_{1\tau}/K_{1\mu})$ and so intrinsically small.  
This means that the same conditions that are necessary to have
strong thermal leptogenesis are also sufficient to protect it from the potential disruptive effect of flavour coupling.
In addition flavour coupling is such that that the flavoured decay parameters responsible for the washout of the pre-existing
electron and muon asymmetries are not the standard $K_{1e}$ and $K_{1\mu}$ but these are replaced by $K_{1e}''$ and $K_{1\mu}''$.
The origin of the lower bound on $m_1$ was studied in detail in \cite{sophie,decrypting} and it was due to the fact that 
$K_{1e} \simeq m_{ee}/m_{\star} \simeq 0.8\,m_1/m_{ee}$. For an initial electronic pre-existing asymmetry $\sim 10^{-4}$,
 the strong thermal conditions implies $K_{1e} \gtrsim 8$ that translates immediately into  $m_{ee} \gtrsim 8\,{\rm meV}$
 and $m_1 \gtrsim 9\,{\rm meV}$. However, when flavour coupling is included the washout of the electron asymmetry is described
 by $K_{1e}''$ that is a linear combination primarily of $K_{1e}$ but with a small contribution from 
 $K_{1\mu} \simeq (m_{\rm sol}s^2_{12} + m_{\rm atm}s_{13}c_{23})^2/m_{\star}m_{\rm sol} s^2_{12} \simeq 25$.
 This contamination from a large $K_{1\mu}$ makes possible to have smaller values of $K_{1e}$ and, therefore, of $m_{ee}$ and $m_1$.

\section{Final remarks}

We could benefit of the analytical description of SO10INLEP given in \cite{full} to re-derive the allowed region in the space of low energy neutrino parameters
including flavour coupling. The results confirm the broad picture already discussed in previous papers but
there are also interesting new aspects introduced by flavour coupling.  
Moreover, in general, we also noticed how we are currently entering a new stage for the prospect of testing SO10INLEP.
Let us briefly summarise our main results.
\begin{itemize}
\item The cosmological upper bound Eq.~(\ref{upperbm1}) definitively rules out quasi-degenerate neutrinos and the latest upper bounds from $0\nu\b\b$ and tritium
decay experiments confirm this so far. Absolute neutrino mass scale experiments are then starting to enter the most interesting absolute neutrino mass 
range for the possibility of testing SO10INLEP, since for $10\,{\rm meV} \lesssim m_1 \lesssim 30\,{\rm meV}$ there is a stringent upper bound on the
atmospheric mixing angle that excludes second octant. 
\item An interesting new feature introduced by flavour coupling is that new muon-dominated solutions 
appear in a portion of the space of parameters that would be otherwise excluded if flavour coupling is not taken into account.  There are also some completely new
electron-dominated solutions appearing (inside an already allowed region of parameter space) 
but these are quite marginal and do not seem to have a particular interest. 
\item We have confirmed the existence of ST-SO10INLEP solutions even when flavour coupling is taken into account. The 
allowed region gets even larger and, importantly, the lower bound on the absolute neutrino mass scale gets more relaxed. The washout of a large 
initial pre-existing asymmetry $N^{\rm p,i}_{B-L} \sim 10^{-3}$ requires $m_{ee} (m_1)\gtrsim 4 (3)\,{\rm meV}$. This is still large enough
that future generation $0\nu\beta\beta$ experiments will be able to test it.  The fact that ST-SO10INLEP is protected by flavour coupling is
somehow a built-in feature: since $K_{1\tau} \lesssim 1$, this necessarily limits the possibility to have a too large leak of the
pre-existing asymmetry from the $e$ and $\mu$ flavours into the $\tau$ flavour so that the pre-existing asymmetry can be still
efficiently washed-out for a subset of the SO10INLEP solutions. This is an important result of our paper since it shows that the
ST-SO10INLEP scenario is stable under perturbations introduced by a more sophisticate calculation of the asymmetry.
\item Our procedure, providing an approach toward a map of $SO(10)$-inspired models, can also be useful within the quest for a realistic grandunified model
since it gives a complementary analytical insight and a global picture of the solutions. Indeed it has provided quite a useful insight in connection with a recent  
identification of a realistic fit of fermion parameters within a minimal $SO(10)$ scenario
that is also able to reproduce the observed baryon asymmetry with $N_2$-leptogenesis \cite{Babu:2024ahk}.
This falls within the category of SO1OINLEP scenarios but extended to allow for a large, in fact maximal, 
value of $\theta_{23}^L$.\footnote{Recently, it was also emphasized the role that a triplet Higgs might have
in $SO(10)$-leptogenesis and in the search of realistic minimal $SO(10)$-models \cite{Fong:2025aya}. 
For a recent mini-review on studies embedding high scale thermal leptogenesis within
grandunified models see \cite{Malinsky:2025jwg}.}
\end{itemize}
In conclusion, our results represent the first step toward a systematic understanding of theoretical uncertainties  within SO10INLEP that is particularly important in view of future experimental expected progress both in long baseline neutrino oscillation experiments and absolute neutrino mass scale experiments. 
It is a very exciting time for the possibility of testing SO10INLEP. Both long baseline and absolute neutrino mass scale experiments will gradually test its predictions and constraints. In particular, $0\nu\beta\beta$ experiments are finally entering the mass range that is predicted
by the strong thermal version of SO10INLEP. An account of flavour coupling has confirmed and even strengthened the picture, making the predictions more robust. At the same time, it also provides useful information for the optimization of a successful strategy toward the identification of a grandunified realistic model. 


\vspace{-1mm}
\subsection*{Acknowledgments}

We acknowledge financial support from the STFC Consolidated Grant ST/T000775/1.
PDB also acknowledges support from the European Union’s Horizon 2020 Europe research and innovation programme under  
the Marie Sk\l odowska-Curie grant agreement HIDDeN European  ITN project (H2020-MSCA-ITN2019//860881-HIDDeN).


\begin{thebibliography}{99}

\bibitem{fy}
M.~Fukugita and T.~Yanagida,
  {\em Baryogenesis Without Grand Unification}, Phys.\ Lett.\ B {\bf 174} (1986) 45.

\bibitem{seesaw}
P.~Minkowski,
  {\em $\mu\to e \g$ At A Rate Of One Out Of 1-Billion Muon Decays?},
  Phys.\ Lett.\  B {\bf 67} (1977) 421;
T. Yanagida, 
{\em Horizontal gauge symmetry and masses of neutrinos},
  Conf.\ Proc.\ C {\bf 7902131} (1979) 95.
 Proceedings of the Workshop on Unified Theory and Baryon Number
of the Universe, eds. O. Sawada and A. Sugamoto (KEK, 1979) p.95;
  P.~Ramond, 
Invited talk given at Conference: C79-02-25
(Feb 1979) p.265-280, CALT-68-709,
 {\em The Family Group in Grand Unified Theories},
  hep-ph/9809459;
 M. Gell-Mann,
P. Ramond and R. Slansky, in Supergravity, eds. P. van Niewwenhuizen and D.
Freedman (North Holland, Amsterdam, 1979) Conf.Proc. C790927 p.315, PRINT-80-0576;
 R.~Barbieri, D.~V.~Nanopoulos, G.~Morchio and F.~Strocchi,
  {\em Neutrino Masses in Grand Unified Theories},
  Phys.\ Lett.\  {\bf 90B} (1980) 91.
R.~N.~Mohapatra and G.~Senjanovic,
  {\em Neutrino Mass and Spontaneous Parity Nonconservation},
  Phys.\ Rev.\ Lett.\  {\bf 44} (1980) 912.

\bibitem{Blanchet:2012bk}
S.~Blanchet and P.~Di Bari,
{\em The minimal scenario of leptogenesis},
New J. Phys. \textbf{14} (2012), 125012
[arXiv:1211.0512 [hep-ph]].

\bibitem{LIGOScientific:2016aoc}
B.~P.~Abbott \textit{et al.} [LIGO Scientific and Virgo],
{\em Observation of Gravitational Waves from a Binary Black Hole Merger},
Phys. Rev. Lett. \textbf{116} (2016) no.6, 061102
[arXiv:1602.03837 [gr-qc]].

\bibitem{Dror:2019syi}
J.~A.~Dror, T.~Hiramatsu, K.~Kohri, H.~Murayama and G.~White,
{\em Testing the Seesaw Mechanism and Leptogenesis with Gravitational Waves},
Phys. Rev. Lett. \textbf{124} (2020) no.4, 041804
[arXiv:1908.03227 [hep-ph]].

\bibitem{DiBari:2020bvn}
P.~Di Bari, D.~Marfatia and Y.~L.~Zhou,
{\em Gravitational waves from neutrino mass and dark matter genesis},
Phys. Rev. D \textbf{102} (2020) no.9, 095017
[arXiv:2001.07637 [hep-ph]].

\bibitem{DiBari:2021dri}
P.~Di Bari, D.~Marfatia and Y.~L.~Zhou,
{\em Gravitational waves from first-order phase transitions in Majoron models of neutrino mass},
JHEP \textbf{10} (2021), 193
[arXiv:2106.00025 [hep-ph]].

\bibitem{Fu:2022lrn}
B.~Fu, S.~F.~King, L.~Marsili, S.~Pascoli, J.~Turner and Y.~L.~Zhou,
{\em A predictive and testable unified theory of fermion masses, mixing and leptogenesis},
JHEP \textbf{11} (2022), 072
[arXiv:2209.00021 [hep-ph]].

\bibitem{SO10inspired}
A.~Y.~Smirnov,
 {\em Seesaw enhancement of lepton mixing},
  Phys.\ Rev.\ D {\bf 48} (1993) 3264
  [hep-ph/9304205];
W.~Buchmuller and M.~Plumacher,
  {\em Baryon asymmetry and neutrino mixing},
  Phys.\ Lett.\ B {\bf 389} (1996) 73 [hep-ph/9608308];
E.~Nezri and J.~Orloff,
  {\em Neutrino oscillations versus leptogenesis in SO(10) models},
  JHEP {\bf 0304} (2003) 020
  [hep-ph/0004227];
F.~Buccella, D.~Falcone and F.~Tramontano,
  {\em Baryogenesis via leptogenesis in SO(10) models},
  Phys.\ Lett.\ B {\bf 524} (2002) 241 [hep-ph/0108172];
G.~C.~Branco, R.~Gonzalez Felipe, F.~R.~Joaquim and M.~N.~Rebelo,
  {\em Leptogenesis, CP violation and neutrino data: What can we learn?},
  Nucl.\ Phys.\ B {\bf 640} (2002) 202 [hep-ph/0202030];
 
\bibitem{Akhmedov:2003dg}  
E.~K.~Akhmedov, M.~Frigerio and A.~Y.~Smirnov,
{\em Probing the seesaw mechanism with neutrino data and leptogenesis},
JHEP \textbf{09} (2003), 021
[arXiv:hep-ph/0305322 [hep-ph]].
  

 \bibitem{compact}
 F.~Buccella, D.~Falcone, C.~S.~Fong, E.~Nardi and G.~Ricciardi,
 {\em Squeezing out predictions with leptogenesis from SO(10)},
  Phys.\ Rev.\ D {\bf 86} (2012) 035012
  [arXiv:1203.0829 [hep-ph]].

\bibitem{decrypting}
P.~Di Bari, L.~Marzola and M.~Re Fiorentin,
  {\em Decrypting $SO(10)$-inspired leptogenesis},
  Nucl.\ Phys.\ B {\bf 893} (2015) 122
  [arXiv:1411.5478 [hep-ph]].

\bibitem{di}
S.~Davidson and A.~Ibarra,
  {\em A Lower bound on the right-handed neutrino mass from leptogenesis},
  Phys.\ Lett.\ B {\bf 535} (2002) 25
  [hep-ph/0202239].
  
\bibitem{cmb}
W.~Buchmuller, P.~Di Bari and M.~Plumacher,
  {\em Cosmic microwave background, matter - antimatter asymmetry and neutrino masses},
  Nucl.\ Phys.\ B {\bf 643} (2002) 367
   Erratum: [Nucl.\ Phys.\ B {\bf 793} (2008) 362]
  [hep-ph/0205349].
  
 
\bibitem{geometry} 
P.~Di Bari, {\em Seesaw geometry and leptogenesis},
  Nucl.\ Phys.\ B {\bf 727} (2005) 318
  [hep-ph/0502082].

  \bibitem{riotto1}
P.~Di Bari and A.~Riotto,
  {\em Successful type I Leptogenesis with SO(10)-inspired mass relations},
  Phys.\ Lett.\ B {\bf 671} (2009) 462
  [arXiv:0809.2285 [hep-ph]].

\bibitem{flavoureffects}
 A.~Abada, S.~Davidson, F.~-X.~Josse-Michaux, M.~Losada and A.~Riotto,
  {\em Flavor issues in leptogenesis},
  JCAP {\bf 0604} (2006) 004;
E.~Nardi, Y.~Nir, E.~Roulet and J.~Racker,
 {\em The Importance of flavor in leptogenesis},
  JHEP {\bf 0601} (2006) 164.
  

\bibitem{flavourlep}
S.~Blanchet and P.~Di Bari,
  {\em Flavor effects on leptogenesis predictions},
  JCAP {\bf 0703} (2007) 018
  [hep-ph/0607330].
  
  
  
\bibitem{vives}
O.~Vives,
 {\em Flavor dependence of CP asymmetries and thermal 
 leptogenesis with strong right-handed neutrino mass hierarchy},
  Phys.\ Rev.\ D {\bf 73} (2006) 073006
  [hep-ph/0512160].

\bibitem{riotto2}
P.~Di Bari and A.~Riotto,
  {\em Testing SO(10)-inspired leptogenesis with low energy neutrino experiments},
  JCAP {\bf 1104} (2011) 037
  [arXiv:1012.2343 [hep-ph]].


\bibitem{susy}
P.~Di Bari and M.~Re Fiorentin,
 {\em Supersymmetric $SO(10)$-inspired leptogenesis and a new $N_2$-dominated scenario},
  JCAP {\bf 1603} (2016) 039
  [arXiv:1512.06739 [hep-ph]].

\bibitem{DOnofrio:2014rug}
M.~D'Onofrio, K.~Rummukainen and A.~Tranberg,
{\em Sphaleron Rate in the Minimal Standard Model},
Phys. Rev. Lett. \textbf{113} (2014) no.14, 141602
[arXiv:1404.3565 [hep-ph]].


\bibitem{nufit24}
I.~Esteban, M.~C.~Gonzalez-Garcia, M.~Maltoni, I.~Martinez-Soler, J.~P.~Pinheiro and T.~Schwetz,
[arXiv:2410.05380 [hep-ph]].
  


\bibitem{Planck:2018vyg}
N.~Aghanim \textit{et al.} [Planck],
{\em Planck 2018 results. VI. Cosmological parameters},
Astron. Astrophys. \textbf{641} (2020), A6
[erratum: Astron. Astrophys. \textbf{652} (2021), C4]
[arXiv:1807.06209 [astro-ph.CO]].


\bibitem{Allali:2024aiv}
I.~J.~Allali and A.~Notari,
[arXiv:2406.14554 [astro-ph.CO]].

\bibitem{Naredo-Tuero:2024sgf}
D.~Naredo-Tuero, M.~Escudero, E.~Fern\'andez-Mart\'\i{}nez, X.~Marcano and V.~Poulin,
{\em Living at the Edge: A Critical Look at the Cosmological Neutrino Mass Bound},
[arXiv:2407.13831 [astro-ph.CO]].

\bibitem{DESI:2025ejh}
W.~Elbers \textit{et al.} [DESI],
{\em Constraints on Neutrino Physics from DESI DR2 BAO and DR1 Full Shape},
[arXiv:2503.14744 [astro-ph.CO]].

\bibitem{Super-Kamiokande:2023ahc}
T.~Wester \textit{et al.} [Super-Kamiokande],
{\em Atmospheric neutrino oscillation analysis with neutron tagging and an expanded fiducial volume in Super-Kamiokande I\textendash{}V},
Phys. Rev. D \textbf{109} (2024) no.7, 072014
[arXiv:2311.05105 [hep-ex]].
 


\bibitem{DiBari:2020plh}
P.~Di Bari and R.~Samanta, {\em The $SO(10)$-inspired leptogenesis timely opportunity},
JHEP \textbf{08} (2020), 124
[arXiv:2005.03057 [hep-ph]].
 
 
 \bibitem{ST}
P.~Di Bari and L.~Marzola,
  {\em SO(10)-inspired solution to the problem of the initial conditions in leptogenesis},
  Nucl.\ Phys.\ B {\bf 877} (2013) 719
  [arXiv:1308.1107 [hep-ph]].

\bibitem{problem}
E.~Bertuzzo, P.~Di Bari and L.~Marzola,
{\em The problem of the initial conditions in flavoured leptogenesis and the tauon $N_2$-dominated scenario},
  Nucl.\ Phys.\ B {\bf 849} (2011) 521
  [arXiv:1007.1641 [hep-ph]].
  
\bibitem{sophie}
P.~Di Bari, S.~King and M.~Re Fiorentin,
  {\em Strong thermal leptogenesis and the absolute neutrino mass scale},
  JCAP {\bf 1403} (2014) 050
  [arXiv:1401.6185 [hep-ph]].


\bibitem{full}
P.~Di Bari and M.~Re Fiorentin,
  {\em A full analytic solution of $SO(10)$-inspired leptogenesis},
  JHEP {\bf 1710} (2017) 029
  [arXiv:1705.01935 [hep-ph]].



\bibitem{Barbieri:1999ma}
R.~Barbieri, P.~Creminelli, A.~Strumia and N.~Tetradis,
{\em Baryogenesis through leptogenesis},
Nucl. Phys. B \textbf{575} (2000), 61-77
[arXiv:hep-ph/9911315 [hep-ph]].

\bibitem{Buchmuller:2001sr}
W.~Buchmuller and M.~Plumacher,
{\em Spectator processes and baryogenesis},
Phys. Lett. B \textbf{511} (2001), 74-76
[arXiv:hep-ph/0104189 [hep-ph]].



\bibitem{fuller}
S.~Antusch, P.~Di Bari, D.~A.~Jones and S.~F.~King,
 {\em A fuller flavour treatment of $N_2$-dominated leptogenesis},
  Nucl.\ Phys.\ B {\bf 856} (2012) 180
  [arXiv:1003.5132 [hep-ph]].


\bibitem{Garbrecht:2014kda}
B.~Garbrecht and P.~Schwaller,
{\em Spectator Effects during Leptogenesis in the Strong Washout Regime},
JCAP \textbf{10} (2014), 012
[arXiv:1404.2915 [hep-ph]].

\bibitem{Garbrecht:2019zaa}
B.~Garbrecht, P.~Klose and C.~Tamarit,
{\em Relativistic and spectator effects in leptogenesis with heavy sterile neutrinos},
JHEP \textbf{02} (2020), 117
[arXiv:1904.09956 [hep-ph]].

\bibitem{Garayoa:2009my}
J.~Garayoa, S.~Pastor, T.~Pinto, N.~Rius and O.~Vives,
{\em On the full Boltzmann equations for Leptogenesis},
JCAP \textbf{09} (2009), 035
[arXiv:0905.4834 [hep-ph]].

\bibitem{Ballett:2016daj}
  P.~Ballett, S.~F.~King, S.~Pascoli, N.~W.~Prouse and T.~Wang,
  {\em Sensitivities and synergies of DUNE and T2HK},
  Phys.\ Rev.\ D {\bf 96} (2017) no.3,  033003
  [arXiv:1612.07275 [hep-ph]].


\bibitem{chianese}
M.~Chianese and P.~Di Bari,
 {\em Strong thermal $SO(10)$-inspired leptogenesis in the light of recent results from long-baseline neutrino experiments}, JHEP {\bf 1805} (2018) 073
  [arXiv:1802.07690 [hep-ph]].


\bibitem{DiBari:2015oca}
P.~Di Bari and S.~F.~King,
 {\em Successful $N_2$ leptogenesis with flavour coupling effects in realistic unified models},
  JCAP {\bf 1510} (2015) no.10,  008
  [arXiv:1507.06431];



\bibitem{KamLAND-Zen:2024eml}
S.~Abe \textit{et al.} [KamLAND-Zen],
{\em Search for Majorana Neutrinos with the Complete KamLAND-Zen Dataset},
[arXiv:2406.11438 [hep-ex]].
  
 \bibitem{Katrin:2024tvg}
M.~Aker \textit{et al.} [Katrin],
{\em Direct neutrino-mass measurement based on 259 days of KATRIN data},
[arXiv:2406.13516 [nucl-ex]].
  
\bibitem{fusaokakoide}
H.~Fusaoka and Y.~Koide,
{\em Updated estimate of running quark masses},
Phys.\ Rev.\ D \textbf{57} (1998), 3986-4001
[arXiv:hep-ph/9712201 [hep-ph]].
 
\bibitem{Casas:2001sr}
J.~A.~Casas and A.~Ibarra,
{\em Oscillating neutrinos and $\mu \to e, \gamma$},
Nucl. Phys. B \textbf{618} (2001), 171-204
[arXiv:hep-ph/0103065 [hep-ph]].
 

\bibitem{Buchmuller:2004nz}
W.~Buchmuller, P.~Di Bari and M.~Plumacher,
{\em Leptogenesis for pedestrians},
Annals Phys. \textbf{315} (2005), 305-351
[arXiv:hep-ph/0401240 [hep-ph]].
 
 
 
\bibitem{manton}
F.~R.~Klinkhamer and N.~S.~Manton,
{\em A Saddle Point Solution in the Weinberg-Salam Theory},
Phys. Rev. D \textbf{30} (1984), 2212
 
\bibitem{kuzmin}
V.~A.~Kuzmin, V.~A.~Rubakov and M.~E.~Shaposhnikov,
{\em On the Anomalous Electroweak Baryon Number Nonconservation in the Early Universe},
Phys. Lett. B \textbf{155} (1985), 36.


\bibitem{klebnikov}
S.~Y.~Khlebnikov and M.~E.~Shaposhnikov,
{\em The Statistical Theory of Anomalous Fermion Number Nonconservation},
Nucl. Phys. B \textbf{308} (1988), 885-912;


\bibitem{Harvey:1990qw}
J.~A.~Harvey and M.~S.~Turner,
{\em Cosmological baryon and lepton number in the presence of electroweak fermion number violation},
Phys. Rev. D \textbf{42} (1990), 3344-3349.



\bibitem{bounds}
S.~Blanchet and P.~Di Bari,
  {\em New aspects of leptogenesis bounds},
  Nucl.\ Phys.\ B {\bf 807} (2009) 155
  [arXiv:0807.0743 [hep-ph]].


\bibitem{density}
 S.~Blanchet, P.~Di Bari, D.~A.~Jones and L.~Marzola,
  {\em Leptogenesis with heavy neutrino flavours: from density matrix to Boltzmann equations},
  JCAP {\bf 1301} (2013) 041
  [arXiv:1112.4528 [hep-ph]].

\bibitem{crv}
  L.~Covi, E.~Roulet and F.~Vissani,
  {\em CP violating decays in leptogenesis scenarios},
  Phys.\ Lett.\ B {\bf 384} (1996) 169 [hep-ph/9605319].
  


\bibitem{Davidson:2008bu}
S.~Davidson, E.~Nardi and Y.~Nir, {\em Leptogenesis},
Phys. Rept. \textbf{466} (2008), 105-177
[arXiv:0802.2962 [hep-ph]].


\bibitem{Engelhard:2006yg}
G.~Engelhard, Y.~Grossman, E.~Nardi and Y.~Nir,
{\em The Importance of N2 leptogenesis},
Phys. Rev. Lett. \textbf{99} (2007), 081802
[arXiv:hep-ph/0612187 [hep-ph]].

\bibitem{Babu:2024ahk}
K.~S.~Babu, P.~Di Bari, C.~S.~Fong and S.~Saad,
{\em Leptogenesis in SO(10) with minimal Yukawa sector},
JHEP \textbf{10} (2024), 190
[arXiv:2409.03840 [hep-ph]].

\bibitem{Fong:2025aya}
C.~S.~Fong and K.~M.~Patel,
{\em Electroweak Triplet Scalar Contribution to $SO(10)$ Leptogenesis},
[arXiv:2505.06391 [hep-ph]].

\bibitem{Malinsky:2025jwg}
M.~Malinsk{\'y},
{\em Thermal leptogenesis in minimal unified models},
[arXiv:2506.23117 [hep-ph]].

\end{thebibliography}
\end{document}